\documentclass[
  reprint,
  aps,
  prd,
  superscriptaddress,
  showkeys,
  nofootinbib,
]{revtex4-2}

\usepackage{graphicx}
\usepackage{dcolumn}
\usepackage[T1]{fontenc}
\usepackage{booktabs}
\usepackage{multirow}
\usepackage{textalpha}
\usepackage{siunitx}
\usepackage{bm}
\usepackage[dvipsnames]{xcolor}
\usepackage{orcidlink}
\usepackage[normalem]{ulem}
\usepackage{amsmath}
\usepackage{graphicx}
\usepackage[noabbre]{}
\usepackage{lineno}

\newcommand{\lya}{Ly$\alpha$}
\newcommand{\hMpc}{h^{-1}\text{Mpc}}

\newcommand{\clpt}{\texttt{CoLoRe 2LPT}}
\newcommand{\abacus}{\texttt{AbacusSummit}}
\newcommand{\phip}{$\phi_\mathrm{p}$}
\newcommand{\phis}{$\phi_\mathrm{s}$}
\newcommand{\phif}{$\phi_\mathrm{f}$}
\newcommand{\alphap}{$\alpha_\mathrm{p}$}
\newcommand{\alphas}{$\alpha_\mathrm{s}$}
\newcommand{\alphaf}{$\alpha_\mathrm{f}$}
\newcommand{\fsig}{$f\sigma_8$}
\newcommand{\lcdm}{$\Lambda$CDM}
\newcommand{\ciii}{C\,{\textsc{iii}}}
\newcommand{\civ}{C\,{\textsc{iv}}}
\newcommand{\siii}{Si\,{\textsc{ii}}}
\newcommand{\siiii}{Si\,{\textsc{iii}}}
\newcommand{\mgii}{Mg\,{\textsc{ii}}}
\newcommand{\heii}{He\,{\textsc{ii}}}

\begin{document}

\preprint{APS/123-QED}

\title{Validation of the DESI DR2 \texorpdfstring{Ly$\alpha$}{Lyα} forest full-shape analysis}

\author{M.~Herbold\orcidlink{0009-0000-8112-765X}}
\affiliation{Center for Cosmology and AstroParticle Physics, The Ohio State University, 191 West Woodruff Avenue, Columbus, OH 43210, USA}
\affiliation{Department of Physics, The Ohio State University, 191 West Woodruff Avenue, Columbus, OH 43210, USA}
\affiliation{The Ohio State University, Columbus, 43210 OH, USA}

\author{A.~Cuceu\orcidlink{0000-0002-2169-0595}}
\affiliation{Lawrence Berkeley National Laboratory, 1 Cyclotron Road, Berkeley, CA 94720, USA}
\affiliation{NASA Einstein Fellow}

\author{P.~Martini\orcidlink{0000-0002-4279-4182}}
\affiliation{Center for Cosmology and AstroParticle Physics, The Ohio State University, 191 West Woodruff Avenue, Columbus, OH 43210, USA}
\affiliation{Department of Astronomy, The Ohio State University, 4055 McPherson Laboratory, 140 W 18th Avenue, Columbus, OH 43210, USA}
\affiliation{The Ohio State University, Columbus, 43210 OH, USA}

\author{H.~K.~Herrera-Alcantar\orcidlink{0000-0002-9136-9609}}
\affiliation{Institut d'Astrophysique de Paris. 98 bis boulevard Arago. 75014 Paris, France}
\affiliation{IRFU, CEA, Universit\'{e} Paris-Saclay, F-91191 Gif-sur-Yvette, France}

\author{J.~Guy\orcidlink{0000-0001-9822-6793}}
\affiliation{Lawrence Berkeley National Laboratory, 1 Cyclotron Road, Berkeley, CA 94720, USA}

\author{C.~Gordon\orcidlink{0000-0003-2561-5733}}
\affiliation{Department of Physics \& Astronomy, University College London, Gower Street, London, WC1E 6BT, UK}

\author{O.~Manasoiu}
\affiliation{Department of Physics \& Astronomy, University College London, Gower Street, London, WC1E 6BT, UK}

\author{J.~Aguilar}
\affiliation{Lawrence Berkeley National Laboratory, 1 Cyclotron Road, Berkeley, CA 94720, USA}

\author{S.~Ahlen\orcidlink{0000-0001-6098-7247}}
\affiliation{Department of Physics, Boston University, 590 Commonwealth Avenue, Boston, MA 02215 USA}

\author{O.~Alves}
\affiliation{University of Michigan, 500 S. State Street, Ann Arbor, MI 48109, USA}

% ...

\collaboration{DESI Collaboration}

\date{\today}

\begin{abstract}
    We present the validation of the Dark Energy Spectroscopic Instrument (DESI) Data Release 2 (DR2) Lyman-$\alpha$ forest full-shape analysis. This analysis combines measurements of the three-dimensional Lyman-$\alpha$ forest auto-correlations, as well as cross-correlations with quasar positions, to extract information from both the baryon acoustic oscillation (BAO) feature and the broadband clustering signal. We primarily focus on the Alcock-Paczynski (AP) measurement from these correlations. Compared to the DESI DR1 full-shape analysis, the DR2 validation employs substantially larger and more realistic synthetic datasets (mocks), including \clpt\ and \abacus\ Lyman-$\alpha$ forest simulations, which include more realistic small-scale Lyman-$\alpha$ and quasar clustering. We additionally incorporate several improvements into the modeling framework, most notably analytic marginalization over small-scales ($<10\,\hMpc$) that could otherwise bias larger scales due to the distortion matrix, and the impact of ultraviolet background fluctuations. The validation program was developed prior to unblinding to establish the robustness of the final measurements. We define quantitative validation criteria for the cosmological parameters of interest, and evaluate them using hundreds of mock realizations. We further test the analysis through independent fits to the auto- and cross-correlation functions, multiple splits of the quasar and forest catalogs, and apply a large suite of analysis and modeling variations to both mock and blinded observational data. We find that the BAO and AP parameters satisfy all validation requirements and remain stable across the full suite of tests. In contrast, mock studies reveal a significant bias in the inferred growth-rate parameter \fsig, which leads us to exclude this measurement from the final DR2 analysis. The agreement observed across mock challenges, data splits, and robustness tests demonstrates that the DR2 Lyman-$\alpha$ full-shape analysis provides a reliable and substantially improved broadband AP measurement relative to previous Lyman-$\alpha$ forest studies.
    
    \textbf{ArXiv ePrint: } \href{https://arxiv.org}{xxxx.xxxxx}

\end{abstract}


\maketitle

\section{Introduction}\label{Section: Introduction}

Understanding the origin and nature of the accelerating expansion of the universe and the evolution of cosmic structure are some of the outstanding questions of modern cosmology. Surveys that probe large scale structure provide a unique and powerful approach to addressing these questions by mapping the spatial distribution of matter across cosmic time, as well as measuring the imprint of various physical processes, such as through the clustering of galaxies and intergalactic gas. The most common measures of large-scale clustering are power spectra and correlation functions, which encode information about the expansion history of the universe, structure growth, and the underlying cosmology. Extracting and understanding these data with growing precision requires increasingly large datasets, as well as robust modeling frameworks that are capable of capturing the full statistical information of these measurements. 

The Dark Energy Spectroscopic Instrument (DESI) is a Stage-IV spectroscopic survey designed to investigate the nature of dark energy and the growth of cosmic structure through precision measurements of the large-scale distribution of matter \cite{2006_dark_energy_report, DESI2016a.Science}. DESI is in the midst of an eight-year survey that aims to collect spectra from over 60 million galaxies and quasars in its 17000 deg$^2$ footprint, producing the largest three-dimensional spectroscopic map of the Universe to date \cite{Snowmass2013.Levi, SurveyOps.Schlafly.2023}.. The first DESI Data Release (DR1) included measurements from the first 13 months of the main survey \cite{DESI2024.I.DR1}. Those observations were a key part of the most statistically significant measurements of the equation of state of dark energy to date \cite{DESI2024.VI.KP7A, DESI2024.VII.KP7B} based on measurements of baryon acoustic oscillations (BAO) and redshift space distortions from galaxies, quasars, and the Lyman-$\alpha$ (\lya) forest \cite{DESI2024.III.KP4,DESI2024.IV.KP6}. The second DESI Data Release (DR2) includes measurements from the first three years of the main survey \cite{DESI.DR2.DR2}. The first DESI DR2 results published last year \cite{DESI.DR2.BAO.lya,DESI.DR2.BAO.cosmo} extended the DR1 results with more statistically significant BAO measurements and analysis improvements. The present paper validates additional analysis of the DR2 correlation function, which is presented in the DESI DR2 \lya\ forest full-shape paper \cite{Cuceu:2026}. 

The \lya\ forest provides a unique probe of large-scale structure at high redshift by tracing the absorption of light from distant quasars by foreground neutral hydrogen in the intergalactic medium (IGM). Because each quasar spectrum traces the matter density field along a given line-of-sight, collections of \lya\ forest spectra provide a three-dimensional map of matter fluctuations at redshifts $z \gtrsim 2$, where measurements using traditional tracers become increasingly challenging \cite{McQuinn2016}. The \lya\ forest has therefore played an important role in extending measurements in precision cosmology to higher redshifts than are accessible to galaxy surveys, such as with measurements of the baryon acoustic oscillation (BAO) scale from both the \lya\ forest flux auto-correlation and its cross-correlation with the positions of quasars. This technique was originally developed using data from the Baryon Oscillation Spectroscopic Survey (BOSS) as part of the Sloan Digital Sky Survey (SDSS) \cite{Dawson2013, Busca_2013, Slosar_2013} and extended BOSS (eBOSS) \cite{Dawson2016, dMdB2020}. Those early measurements established the \lya\ forest as a powerful probe for constraining the expansion history of the universe, and recent DESI observations have further improved the statistical precision of these measurements \cite{DESI2024.IV.KP6, DESI.DR2.BAO.lya}. 

However, the \lya\ forest contains additional cosmological information beyond the BAO scale. This information can be accessed through the full shape of the clustering signal, including information from the broadband shape of the correlation functions and the anisotropic distortions induced by structure growth. These measurements provide sensitivity to both the geometry of the universe through the Alcock-Paczynski (AP) effect and the growth of cosmic structure through redshift-space distortions (RSD). Galaxy full-shape analyses have successfully used RSD to measure \fsig\ \cite{Reid_2012, Alam_2017, Alam_2021, DESI2024.VII.KP7B}, however, extracting cosmological information from RSD is more challenging in the \lya\ forest than in galaxy surveys. The \lya\ forest is affected by an unknown velocity divergence bias, due to the RSD sensitivity to the growth rate of cosmic structure and the amplitude of matter fluctuations, that strongly degrades constraints on \fsig\ \cite{McDonald_2023, Seljak_2012, Chen_2021, Ivanov_2024}. This degeneracy between the \lya\ and QSO RSD signals in the \lya-QSO cross-correlation can be partially broken by jointly fitting the \lya\ auto- and cross-correlation functions \cite{Cuceu2021}. Furthermore, unlike galaxy full-shape analyses, which constrain a single isotropic dilation parameter, the \lya\ full-shape analysis separates the isotropic AP scaling into peak (\alphap) and broadband (smooth) (\alphas) components. Marginalizing over \alphas reduces sensitivity to isotropic broadband-specific systematics while preserving the cosmological information contained in the BAO peak. The DESI DR2 \lya\ key paper \cite{Cuceu:2026} on the full-shape of the \lya\ correlation function builds on the DESI DR1 \lya\ forest full-shape measurement \cite{Cuceu:2025}, by utilizing both the increased statistical power and improved data quality of DR2, as well as a number of improvements in the modeling and analysis choices.

The increased cosmological information provided by full-shape analyses also introduces additional challenges compared to BAO measurements. By utilizing the broadband shape and anisotropic dependence of the correlation functions, full-shape analyses extract information from a wider range of scales, and are therefore sensitive to the accuracy of the theoretical modeling, observational systematics, and analysis choices. Establishing the robustness of the resulting cosmological constraints therefore requires a comprehensive validation program beyond the standard measurement pipeline. In this work, we validate the DESI DR2 \lya\ forest full-shape analysis using both synthetic spectra and blinded observational data. Synthetic mock datasets are used to assess the accuracy of the modeling framework, quantify potential parameter biases, and verify the calibration of statistical uncertainties, while blinded data analyses provide an independent assessment of the robustness of the measurements to changes in data selection and analysis choices without knowledge of the final cosmological results. These tests include internal consistency checks using independent data splits,  and alternative analysis configurations designed to verify the robustness of the pipeline and identify potential sensitivities to modeling assumptions at various stages of the analysis. The goal of this work is to demonstrate the reliability and robustness of the DESI DR2 \lya\ forest full-shape measurements and the subsequent cosmological interpretation presented in the companion analysis paper \cite{Cuceu:2026}.
    
The remainder of this paper is organized as follows. In Section \ref{section: data}, we describe the DESI DR2 dataset and the catalogs used in this analysis. Section \ref{section: analysis} summarizes the analysis methodology, including the measurement and modeling of the \lya\ correlation functions, the parameterization and fitting of cosmological constraints, and the baseline model used throughout this work. In Section \ref{Section: Mocks}, we present validation tests performed using synthetic mock datasets, including assessments of parameter biases, uncertainty estimates, and model performance. Section \ref{section: data validation} presents validation tests performed on blinded observational data, including catalog splits and alternative analysis choices designed to assess the robustness of the measurements. In Section \ref{Section: Discussion}, we discuss the primary findings of the validation campaign and investigate the origin of any notable discrepancies observed in the tests. Finally, Section \ref{Section: Conclusion} summarizes our findings and the implications for the DESI DR2 \lya\ full-shape cosmological analysis.

\section{Data} \label{section: data}

The DESI survey is conducted with the 4-meter Mayall Telescope at Kitt Peak National Observatory and a highly multiplexed fiber spectrograph. DESI combines a wide-field optical corrector, 5000 robotic fiber positioners, and a high-throughput spectroscopic system capable of simultaneously obtaining thousands of spectra across a three-degree diameter field of view \cite{DESI2016b.Instr}. This unique combination makes DESI extraordinarily efficient, sometimes measuring more than 100,000 galaxy spectra per night. A number of papers describe the DESI instrumentation and science goals \cite{DESI2022.KP1.Instr}, the focal plane system \cite{FocalPlane.Silber.2023}, optical corrector \cite{Corrector.Miller.2023}, and fiber system \cite{FiberSystem.Poppett.2024}.

DESI targets were selected from the DESI Legacy Imaging Surveys \cite{BASS.Zou.2017,LS.Overview.Dey.2019}. The quasar target selection algorithms are described in \cite{QSOPrelim.Yeche.2020, QSO.TS.Chaussidon.2023}, are integrated into the target selection pipeline \cite{TS.Pipeline.Myers.2023}, and were extensively tested during the survey validation period \cite{DESI2023a.KP1.SV} before the main survey began. This early survey validation period combined with the first two months of observations resulted in the DESI Early Data Release (EDR) \cite{DESI2023b.KP1.EDR}, which gave rise to many BAO scientific results using galaxies, quasars, and the \lya\ forest \cite{BAO.EDR.Moon.2023, Gordon2023}. The EDR combined with the first year of DESI observations resulted in data release 1 (DR1) \cite{DESI2024.I.DR1}, which supported many key science papers. Those included the measurement and validation of the two-point clustering of galaxies and quasars \cite{DESI2024.II.KP3}, BAO measurements of galaxies and quasars \cite{DESI2024.III.KP4} and the \lya\ forest \cite{DESI2024.IV.KP6}, and the full-shape of the correlation functions of galaxies and quasars \cite{DESI2024.V.KP5}. Additional DR1 papers explored the cosmological interpretation of the BAO measurements \cite{DESI2024.VI.KP7A}, \lya\ full-shape \cite{Cuceu:2025}, and galaxy and quasar measurements of both BAO and full-shape information \cite{DESI2024.VII.KP7B}. 
    
In this work, we utilize data from the first three years of the DESI survey, which will be publicly released as data release 2 (DR2). In addition to a much larger footprint (comprising $\sim 70\%$ of the full dark time survey), DR2 $z>2$ quasars are re-observed on average three times (compared to $\sim$1-2 times in DR1), resulting in higher SNR per quasar. It is worth noting that DR2 will be released in two versions, \textsc{kibo} and \textsc{loa}. \textsc{Kibo} is the predecessor to \textsc{loa}, and was used for the majority of the DR2 BAO analysis validation \cite{DESI.DR2.BAO.lya}. However, the \textsc{loa} release implemented a number of software related fixes such as the weighting in coadded spectra, and is the release used in this analysis.

\subsection{Spectra} \label{section:spectra}
    
DESI spectra are collected by ten three-channel spectrographs, and each channel has a distinct diffraction grating design and resolution range. The blue channel, which is most relevant for \lya\ analyses, has a spectral resolution range that varies from $\sim 2000 - 3000$. The red and infrared channels have resolutions that vary from approximately 3500 - 4500 and 4000 - 5500, respectively. The total wavelength range is from 3600 - 9800 \AA, contributed by the blue (3600-5930 \AA), red (5600-7720 \AA), and infrared (7470-9800 \AA) arms of each spectrograph. Data collected at the observatory are transferred to the National Energy Research Scientific Computing Center (NERSC) for processing and analysis. 

The spectroscopic pipeline \cite{Spectro.Pipeline.Guy.2023} processes the spectrum from each source to produce a 1-D spectrum with 0.8\AA\ per pixel dispersion that is used for subsequent analyses. The spectral data products contain information about masked pixels, instrumental noise, sky noise, and the spectrograph resolution. We estimate target redshifts using \textsc{Redrock} \cite{Redrock.Bailey.2024, Anand_2024}, which fits with spectral templates, including high redshift quasars \cite{Brodzeller_2023}. Additionally, we identify quasars missed by \textsc{redrock} by searching for \mgii\ emission in quasar targets classified as galaxies by \textsc{redrock} \cite{QSO.TS.Chaussidon.2023} and using a convolutional neural network quasar classifier, \textsc{QuasarNet}, which has been developed by \cite{quasar_net_2018} and optimized for DESI spectra \cite{green_2025} based on visual inspection during validation \cite{VIQSO.Alexander.2023}.

Because DESI prioritizes quasars for \lya\ forest analyses, these targets are re-observed on multiple passes over the footprint to increase SNR. Repeated observations of any sources are coadded and organized by \textsc{HEALPix} pixel \cite{Gorski_2005}.  

\subsection{Catalogs} \label{catalogs}

The DESI DR2 quasar catalog is constructed using the same method described in \cite{DESI2024.IV.KP6} and is the same catalog that is used for \cite{DESI.DR2.BAO.lya}. This catalog contains quasars classified by each of the three classifiers that do not have problems identified by the spectroscopic pipeline. The DR2 quasar catalog has 1,289,874 quasars with $z > 1.77$ and 824,989 quasars with $z > 2.09$. 

We construct two auxiliary catalogs prior to measuring the \lya\ forest: a Damped \lya\ Absorption (DLA) system catalog and a Broad Absorption Line (BAL) quasar catalog. DLA systems have neutral hydrogen densities $\log_{10} \left( N_{HI}\right)>20.3\,\mathrm{cm}^{-2}$, which lead to wide damping wings that can impact the continuum level and complicate correlation modeling due to increased clustering \cite{Font_Ribera_2012}. DR2 \lya\ analyses utilize three methods to identify and characterize DLAs: a convolutional neural network \cite{Wang2022}, Gaussian Process \cite{Ho21}, and template-based approach \cite{brodzeller2025}. The DR2 full-shape paper \cite{Cuceu:2026} briefly describes the construction of the DLA catalog with an emphasis on changes relative to DR2 \lya\ BAO. BAL quasars contain absorption troughs that are associated with many emission features \cite{Ennesser_2022}, which can add absorption unrelated to the IGM matter distribution and introduce redshift errors \cite{garcia_2023, 2023arXiv230903434F}. BAL troughs associated with emission lines in the forest region are identified and masked based on the velocities of the \civ\ absorption troughs. This method is described in detail in a DESI DR1 supporting paper \cite{KP6s9-Martini}. 

\subsection{Continuum Fitting}

The final part of the data selection process involves measuring flux decrements in the forest region. We restrict our measurement to 3600 - 5772 \AA\ in the observed frame, where the lower bound is set by the minimum wavelength of the blue spectrograph channel, and the upper bound corresponds to the central overlap region of the blue and red spectrograph channels. We measure the \lya\ absorption of the forest in two different regions of the rest-frame of the quasar, which we will refer to as Region A (1040 - 1205 \AA), and Region B (920 - 1020 \AA). Because the B region is associated with higher redshift quasars, these spectra typically have lower SNR. Some of Region B also includes higher order Lyman lines. We therefore analyze the A and B region correlations separately. 

To measure the fluctuations, we first mask regions of the forest associated with bad pixels (e.g., cosmic rays) or astrophysical contaminants (e.g., BALs, DLAs). Additionally, forests with less than the minimum path length ($< 120$\,\AA) to fit the continuum are discarded. Then we apply a small correction to the spectro-photometric calibration and instrumental noise estimates by calculating the mean transmitted flux in the rest-frame \ciii\ region (1600 - 1850 \AA). Finally, the remainder of the analysis utilizes the transmitted flux field $\delta_q(\lambda)$, which relates the observed and expected flux: 

    \begin{equation}
        \delta_q(\lambda) = \frac{f_q(\lambda)}{\overline{F}(\lambda)C_q(\lambda)} -1
    \end{equation}    

\noindent
where $\overline{F}(\lambda)$ is the mean transmitted flux and $C_q(\lambda)$ is the unabsorbed continuum of quasar $q$. The continuum-fitting procedure assumes a universal mean continuum shape as a function of rest-frame wavelength and models quasar-to-quasar variations through two nuisance parameters that account for spectral diversity and luminosity dependence. The mean continuum shape and nuisance parameters are determined simultaneously through a maximum-likelihood fit that incorporates both instrumental noise and the intrinsic variance of the \lya\ forest. Further details of the \ciii\ correction and continuum-estimation procedure are given in \cite{Ramirez2024}.

\section{Analysis} \label{section: analysis}

\subsection{Measurement of Correlations}

Using the catalog of quasar positions and redshifts, and the transmitted flux field (deltas) of the A and B regions described previously, we measure four correlation functions following \cite{dMdB2020}. The correlations include: the auto-correlation of the \lya\ forest measured in the A and B regions ($\mathrm{Ly}\alpha(\mathrm{A})\times\mathrm{Ly}\alpha(\mathrm{A})$ and $\mathrm{Ly}\alpha(\mathrm{A})\times\mathrm{Ly}\alpha(\mathrm{B})$, respectively), and the cross-correlation of the \lya\ forest measured in the A and B regions with quasar positions ($\mathrm{Ly}\alpha(\mathrm{A})\times\mathrm{QSO}$ and $\mathrm{Ly}\alpha(\mathrm{B})\times\mathrm{QSO}$), respectively). We then use a comoving separation grid along ($\mathrm{r}_\parallel$) and across ($\mathrm{r}_\perp$) the line-of-sight, as well as convert angles and redshifts using fiducial cosmology (flat $\Lambda$CDM from Planck 2018 \cite{Planck2018}). We measure the correlations using a bin size of $4\ \hMpc$ extending to separations of $200 \ \hMpc$. We use the same method to measure correlations as in DR1 and DR2 BAO \cite{DESI2024.IV.KP6, DESI.DR2.BAO.lya}, which is detailed in \cite{Gordon2023} and builds on previous work by BOSS and eBOSS \cite{Bautista2017, dMdB2020}. 

We then compute the covariance matrix using the  HEALPix pixel subsampling method of \cite{dMdB2020}, where the size of a HEALPix pixel is about $(250\ \hMpc)^2$ at an effective redshift $z\sim2.3$ for a choice of $\mathrm{NSIDE=16}$. The correlation functions are measured independently in unique sub-samples of the sky and combined using weighted averages, with the covariance estimated from the scatter among the regional measurements (i.e. replacing the unknown covariance of a given sub-sample with the square of its difference with the computed average). We then smooth the covariance by averaging the correlation-matrix elements that share line-of-sight transverse separations, and replacing all non-diagonal elements by their corresponding average, which is detailed in \cite{KP6s6-Cuceu}. We use this sub-sampling technique to measure the covariance of the full data set, composed of all four correlation functions. 

\subsection{Modeling of Correlations}

In this section, we briefly describe how the correlations and main contaminants are modeled, which are described in greater detail in \cite{Gordon2023, KP6s5-Guy, KP6s6-Cuceu}. 

The continuum fitting method described in the previous section introduces a distortion of the measured \lya\ fluctuation field \cite{Slosar_2011}. As a result, the measured correlation function is a distorted version of the true underlying correlation function, with the largest impact on modes oriented along the line-of-sight. Following previous DESI analyses \cite{DESI2024.IV.KP6}, and using the formalism introduced by \cite{Bautista2017}, we apply the same transformation to the model using a distortion matrix that relates the undistorted and observed correlation functions. The distortion matrix $D_{MN}$ multiplies the undistorted correlation function model $\xi(r_{||},r_\bot)$ to give the distorted correlation:

    \begin{equation}
        \hat\xi_M = \sum_N D_{MN} \xi_N,
    \end{equation} \label{eq. dmat}

\noindent
where $N$ are the undistorted model bins, and $M$ are the data bins. The auto- and cross-correlation elements are given by equations (21) and (22) of \cite{dMdB2020}, respectively. It is worth noting that the comoving separation bin size of the model ($2\ \hMpc$) is half as large as the bin size used for the data ($4\ \hMpc$), and the line-of-sight separation is extended to $300\ \hMpc$ rather than $200 \ \hMpc$. 

The distortion matrix is computed from the survey geometry and continuum-fitting weights discussed in the previous section, and it is incorporated into all model predictions prior to fitting and parameter inference. Because the distortion matrix is computationally expensive, we use only a fraction (1\%) of the \lya\ pixels as an approximation \cite{Bautista2017, Gordon2023}. However, in \S \ref{Section: Correlations}, we verify that using larger samples of the data, as well as alternative binning schemes and scale cuts, have negligible effect on the measurement. Additionally, the calculation also accounts for redshift evolution of the \lya\ and quasar clustering amplitudes, which is approximated by a power law. Tests on synthetic data demonstrate that this improved treatment provides a better description of the measured correlations while producing negligible impacts on the recovered cosmological parameters. A more detailed description of the continuum fit distortion can be found in \cite{Busca2025}. 

In addition to \lya\ absorption, the observed flux contains contributions from several foreground absorber transitions along the line-of-sight, which we call metals, such as \siii, \siiii, \civ\ \cite{Gordon2023}. If an absorption feature's true and assumed comoving separations do not match, such as when assuming that a metal feature is from the \lya\ transition, then we observe spurious correlations which appear as peaks in the measured correlation function \cite{DESI2024.IV.KP6}. These lines are especially prominent when stacking spectra centered around strong absorbers \cite{Pieri_2014}. We model this effect using the metal-matrix formalism introduced in previous DESI analyses, such as \cite{DESI2024.IV.KP6, DESI.DR2.BAO.lya}.  

The metal matrix is a collection of weighted pairs, which quantifies how correlations measured under the assumption that the absorber is from the \lya\ transition, are related to the true separations of pairs involving metal transitions. This is incorporated directly into the model prediction, allowing the contribution of each absorber type to be propagated into the observed correlation function. Following the DR2 BAO analysis, we compute the metal matrix as a function of both the line-of-sight and transverse components, which is an improvement over the previous approach which only accounted for the line-of-sight. Although we find that this additional correction has negligible impact on the cosmological constraints, it provides a more accurate representation of metal contamination in the measured correlations.

In addition, the model includes several nuisance contributions known to affect \lya\ forest correlations. Correlated noise introduced during the spectroscopic data processing \cite{Spectro.Pipeline.Guy.2023} is modeled following \cite{DESI2024.IV.KP6} and detailed in \cite{KP6s5-Guy}, where the dominant contribution arises from the sky-background noise, and primarily affects pairs of spectra observed on the same spectrograph. We also account for contamination from high-column-density (HCD) absorbers, including Lyman-limit systems and DLAs, whose broad damping wings introduce additional large-scale correlations along the line-of-sight \cite{McQuinn_2011, Font-Ribera_2012, Rogers_2018}. Although identified DLAs are masked during data processing (see \S \ref{catalogs}), residual contamination from unidentified HCDs is modeled through a scale-dependent modification of the \lya\ clustering signal (see \cite{tan2025}).

For the \lya\ $\times$ QSO cross-correlations, the model additionally includes the impact of quasar redshift uncertainties, which can smooth the clustering signal along the line-of-sight, as well as introduce a possible systematic redshift offset between quasars and the \lya\ forest reference frame \cite{Youles2022, Gordon2025}. We additionally model the transverse proximity effect following \cite{Font_Ribera_2012}, which accounts for the modification of the surrounding intergalactic medium by ionizing radiation from quasars. Finally, following \cite{Arinyo2015}, we incorporate a small-scale nonlinear correction to account for structure growth, peculiar velocities, and thermal broadening of the intergalactic medium. 
    
    \subsection{Fitting Alcock-Paczynski and Growth Rate}

The Alcock-Paczynski (AP) effect \cite{AlcockPaczynski1979} arises when angular separations and redshift differences are converted into comoving distances using an incorrect assumed cosmological model. In this case, intrinsically isotropic clustering appears anisotropically distorted along and across the line-of-sight. Deviations between the true and fiducial cosmologies can be modeled through anisotropic coordinate rescaling, which are related to the underlying cosmological distances. Following the convention introduced by \cite{Cuceu2021} adopted in the DR1 full-shape analysis \cite{Cuceu:2025}, we characterize the AP effect using the parameter:

    \begin{align}
    \phi \equiv \frac{q_\perp}{q_\parallel} = \frac{D_M H}{[D_M H]_{\rm fid}},
    \end{align}

where $q_\perp$ and $q_\parallel$ rescale coordinates along and across the line-of-sight, $D_{M}$ is the comoving angular diameter distance, and $H$ is the Hubble parameter. Where $\phi$ parametrizes the anisotropic rescaling of the correlation function, $\alpha\ \equiv \sqrt{q_{\perp}q_{\parallel}}$ parametrizes an isotropic rescaling. 

Additionally, $\phi$ and $\alpha$ contain both a peak and a smooth component, which we denote (\phip\, \phis\ ), and (\alphap\, \alphas\ ), respectively. Unlike BAO measurements, which extract AP information primarily from the anisotropic scaling of the BAO feature, the full-shape analysis additionally exploits the broadband anisotropy of the Lyman-$\alpha$ forest correlation functions, which enables significantly tighter AP constraints from the same dataset. We separately parameterize the AP information contained in the BAO peak feature and broadband clustering signal through the parameters \phip\ and \phis\, respectively. Although both the peak and smooth components are affected by $\phi$ in the same way, they are treated separately to better understand the cosmological value of rescaling the broadband separately.  However, the smooth component of $\alpha$ is less constrained, as it is likely not the only feature contributing to the isotropic scale of the broadband. Therefore, we do not focus on the measurement or interpretation of \alphas\ alone. 
    
In addition to AP information, the anisotropy of the correlation functions is sensitive to RSD, whose amplitude is governed by the growth-rate parameter \fsig. The model therefore simultaneously fits the AP and RSD contributions, together with the nuisance parameters. However, extensive tests on mock datasets (see \S \ref{Section: Mocks}) revealed a significant bias in the recovered \fsig\ measurement, and blinded data validation tests (see Appendix \ref{Appendix: Growth Rate}) further demonstrated the parameter's instability under reasonable analysis modeling choices. This parameter was consequently removed from the final DR2 cosmological analysis. The BAO parameters (\phip\, \alphap) and the AP parameter \phis\, however, were found to be robust across all validation tests and remain the primary focus of this work. 

\subsection{Changes Since DR2 BAO} \label{section: changes since dr2 bao}
    
This analysis uses the same DESI DR2 dataset used for the DR2 \lya\ BAO analysis \cite{DESI2024.IV.KP6}, which is detailed in \S \ref{section: data}. However, there are some changes to the catalogs and model used in this analysis. First, we use an improved DLA catalog, which is described in the DESI DR2 \lya\ forest full-shape paper \cite{Cuceu:2026}, and whose impact is assessed in \S\ref{Section: Fluctuations}. Second, we use a slightly modified version of the BAL catalog after modest code changes focused on increased computational efficiency of the BAL finding tool. Finally, we adopted two significant modeling changes, which are detailed below. 

The first significant change to our model since the DR2 \lya\ BAO analysis is marginalization over the small scales of the undistorted model. This effectively introduces scale cuts at the level of undistorted correlations rather than at the level of the fitted data, as in previous \lya\ analyses. While these scales are below the range of our model fit, they nevertheless impact larger separations due to distortion introduced by our continuum-fitting procedure. This distortion mixes information from these small scales into the larger-scale correlation function, meaning that conventional scale cuts do not fully remove their influence.This small-scale contamination can cause biases in the measured parameters, which we have identified by running fits with the same scale cuts on both true-continuum and continuum-fitted mocks and comparing the results. To eliminate this effect, we analytically marginalize over the undistorted correlation function at $r<r_{\rm min}$, where $r_{\rm min} \geq 30\, \hMpc$ and $40\,\hMpc$ for the auto- and cross-correlations, respectively, in the baseline analysis (see \S \ref{subsec: mock_stack_res} for validation of these scale-cuts). This is achieved by introducing a set of linear template amplitudes, and incorporating their uncertainty into a modified covariance matrix while retaining the constraining power of the excluded small-scale data. This procedure removes biases arising from continuum-fitting distortions without increasing the effective number of degrees of freedom in the fit. The complete formalism and implementation are described in \cite{Cuceu:2026}. 

The second major change relative to the DR2 \lya\ BAO analysis is the inclusion of ultraviolet background (UVB) fluctuations. Previous analyses assumed a homogeneous ionizing UV background, but fluctuations in the UV radiation field are expected to produce a scale-dependent modification to the \lya\ bias on scales comparable to the mean free path of ionizing photons. During validation of our DR2 full-shape analysis, we found statistically significant evidence for this effect, motivating its inclusion in the baseline model. We model UVB fluctuations following \cite{Gontcho2014}, introducing a scale-dependent correction to the \lya\ bias while treating the UVB response parameter as a free parameter and fixing the remaining quantities to their fiducial values. Although this model significantly improves the quality of the fit, it has only a negligible impact on the recovered AP parameters, as demonstrated by the robustness tests presented in \S \ref{Section: Model Variations}. We also investigated more complex UVB models, including the source shot noise term of the UVB sources and the effect of \heii\ reionization, but found no statistically significant improvement. The complete formalism is described in \cite{Cuceu:2026}.

\subsection{Baseline Model} \label{section: Baseline}

The baseline DR2 full-shape analysis builds upon the models developed for the DESI DR2 BAO analysis \cite{DESI.DR2.BAO.lya} and the DR1 full-shape analysis \cite{Cuceu:2025}, while incorporating the methodological improvements described in \S \ref{section: changes since dr2 bao}. A complete description of the baseline model is provided in the DESI DR2 \lya\ forest full-shape paper \cite{Cuceu:2026}, including the full likelihood, parameterization, and modeling assumptions (see in particular Eq.9 for the non-linear \lya\ forest correction). Here we summarize only the baseline analysis choices that provide the reference point for the validation tests presented throughout the remainder of this work.

We fit the \lya\ auto- and \lya-QSO cross-correlations over the range $30<r<200\ \hMpc$ and $40<r<200\ \hMpc$, respectively, simultaneously sampling independent peak (\alphap, \phip) and smooth (\alphas, \phis) parameters together with the growth rate $f$, while fixing $\sigma_8$ to its fiducial value so that \fsig\ is obtained as a derived parameter following \cite{GilMarin2020},  including the appropriate coordinate-rescaling correction. Gaussian priors are applied to the nuisance parameters inherited from the DR2 BAO analysis, with an updated prior on $L_{\rm HCD}$, while the non-linear \lya\ forest correction is modeled using the Arinyo formalism \cite{Arinyo2015}. In the baseline analysis, the Arinyo nuisance parameters are fixed to values calibrated using \textsc{ForestFlow} simulations of the DESI DR1 P1D analysis \cite{ForestFlow_2025,2026arXiv260121432C}, and we validate the robustness of this choice with a number of tests in \S \ref{Section: Alternative Analyses}. Baseline posterior constraints are obtained with the \textsc{PolyChord} nested sampler, while the majority of mock and blinded robustness tests use the computationally cheaper \textsc{iminuit} fitter, which we find provides equivalent best-fit parameter shifts for the purposes of the validation tests. Further discussion on the comparison of sampler and fitter results can be found in Appendix C of \cite{Cuceu:2025}. 

    \begin{figure}
        \centering
        \includegraphics[width=0.9\columnwidth,keepaspectratio]{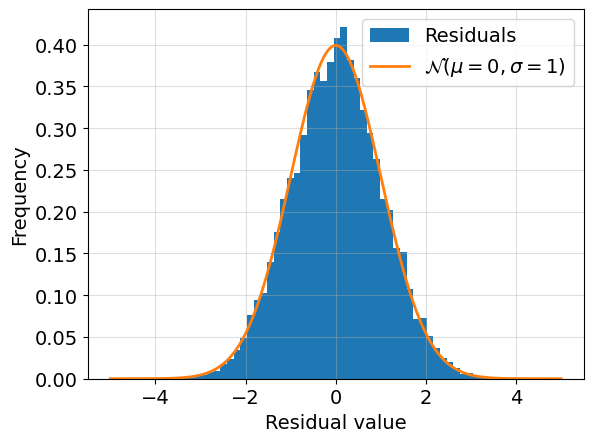}
        \caption{
        Comparison between the histogram of the distribution of normalized residuals of our measured correlation functions and best fitting model, with a normal Gaussian distribution with $\mu=0$ and $\sigma=1$. The distribution of residuals does not display significant deviation from Gaussianity, demonstrating that the model accurately describes the statistical properties of the observed correlations over the scales included in the analysis. 
        }
        \label{fig:residuals}
    \end{figure}

Before assessing the robustness of the analysis to alternative modeling choices, we first verify that our baseline model provides an accurate description of the DR2 data. The overall goodness-of-fit of the baseline model yields $\chi^2 = 11492.0$ for $(11388-21)$ degrees of freedom, corresponding to a reduced $\chi^2 = 1.011$ and probability to exceed (PTE) of 0.20, indicating an overall fit consistent with statistical expectations. While the reduced $\chi^2$ only summarizes the overall agreement between our model and the data, we also investigate whether the residuals behave as expected under the assumed likelihood. Following previous large-scale structure analyses such as \cite{Loureiro_2019}, we examine the distribution of covariance-normalized residuals compared to a normal Gaussian distribution. We use the full covariance matrix to transform the residual vector, which is the difference between the measured correlation functions and our best-fitting model, into a set of uncorrelated, unit-variance residuals. If the covariance model accurately describes the data, and the Gaussian likelihood provides a sufficient statistical model, then the normalized residuals are expected to follow a standard normal distribution $\mathcal{N}(0,1)$ \cite{andrae_2010}. In Figure \ref{fig:residuals}, we compare the normalized residuals for the combined auto- and cross- correlations with a standard normal distribution, and find no significant deviation from Gaussianity. This indicates that the model provides an adequate description of the complete data vector, and provides a complimentary diagnostic to the overall reduced $\chi^2$. Having demonstrated the validity of the baseline model, the following section investigates its robustness to reasonable alternative modeling choices at various stages of the analysis pipeline. 

\section{Validation with Synthetic Spectra}\label{Section: Mocks} 
    
We perform extensive tests on synthetic spectra (mocks), providing an end-to-end validation of the analysis pipeline against a known underlying truth under controlled conditions. Our primary validation dataset consists of the \clpt\ mocks described in \S \ref{subsec: mock data sets}.

We first use high signal-to-noise stacked mock correlations to validate the baseline $r_{\mathrm{min}}$ scale cuts, and verify that with these cuts, the model accurately recovers the input cosmological parameters without significant bias (\S \ref{subsec: mock_stack_res}). We then analyze the ensemble of individual mock realizations to assess whether the recovered parameters and their estimated uncertainties exhibit the expected statistical behavior (\S \ref{subsec: mock_pop}). Appendix \ref{Appendix: Fiducial Cosmology} presents additional tests of the robustness of our results to changes in the assumed fiducial cosmology.
    
\subsection{Mock Datasets} \label{subsec: mock data sets}

Our primary validation suite consists of the \clpt\ mocks, a new generation of \lya\ forest light-cone cosmological simulations based on the \texttt{CoLoRe} framework~\cite{Ramirez2022}, which extends the original log-normal prescription by adopting a second-order Lagrangian perturbation theory (2LPT) approximation. These simulations are presented and extensively validated in \cite{RuizHerrera:2026}. These mocks are designed to reproduce the DESI DR2 \lya\ dataset in terms of quasar number density, footprint, and signal-to-noise ratio. The synthetic spectra additionally include astrophysical contaminants affecting the \lya\ forest, including metal absorption, BAL features, and DLA systems. The mock generation procedure is described in detail in \cite{2024arXiv240100303H, Cuceu:2025, Casas2025}, and we adopt the same methodology here. We include random Gaussian redshift uncertainties of $\sigma_v = 400\, \mathrm{km\, s^{-1}}$ at the level of the mock catalogs to emulate statistical errors of the redshift measurement pipeline. However, unlike \cite{Casas2025}, these redshift errors are not propagated into the continuum-fitting step. \footnote{In the current implementation, the effect of these redshift uncertainties in the mocks is possibly overestimated (as discussed in \cite{Gordon2025}) and likely unrealistically modeled as a Gaussian. Because the resulting spurious correlations are currently un-modeled in our baseline analysis, we instead treat this effect as a systematic whose impact we quantify using an analysis variation. We additionally tested modeling with a Lorentzian rather than a Gaussian, and found no significant impact on AP.}

While the \clpt\ mocks reproduce the observed \lya\ forest and quasar clustering with good fidelity, their underlying density field is generated using an approximate perturbative approach and therefore does not fully capture the non-linear evolution of structure on the smallest scales. To assess the robustness of our model and adopted scale cuts to non-linear effects, we complement the \clpt\ validation suite with \lya\ forest mocks based on the \abacus\ $N$-body simulations, which naturally capture the fully non-linear evolution of the matter density field. A detailed description of the mock generation and optical depth models is provided in \cite{Hadzhiyska2023, 2025MNRAS.540.1960H}, and a comprehensive discussion of how we use these mocks in the full-shape analysis is deferred to the Key Paper \cite{Cuceu:2026}.

\subsection{Validating model and scale cuts} \label{subsec: mock_stack_res}

We begin by validating the $r_{\mathrm{min}}$ scale cuts that inform our baseline analysis. We use the stack of 400 \clpt\ realizations, which combines the mock correlation functions into a single high signal-to-noise measurement with an effective uncertainty substantially smaller than that of the DESI DR2 data. This reduced statistical uncertainty provides a sensitive probe of potential parameter biases with high precision.

As part of our unblinding requirements, we establish that any bias must be smaller than one third of the statistical uncertainty of the combined fit to the four DESI DR2 correlations. During the scale cut scan, this threshold is evaluated at each $r_{\mathrm{min}}$, since the statistical uncertainty depends on the range of scales included in the fit. We adopt, for each correlation, the smallest $r_{\mathrm{min}}$ for which the recovered parameters remain stable and within the threshold. 

We follow the same baseline model setup described in \S \ref{section: Baseline}, with a few exceptions specific to the mocks. We include a sinc damping term to account for the finite resolution of the mock grid, fixing the transverse smoothing scale to $2.4\, \hMpc$, corresponding to the grid cell size, and the line-of-sight smoothing scale to $2.4(1+f)=4.73\,\hMpc$, to account for the stretching induced by redshift-space distortions.\footnote{We have also tested allowing the line-of-sight smoothing to be independent of the transverse smoothing using an extra free parameter, and did not find a significant impact on our results.} Because the metal contamination in the synthetic spectra is implemented as a rescaling of the \lya\ forest transmission (see Section 2.3.3 of \cite{2024arXiv240100303H}), the metal absorption is expected to inherit the same redshift-space distortion behavior as the \lya\ field; we therefore allow an additional common metal RSD parameter to vary freely. Finally, since the \clpt\ mocks do not include UV background fluctuations, \civ\ absorption, sky-subtraction residuals, or the quasar proximity effect, we remove the corresponding components from the model. We have verified that including any of these components in the model has a negligible impact on the results.

\begin{figure}
    \centering
    \begin{minipage}{\linewidth}
      \includegraphics[width=1\linewidth]{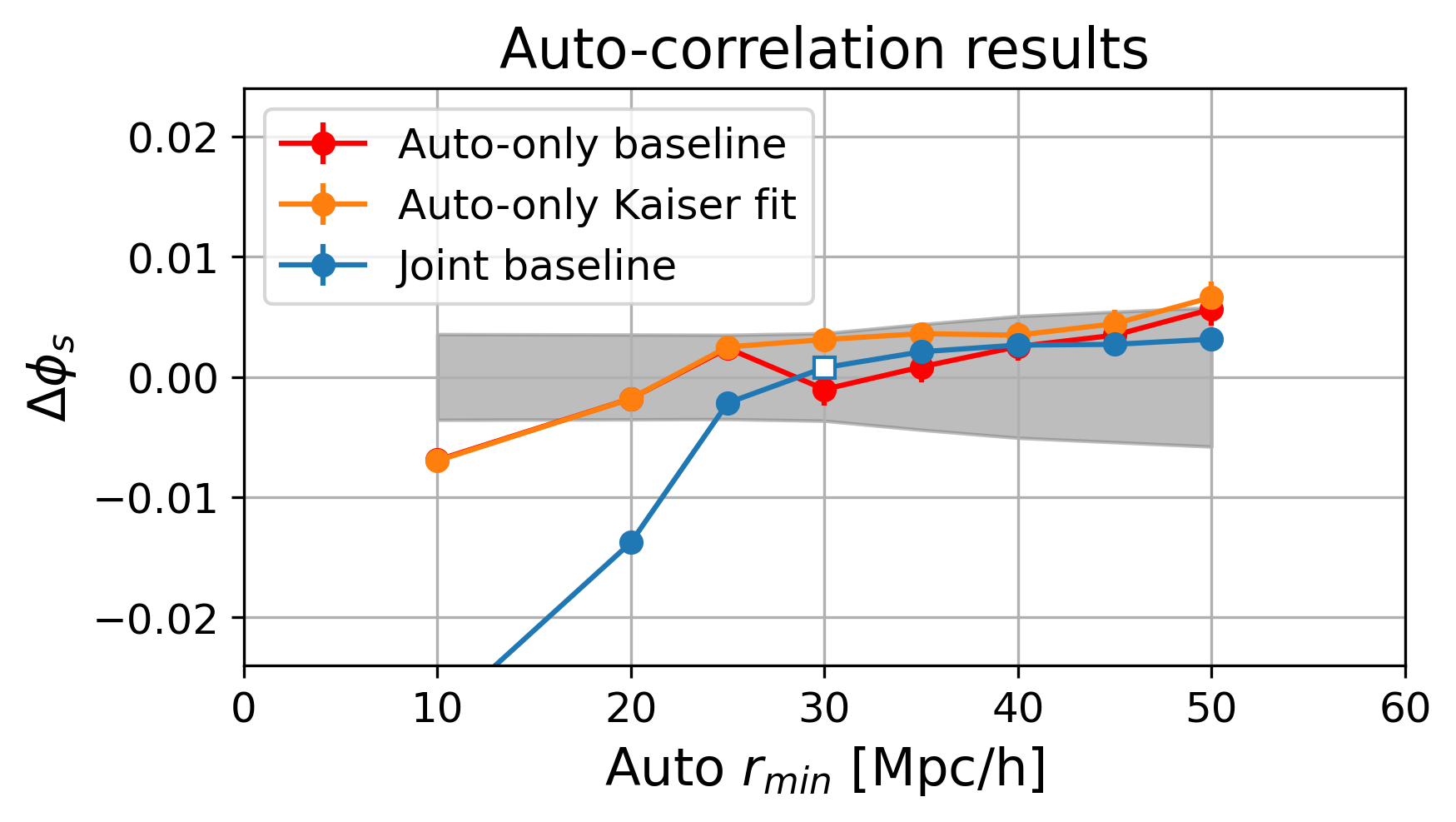}
    \end{minipage}
    \hfill
    
    \vspace{0.5em} 

    \begin{minipage}{\linewidth}
      \centering
      \includegraphics[width=1\linewidth]{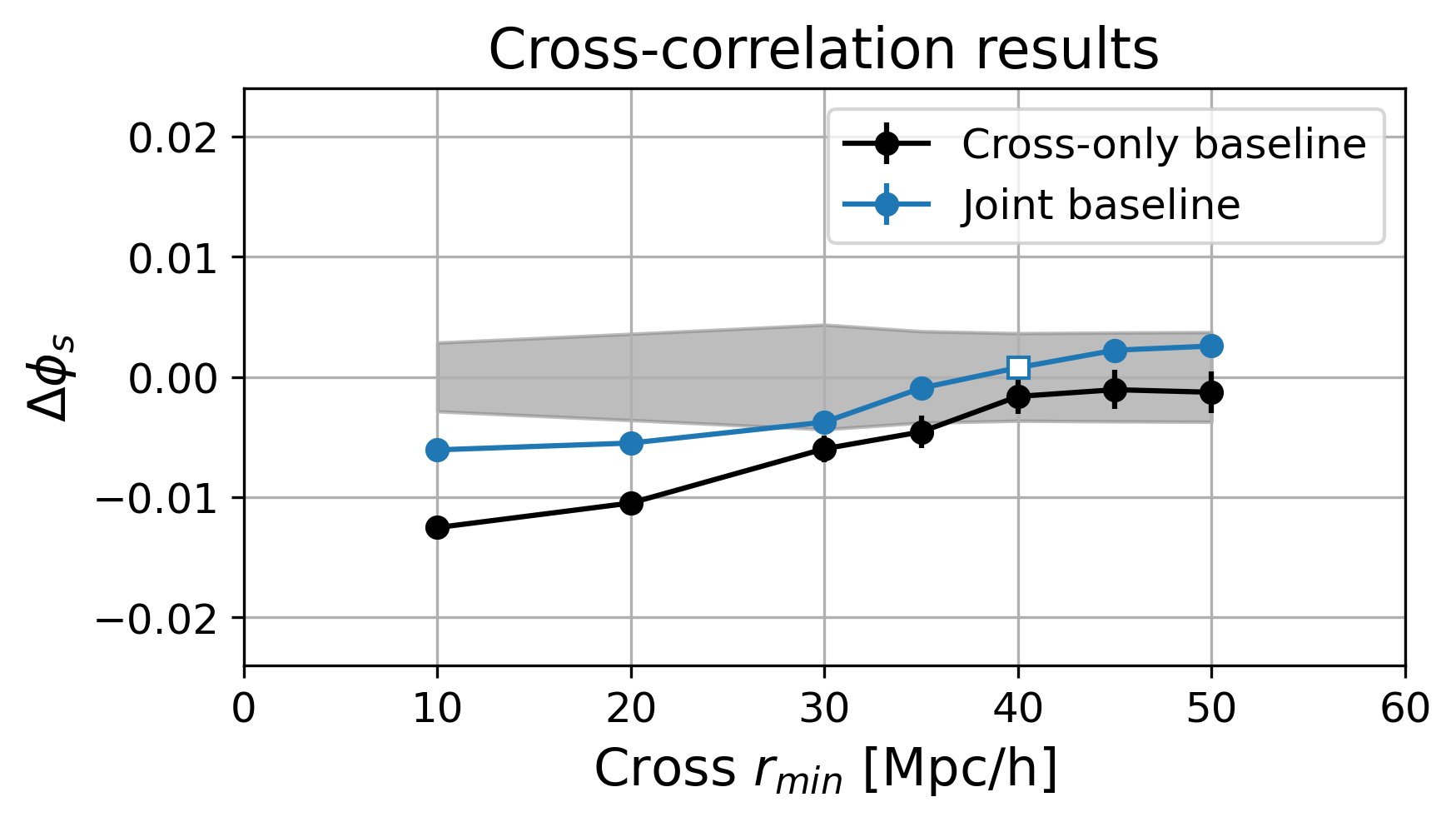}
    \end{minipage}
    \caption{
        Broadband AP constraints relative to the truth as a function of the minimum separation ($r_{\mathrm{min}}$) used in the analysis, measured from the stacked correlations of 400 \clpt\ mocks. The gray bands indicate the $1/3\,\sigma$ threshold obtained from the joint analysis of all four DESI DR2 correlations. (Top) Results from the two auto-correlations with the baseline model (red) and a Kaiser-only model (orange). We also show the results from the joint analysis of all four correlations (blue), where the cross-correlation has fixed $r_\mathrm{{min}}=40\, \hMpc$.
        (Bottom) Results from the two cross-correlations (black), and from the joint analysis with the auto-correlation having fixed $r_\mathrm{{min}}=30\,\hMpc$ (blue). 
        The baseline result is shown as an empty square.}
    \label{fig:mock_scale_cuts}
\end{figure}

The impact of scale cuts is shown in Figure \ref{fig:mock_scale_cuts}, which presents the dependence of \phis\ relative to the truth on $r_{\mathrm{min}}$ for the auto-correlation (top panel) and the cross-correlation (bottom panel). For the auto-correlation, we compare a Kaiser-only model with a model that includes the Arinyo small-scale corrections, where the Arinyo parameters are sampled with Gaussian priors. We find no significant differences between these two approaches. The best-fit Arinyo $q_1$ parameter is generally poorly constrained and consistent with zero.

As expected, the parameter biases increase as $r_{\mathrm{min}}$ is reduced. The recovered parameters satisfy our validation threshold of remaining within one third of the statistical uncertainty for $ r_{\mathrm{min}} \geq 25\,\hMpc$ in the auto-correlation, and $r_{\mathrm{min}} \geq 40\,\hMpc$ in the cross-correlation using \clpt\ mocks. We have also performed the same validation using mocks based on the \abacus\ $N$-body simulations \citep{Hadzhiyska2023,2025MNRAS.540.1960H}, which incorporate fully non-linear matter clustering and yield broadly consistent results, with all models converging for  $ r_{\mathrm{min}} \geq 30\,\hMpc$ for a joint fit; these tests are presented in detail in the \lya\ full-shape analysis Key Paper \cite{Cuceu:2026}. Taken together, these results support the conservative scale cuts of $r_{\mathrm{min}}=30\,\hMpc$ for the auto-correlation and $r_{\mathrm{min}}=40\,\hMpc$ for the cross-correlation adopted for our baseline analysis (\S \ref{section: Baseline}).

With the baseline scale cuts validated, we turn to the model itself to assess possible systematic biases in the recovered parameters. As before, we adopt a threshold of one-third of the statistical uncertainty, which, for the adopted scale cuts, corresponds to a tolerance of $\sim0.38\,\%$ in \phis. Any bias exceeding this threshold would be investigated further and either corrected or incorporated into the systematic error budget. Although the DR2 BAO analysis has already been validated using earlier mock suites \citep{DESI2024.IV.KP6}, we repeat these tests with the new generation of \clpt\ mocks and the updated analysis methodology.

\begin{figure*}
    \centering
    \includegraphics[width=0.9\textwidth,keepaspectratio]{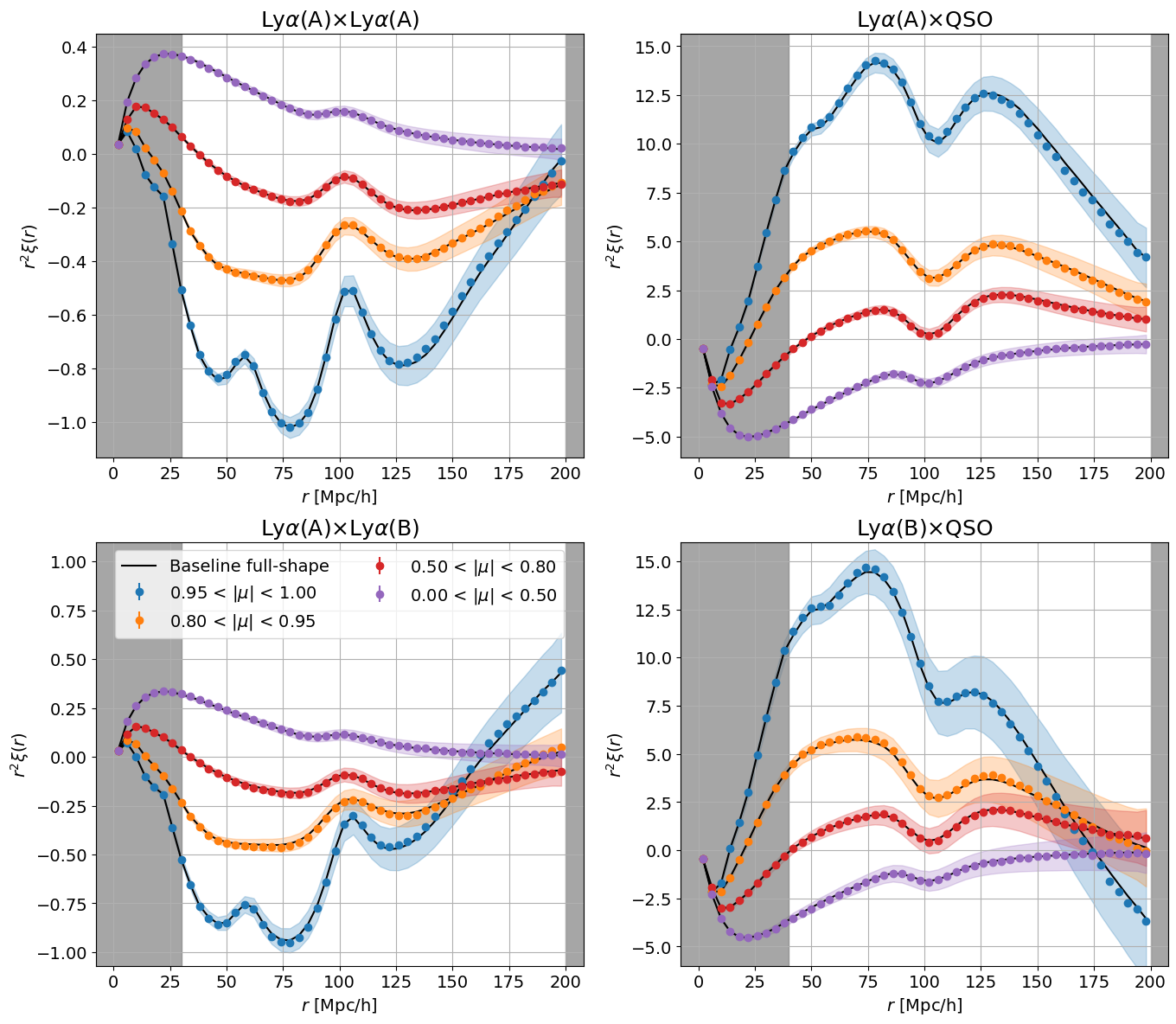}
    \caption{
        Stacks of the four \lya\ correlation functions measured from 400 \clpt\ mocks, along with the best-fit model (black). The shaded regions indicate the uncertainties of the DESI DR2 measurement on data, centered around the best-fit model. The model provides a great fit to the stacked mock correlations when compared to the DR2 uncertainties. The dark-shaded areas highlight the scales not used in this particular fit. 
        }
    \label{fig:mock_wedges}
\end{figure*}

Figure \ref{fig:mock_wedges} shows the correlation wedges measured from the stack of the 400 \clpt\ mocks, together with the best-fit model. The mock measurements closely follow the model across all four correlation functions and lie comfortably within the uncertainty envelope of the DESI DR2 data, confirming that the model provides a good fit to the stacked mock correlations at the precision of the DR2 measurement.

    \begin{figure}
        \centering
        \includegraphics[width=1.0\columnwidth,keepaspectratio]{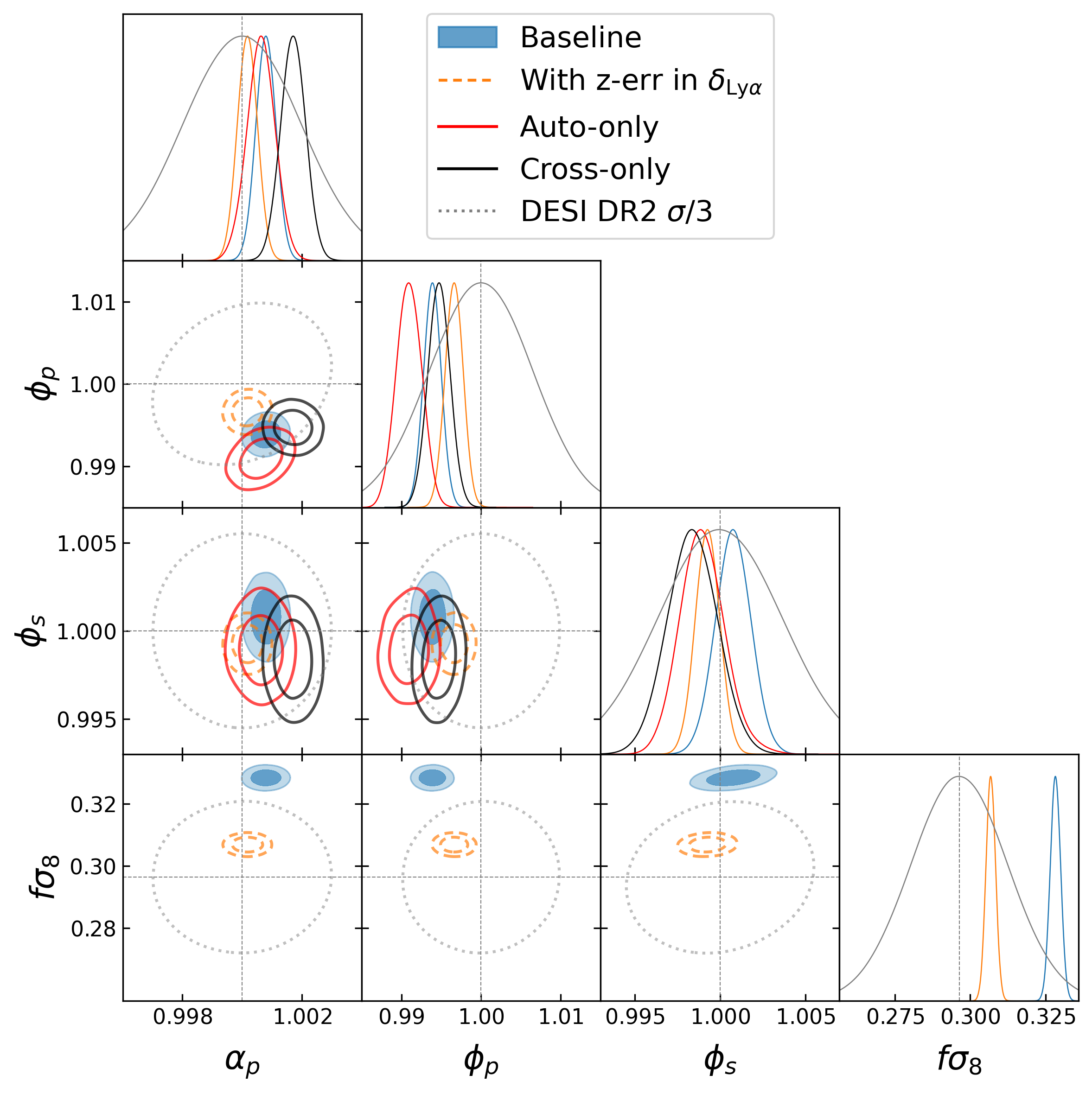}
        \caption{
           Parameter constraints from a joint analysis of all four stacked correlations of 400 \clpt\ mocks (blue). The red and black contours show results for separate fits to the two auto-correlations and the two cross-correlations, respectively. The dashed orange contours show results from the joint analysis with redshift errors included during the continuum-fitting process. The gray dotted contours show the $1/3\; \sigma$ threshold, while the vertical and horizontal dotted lines indicate the true value of each parameter in these mocks. The baseline results are within the thresholds for the broadband AP constraint (\phis) and for the two BAO parameters (\alphap\ and \phip), but show a $\sim0.5\,\sigma$ deviation in \fsig\ relative to the truth.
        }
        \label{fig:mock_stack}
    \end{figure}

Turning to the recovered parameters, Figure \ref{fig:mock_stack} shows the full-shape constraints from the stack of 400 \clpt\ mocks obtained by combining all four \lya\ correlation functions, shown as the blue shaded (baseline) contour. The two BAO parameters, \alphap\ and \phip, as well as the broadband AP parameter, \phis, all lie within the one-third of the statistical uncertainty threshold, shown by the black dotted contour. While \alphap\ and \phis\ are consistent with the truth, \phip\ shows a $0.6\,\%$ bias at $\sim5\,\sigma$ significance. In contrast, we find that the \fsig\ parameter lies outside the threshold, displaying a $10\,\%$ bias at high significance ($\gg5\,\sigma$). We have found no reasonable variation of the model that fully reconciles this measurement with the truth, and extending the scale cuts to $ r_{\mathrm{min}}=70\,\hMpc$ ($60\,\hMpc$) for the cross-correlation (auto-correlation) does not significantly reduce the bias on \fsig. This persistent, statistically significant bias observed in these validation tests was a primary motivation for the decision to de-scope the \fsig\ measurement.

The same figure also shows the results obtained by fitting the auto- and cross-correlations separately. For the cross-correlation-only fit, we fix the growth rate, $f$, and the quasar bias, $b_\mathrm{Q}$, to the truth values used to generate the mocks. The resulting constraints are consistent with each other and with the joint fit. We also test the impact of spurious correlations along the line-of-sight due to redshift errors affecting the continuum-fitting process, as discussed in \cite{Youles2022, Gordon2025}, shown as the dashed orange contours in Figure \ref{fig:mock_stack}. These redshift errors are present in the quasar catalog for the baseline analysis but only affect the cross-correlation measurement, producing a smoothing effect along the line-of-sight. We estimate the systematic error introduced by these spurious correlations from the difference between this variation and the baseline analysis. The measured shifts for the main parameters are $0.06\,\%$ in \alphap, $0.27\,\%$ in \phip, and $0.15\,\%$ in \phis, all within our threshold. However, \fsig\ displays a much larger shift of $\sim7\,\%$. As the impact of the spurious correlations is likely overestimated in our mocks~\citep{Gordon2025}, the observed shifts constitute upper bounds on the systematic errors. The results shown in Figure \ref{fig:mock_scale_cuts} and Figure \ref{fig:mock_stack} are consistent with those from the \abacus\ mocks, which are discussed in \cite{Cuceu:2026}. Therefore, we find that for the purpose of the AP measurements, our baseline model and scale cuts have been validated with the mocks to within the threshold we imposed.

\subsection{Mock population tests} \label{subsec: mock_pop}

In this section we focus on fits on each of the 400 \clpt\ mocks individually. For these fits, the common metal RSD parameter, discussed in \S \ref{subsec: mock_stack_res}, is fixed to the best-fit value obtained from the stack of mocks. We examine the distribution of the recovered parameter uncertainties, as well as the distribution of the parameter values relative to the truth.
    
    \begin{figure}
        \centering
        \includegraphics[width=0.9\columnwidth]{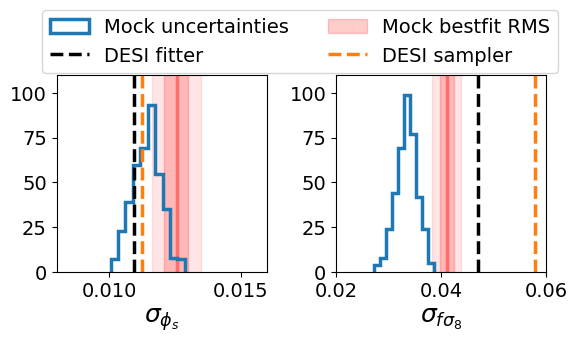}
        \caption{
            Histograms of measurement uncertainties for \phis\ and \fsig\ from the population of \clpt\ mocks. DESI DR2 uncertainties from both the fitter and sampler are shown for comparison, and are consistent in the case of \phis, but differ significantly for \fsig. The best-fit RMS from the 400 mocks is also shown for each parameter. In both cases, this is offset relative to the uncertainty histogram, likely because fitter uncertainties are underestimated, as seen in the difference between the black and orange dashed lines. 
            }
        \label{fig:mock_errors}
    \end{figure}

First, we consider the distribution of parameter uncertainties. Figure \ref{fig:mock_errors} shows, for \phis\ and \fsig, the histogram of uncertainties across the population of \clpt\ mocks (blue), the root mean square (RMS) of the parameter best fits (red shaded region), and the uncertainties measured from the real DESI DR2 data with the fitter (black dashed line) and the sampler (orange dashed line). For \phis, all four quantities are consistent, whereas for \fsig, the data uncertainties are significantly larger than the mock uncertainties.
    
For both parameters (\phis, \fsig), the RMS of the best fit is slightly larger than the mean of the mock uncertainties. This is at least in part because we are using Gaussian-approximated fitter uncertainties, which tend to underestimate the true uncertainty (see \cite{Cuceu2020}), as can be seen from the difference between the fitter and sampler results on the data (black versus orange dashed lines). In the case of \phis, this shift is small, and the RMS remains consistent with the mock uncertainties. Given that the fitter uncertainty from the data is also consistent with the mock uncertainties, we consider that \phis\ uncertainties successfully pass this test.
 
For \fsig, the offset between the best-fit RMS and the mock uncertainties may be related to the offset between the fitter and sampler results on the data. However, the data uncertainty from the fitter is still $\sim40\,\%$ larger than the mock uncertainties for \fsig. This could mean that the mocks are not representative of the data for this measurement, or that either the mocks or the data are affected by a systematic error.

    \begin{figure}
        \centering
        \includegraphics[width=0.9\columnwidth]{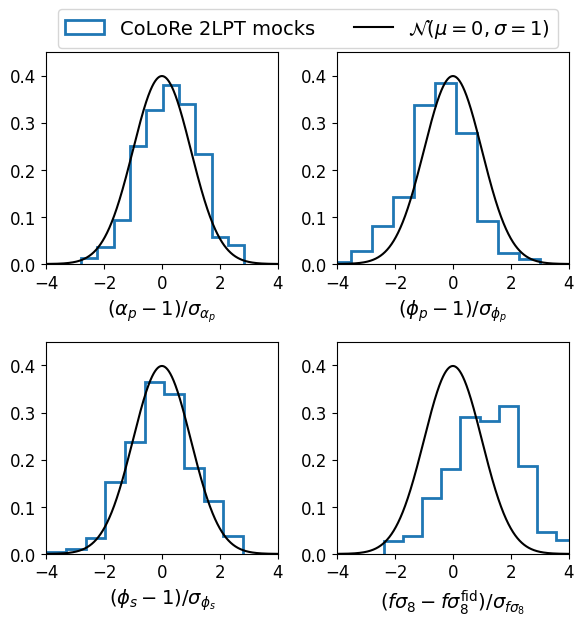}
        \caption{
            Histogram of pull values with respect to the truth for the population of \clpt\ mocks (blue). BAO parameters (\alphap, \phip) are shown in the top row of plots, while \phis\ and \fsig\ are shown in the bottom row. The black lines are given by a Gaussian with null mean and unit variance, which is the target distribution assuming no systematic biases and accurate uncertainties. The first three parameters are broadly consistent with the target Gaussian, except for a small shift in the \phip\ distribution, which is consistent with the $\sim0.6\%$ systematic shift present in the parameter constraints from a joint analysis of all four stacked correlations of 400 \clpt\ mocks. However, \fsig\ displays a significant deviation from the target distribution, both in terms of mean and shape.
        }
        \label{fig:mock_pull}
    \end{figure}

Next, we test the distribution of pulls from the mocks with respect to the truth for the four primary parameters, defined for each mock as the difference between the best-fit value and the truth divided by the Gaussian uncertainty (Figure \ref{fig:mock_pull}). The \phis\ and \alphap\ pulls are consistent with the target Gaussian distribution, $\mathcal{N}(0,1)$, while \phip\ displays a shift consistent with the bias found in the stack of mocks (Figure \ref{fig:mock_stack}). In contrast, \fsig\ is significantly shifted from the target distribution, both in mean and in shape, consistent with the bias found in the stack of mocks (\S \ref{subsec: mock_stack_res}).

To quantify the widths of these pull distributions, we compute their standard deviations relative to the distribution means:
    \begin{align*}
        \sigma\left[(\alpha_p - \overline{\alpha_p})/\sigma_{\alpha_p}\right] &= 0.98 \pm 0.04, \\
        \sigma\left[(\phi_p - \overline{\phi_p})/\sigma_{\phi_p}\right] &= 1.04 \pm 0.04, \\
        \sigma\left[(\phi_s - \overline{\phi_s})/\sigma_{\phi_s}\right] &= 1.10 \pm 0.04, \\
        \sigma\left[(f\sigma_8 - \overline{f\sigma_8})/\sigma_{f\sigma_8}\right] &= 1.2 \pm 0.04,\\
    \end{align*}
where the uncertainties are estimated through bootstrap sampling. The values for \alphap\ and \phip\ are consistent with unity, confirming that the BAO uncertainties are well calibrated. The somewhat high value of $\sigma$\phis, deviating from unity at $\sim2.5\,\sigma$ significance, is at least in part due to the Gaussian approximation of the uncertainties. Tests on the data indicate that the sampler produces a $3\,\%$ larger uncertainty than the fitter, which would reduce the significance of this deviation in mocks to below $2\,\sigma$. Furthermore, this difference varies from one mock to another, with realizations that have larger uncertainties usually exhibiting a larger difference between the fitter and the sampler results as well \cite{Cuceu2020,Cuceu:2025}. Given that our data uncertainty is smaller than the uncertainty of $81\,\%$ of mocks (see Figure \ref{fig:mock_errors}), we expect this difference to be larger than $3\,\%$ for most mocks. Therefore, we conclude that $\sigma$\phis\ is consistent with unity.

A similar explanation may apply to the larger value of \fsig. However, this deviation is detected at a much higher significance level ($\sim5\,\sigma$), so a full validation of the \fsig\ uncertainties would likely require rerunning the sampler on at least a subset of the mocks. Given our decision to de-scope the \fsig\ measurement, we did not pursue this further.

\section{Validation with Blinded Data} \label{section: data validation}

Prior to unblinding, we performed an extensive validation program on the DESI DR2 data. The purpose of these tests is to identify potential systematic effects, assess the robustness of the analysis to reasonable methodological choices, and verify the internal consistency of the dataset before examining the final cosmological constraints. To avoid experimenter bias, the analysis was blinded following the same procedure adopted for the DESI DR1 full-shape analysis \cite{Cuceu:2025}. Random offsets were drawn from Gaussian distributions with zero mean and widths comparable to expected statistical uncertainties, and added within the likelihood to the primary parameters of interest, \phis\ and \fsig. These offsets were generated using a fixed random seed, and were therefore identical across all validation tests, which allowed for meaningful comparison during validation while concealing the true parameter values. We defined a set of thresholds that we required all validation tests to either pass or be well understood under further investigation, before we were allowed to unblind. 

The data validation process was divided into two primary categories: data splits and analysis variations. Data splits test the internal consistency of the dataset by repeating the analysis on subsets of the data and comparing the resulting constraints. Analysis variations, however, test the robustness of the pipeline by modifying choices made at various stages of the analysis. This includes changes made prior to the correlation-function fit, such as the estimation of transmitted-flux fluctuations, the construction of the correlation functions, and the computation of covariance and distortion matrices. This also includes modeling variations, which test the sensitivity of the results to assumptions in the theoretical description of the measured correlations, including alternative scale cuts, priors, and nuisance-parameter treatments. Many of these tests were inherited from the DESI DR2 \lya\ BAO analysis \cite{DESI2024.IV.KP6}, while others were introduced specifically to validate the new modeling components used in the full-shape analysis, such as the inclusion of UVB fluctuations. Additionally, although some of the BAO variations have been considered non-contingent for unblinding, they have been included for posterity in \ref{Appendix: Old Variations}. 

Quantitative unblinding criteria were established for each of these test classes. For tests that compare analysis results from sub-samples of the data, we defined the following threshold: differences in $\chi^2$ between a given set and the baseline analysis exceeding two standard deviations ($p < 0.045$) triggered further investigation. This significance accounts for the number of independent tests and any known correlations between samples. Further investigation involved improving the model, understanding the source of the shift and adding a systematic uncertainty, or performing a detailed investigation with mocks and accounting for uncertainties such as statistical fluctuations using nested sampling. For analysis and modeling variations, we required that shifts in the primary parameter of interest, \phis\, remain below one third of the expected statistical uncertainty of the (blinded) baseline analysis.  This corresponds to a tolerance of $\sim 0.0038$ for \phis. Variations that exceeded this threshold were investigated further, accounting for expected statistical fluctuations arising from changes in sample size, weighting, binning, scale cuts, or other modifications to the analysis. Depending on the outcome of that investigation, a variation could motivate improvements to the baseline methodology, the assignment of a systematic uncertainty, or additional validation using mock datasets. 

\subsection{Data splits}\label{Section: data splits}

The first set of tests presented involves data splits, where the full analysis is performed using subsets of the data. Splitting the data enables checking for various inconsistencies and biases, which may be attributed to statistical fluctuations. 

The list of data split tests are: 
    \begin{itemize}
        \item Auto vs Cross-correlations only
        \item \lya(A) vs \lya(A)+\lya(B) regions 
        \item North vs South
        \item Signal to Noise Ratio
        \item $\text{C}_{\text{IV}}$ Equivalent Width
    \end{itemize}

We consider two types of data splits: those that split the data vector, and those that split the quasar catalog. The first two data splits we discuss are the former. 

\begin{figure}[htbp]
        \centering
        \includegraphics[width=1.0\linewidth]{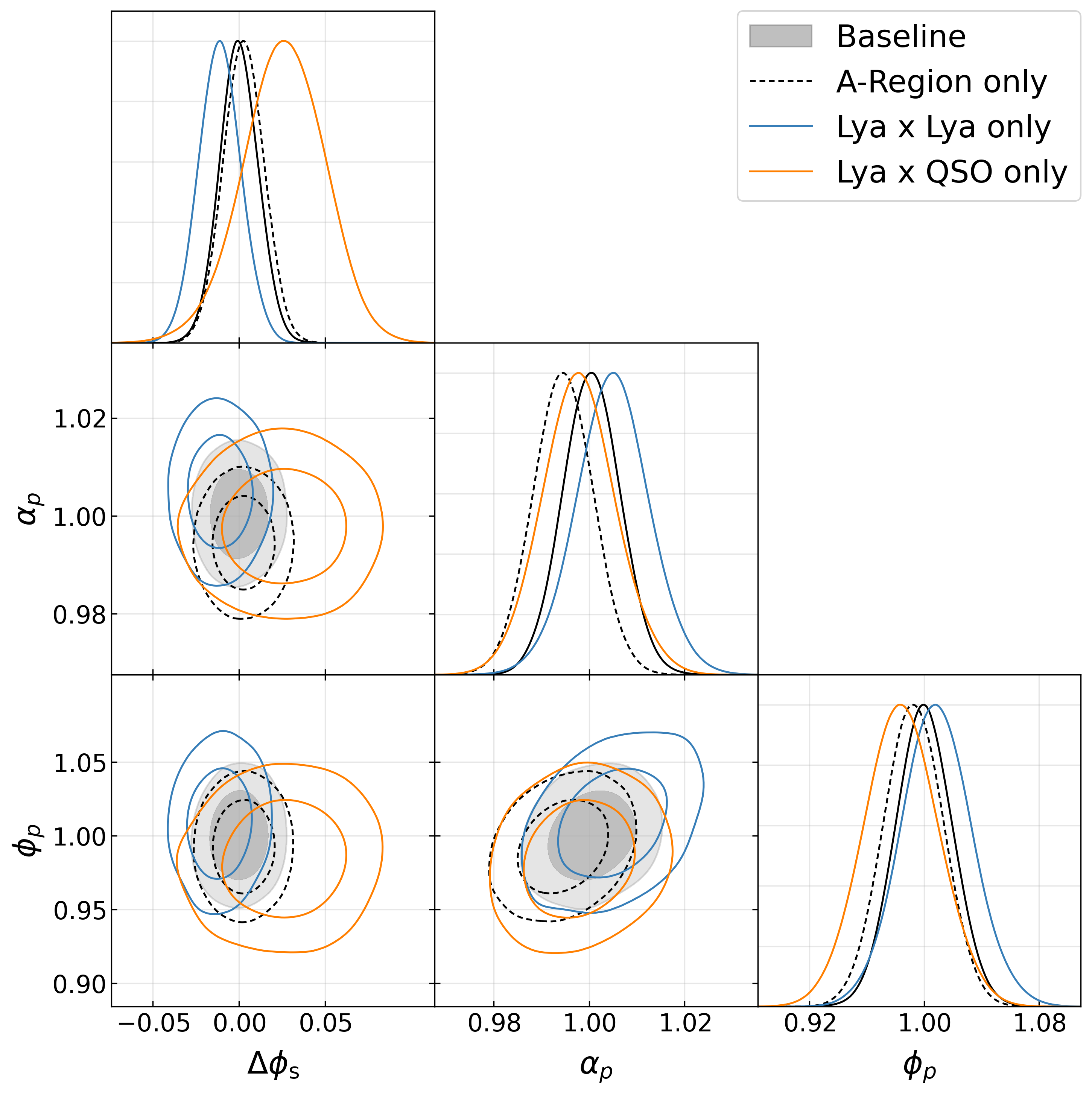}
        \caption{Full-shape and BAO constraints from the two Ly$\alpha$ auto-correlations only (blue) and the constraints from the cross-correlations only (orange), compared to the baseline constraints (gray), which use all four correlations. The Auto and Cross only splits are in good agreement with the baseline results for \phis.
     Also shown are the results from the correlations which only use pixels in the A-region (i.e. \lya(A)$\times$\lya(A), \lya(A)$\times$QSO), which are in good agreement with the baseline results which additionally use the B-region (i.e. \lya(A)$\times$\lya(B), \lya(B)$\times$QSO). 
        }
        \label{fig:auto_cross_split}
    \end{figure}

The first data split that we consider is independently fitting the auto- and cross-correlations, where we fit the two auto-correlations (\lya(A)$\times$\lya(A) and \lya(A)$\times$\lya(B) separately from the two cross-correlations (\lya(A)$\times$QSO and \lya(B)$\times$QSO). Because the auto and cross-correlation joint fit is needed to break the \lya\ and QSO RSD terms, this data split can only measure the AP effect \cite{Cuceu2021}. We show the constraints on the full-shape parameter, \phis, as well as the BAO parameters, \phip\ and \alphap\, in Figure \ref{fig:auto_cross_split}. We find good agreement of the auto- and cross-correlations with the baseline analysis for all parameters of interest. 

The next data split is grouping the correlations that only use pixels in the A region (i.e. \lya(A)$\times$\lya(A), \lya(A)$\times$QSO). The correlations in the A region contain information for both AP and RSD, and are therefore very similar to the baseline analysis, which uses all four correlations. The B region does not contain as much information as the A region for a few reasons, such as additional noise due to higher instances of contamination by other Lyman lines, reduced number of sightlines due to only existing in higher redshift QSOs, and being significantly shorter than the A region which results in fewer pixels per line-of-sight. This data split shows good agreement between using only the A region, and when using the A and B region combined, which is shown in Figure \ref{fig:auto_cross_split}. Additionally, we find that including the B region, as in the baseline analysis, improves constraints on \phis\ by about 9\%. 

The remaining data splits involve splitting at the quasar catalog level, where the full analysis is performed independently on each subset of the catalog. These data splits repeat consistency checks performed by previous \lya\ analyses \cite{DESI2024.IV.KP6, DESI.DR2.BAO.lya, Cuceu:2025}.

\begin{figure}[htbp]
        \centering
        \includegraphics[width=1.0\linewidth]{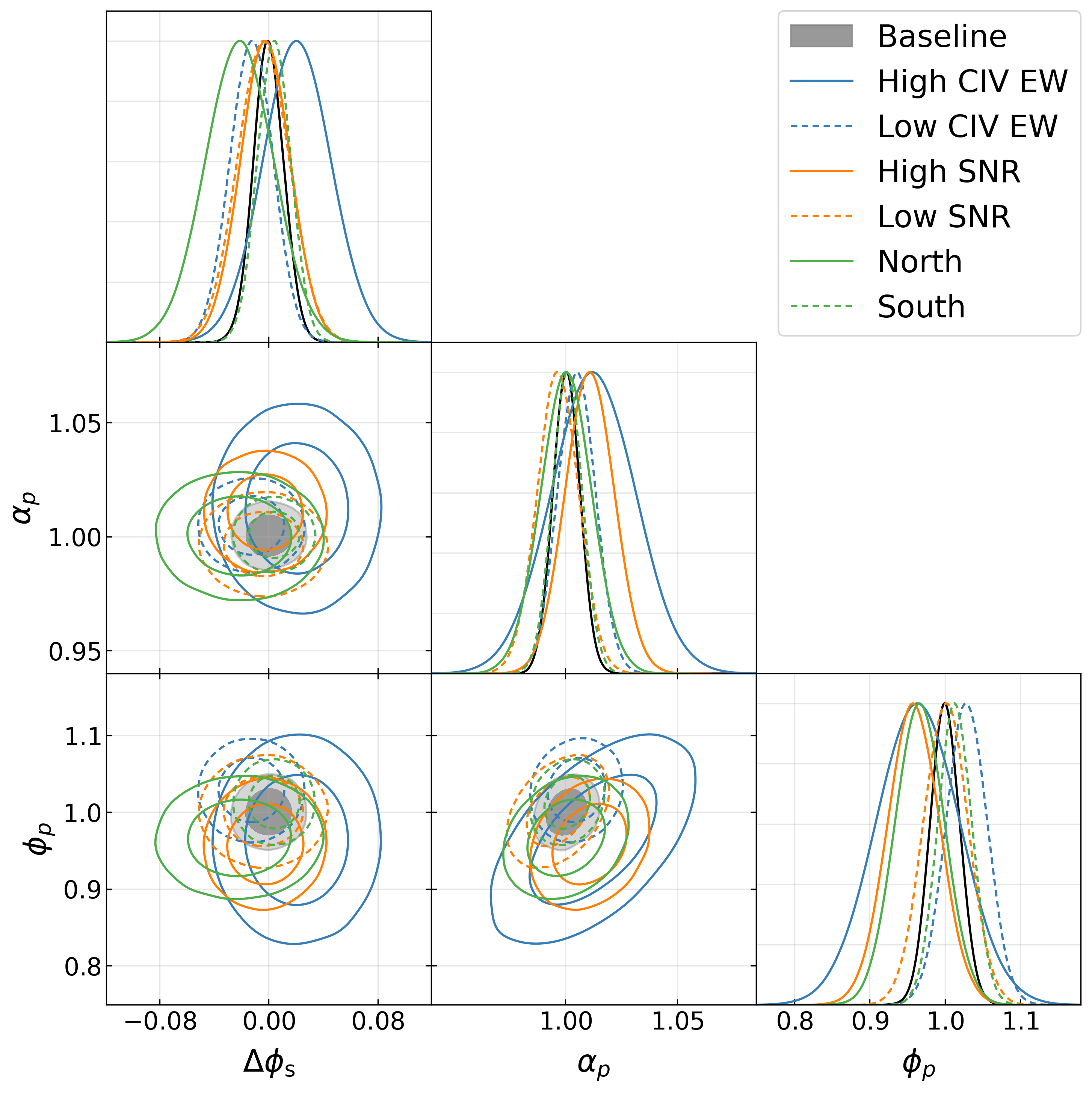}
    \caption{Full-shape and BAO constraints from the baseline analysis (gray), and various data-splits at the catalog level. Catalogs have been split into high (solid) and low (dashed) \civ\ equivalent width in the quasar spectra (blue), high (solid) and low (dashed) signal-to-noise ratios in the quasar spectra (orange), as well as North (solid) and South (dashed) imaging used in the quasar target selection (green). 
    }
    \label{fig:data_splits}
    \end{figure}

The first data split of this category is a geographic target selection split, where we separate quasars targeted with different imaging surveys. Here, "North" refers to imaging performed by the BASS and MzLS surveys, accounting for approximately 18\% of the quasar catalog. "South" refers to imaging performed with the DECam camera, including the entire South Galactic Cap, and the southernmost part of the North Galactic Cap (at declination $\delta < 32.375^\circ$), accounting for the remaining 82\% of the quasars in the sample. We find the size of uncertainties and location of the contours (green) in Figure \ref{fig:data_splits} are consistent with statistical fluctuations attributed to the difference in size of the quasar data subsets.

The next data split separates the quasar catalog in signal-to-noise ratio (SNR), where we define SNR as the mean signal-to-noise over the rest-frame wavelength range, $1420 < \lambda_{\text{rest}} < 1480$ \AA. The "High SNR" sample refers to quasars with a SNR greater than 4.25 and the "Low SNR" refers to those below this threshold. We split the quasar catalog such that the uncertainties in each subset are approximately equivalent, rather than in equal numbers of \lya\ forests. Figure \ref{fig:data_splits} shows that the size of the (orange) contours are comparable in size, and that they are consistent with each other. 

The last data split separates the quasar catalog by the equivalent width (EW) of the \civ\ emission line. We define a "High EW" and "Low EW" sample, split at 39 \AA. This split reflects the average of the measured median \civ\ EW of quasars (37.3 \AA), and the median of $3\sigma$ measurements of the \civ\ emission line (41.6 \AA). The \civ\ EW measurements are performed using \textsc{fastspecfit}\footnote{\url{https://github.com/desihub/fastspecfit}}, a collection of public code for spectral synthesis and emission-line fitting of DESI spectrophotometric data \cite{fastspecfit_2023}. This split is an important validation test, since there is a known anti-correlation between the quasar continuum luminosity and the EW of the emission lines known as the Baldwin effect \cite{Baldwin1977}. As a result of this correlation, the low EW sample measures a slightly higher luminosity and effective redshift quasar sample relative to the high EW sample. We find the uncertainty contours (blue) in Figure \ref{fig:data_splits} are consistent with one another. 

All of the data splits considered here are within the thresholds defined prior to unblinding. Therefore, we find that all data splits are consistent, and do not require further investigation. 

\subsection{Analysis Variations}\label{Section: Alternative Analyses}

We have performed many analysis variations to test the sensitivity of our results to plausible alternative analysis decisions at various stages of the pipeline. We present the shifts in \phis\ produced by these variations in Figures \ref{fig: variations}, \ref{fig:scale_cuts}, and \ref{fig: variations model}, which show parameter values and uncertainties of these tests relative to the baseline analysis. The majority of these tests are inherited from previous BAO \cite{DESI2024.IV.KP6, DESI.DR2.BAO.lya} and full-shape \cite{Cuceu:2025} analyses. However, some variations used in earlier studies have been retired as they no longer represent realistic analysis alternatives (see Appendix \ref{Appendix: Old Variations}), while additional tests have been introduced to assess the robustness of methodological updates implemented for this analysis, such as those discussed in \S \ref{section: changes since dr2 bao}.

\subsubsection{Variations in the Estimation of the Fluctuations}\label{Section: Fluctuations}

The first set of variations impact how we measure the \lya\ fluctuations. The first group of these variations result in changes to the dataset, and so we expect them to exhibit some statistical fluctuations due to differences in sample size. The brown (circle) points in Figure \ref{fig: variations} illustrates how the value of \phis\ changes for each of the following variations: 

\begin{itemize}
    
    \item \textbf{$\lambda_{\rm obs} < 5500$\,\AA}: we only use \lya\ pixels below this observed wavelength, instead of the baseline value of $\lambda_{\rm obs} < 5577$ \AA.  
         
    \item \textbf{$\lambda_{\rm obs} > 3650$\,\AA}: we only use \lya\ pixels above this observed wavelength, instead of the baseline value of $\lambda_{\rm obs} > 3600$\,\AA.  
      
     \item \textbf{$\lambda_{\rm RF} < 1200$\,\AA}: we only use \lya\ pixels below this rest-frame wavelength, instead of the baseline value of $\lambda_{\rm RF} < 1205$ \AA. 
        
     \item \textbf{$> 50$ pixels in forest}: we include lines-of-sight with more than 50 valid \lya\ pixels, while the baseline requirement is at least 150 pixels. 
    
     \item \textbf{only QSO targets}: we only use quasars that were originally targeted as quasars. This variation causes a significant change in the size of the dataset, as this subset represents approximately 88\% of the quasar catalog used in the baseline analysis. 

     \item \textbf{DLA ${\rm SNR}>2$ }: we only mask DLAs identified in quasar spectra with ${\rm SNR\_RED}>2$, as was done in previous BAO analyses, where SNR was measured over the rest-frame wavelength range of $1420 < \lambda_{\mathrm{rest}} < 1480$ \AA, and is calculated as the average SNR per pixel, with pixel width of 0.8 \AA. The baseline full-shape analysis no longer uses a SNR threshold for masking DLAs, as discussed in the DESI DR2 \lya\ forest full-shape paper \cite{Cuceu:2026}. 
        
     \item \textbf{DLA ${\rm SNR}>3$ }: we only mask DLAs identified in quasar spectra with ${\rm SNR\_RED}>3$, whereas the baseline analysis does not use an SNR threshold for masking DLAs. 

     \item \textbf{DR2 BAO DLA catalog}: we use the (older) DLA catalog from the DESI DR2 BAO analysis \cite{DESI2024.IV.KP6}. 

     \item \textbf{weak BALs}: we do not include the \lya\ forest of quasars where we have identified strong BAL features. In particular, we still apply BAL masking as usual, but do not allow the strongest BALs ($AI > 436.5$) to contributing to the delta field. This corresponds to excluding approximately the 50\% percentile of the strongest BALs in DR2 following \cite{KP6s9-Martini}, as was done previously in eBOSS analyses \cite{Ennesser2022}. 

     \item \textbf{no sharp lines mask}: we do not mask sharp lines in the spectra, as discussed in \cite{Ramirez2024}.  In the baseline analysis we mask features related to sky lines and Calcium absorption from the Milky Way interstellar medium.
     
\end{itemize}

The next subset of analysis variations do not result in changes to the dataset. These variations affect the continuum fitting process, the shifts of which are shown by the purple (square) points in Figure \ref{fig: variations}. Since these variations can impact the weights used for continuum fitting and measuring the correlation functions, some statistical fluctuations occur. The list of \lya\ fluctuation estimation variation tests are as follows: 

\begin{itemize}
     \item \textbf{no calibration}: we do not re-calibrate the spectra using the \ciii\ region, as is done in the baseline analysis (and described in detail in \cite{Ramirez2024}). 
     
     \item \textbf{$\eta_{\rm pip} = 1$}: we do not re-calibrate the instrumental noise $\eta$, as is done in the baseline analysis. 
      
    \item \textbf{$\eta_{\rm LSS} = 3.5$}: we use non-optimal weights, and reduce by a factor of two the contribution from the intrinsic \lya\ forest variance to the weights. 
     
     \item \textbf{$\Delta \lambda_F = 2.4$\,\AA}: we rebin flux by coadding pixels in groups of three, prior to continuum fitting. Additionally, we use $\eta_{\rm LSS} = 3.1$, which was found to be optimal for coarser pixelization in \cite{Ramirez2024}. 
\end{itemize}

All of the variations pass our threshold, with the exception of $\Delta \lambda = 2.4$\,\AA, which can be attributed to statistical fluctuations. This is discussed in more detail in \S \ref{Section: Discussion}.

\subsubsection{Variations in the Measurement of Correlations}\label{Section: Correlations}

\begin{figure}[htbp]
        \centering
        \includegraphics[width=0.9\linewidth]{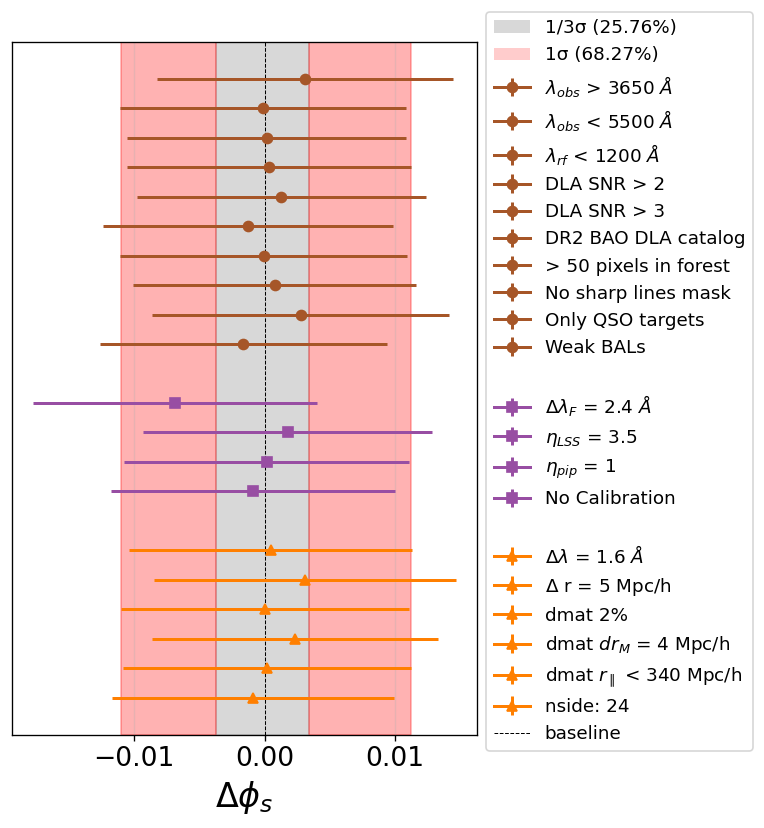}
        \caption{Shifts in the AP measurement from a set of analysis variations. Analysis variations shown here are those that results in changes to the dataset (brown, circle),  those that change the method of estimating Ly$\alpha$ overdensities (purple, square), and those that change the method of computing correlations, covariances, and distortion matrices (orange, triangle). The shaded regions represent the uncertainty in the baseline measurement, at 1 (red) and 1/3 (gray) standard deviation.
        }
        \label{fig: variations}
    \end{figure}

The next set of analysis variations includes changes in the measurement of the correlation functions, covariance matrices, or distortion matrices. The orange (triangle) points in Figure \ref{fig: variations} show the shifts in parameters from these alternative analyses. 

The correlation variation tests are as follows: 

\begin{itemize}
    
    \item \textbf{dmat $r_\parallel < 340 \,\hMpc$}: we model the distortion matrix up to $r_\parallel = 340 \,\hMpc$, whereas the baseline analysis extends the model to $r_\parallel = 300 \,\hMpc$. 
     
    \item \textbf{dmat 2\%}: we use 2\% of the dataset to compute the distortion matrix, compared 1\% to in the baseline analysis.
     
     \item \textbf{dmat $dr_{\textrm  {M}} = 4 \,\hMpc$}: we model the distortion matrix with 4 $\hMpc$ bins for the model, as is used in the measurement of the correlation function, compared to 2 $\hMpc$ in the baseline analysis. The data is still binned in $4 \,\hMpc$ bins. 
               
     \item \textbf{$\Delta \lambda = 1.6$ \AA}: we re-bin the continuum-fitted \lya\ pixels in groups of two (rebinned pixels of 1.6\,\AA), rather than in groups of three (rebinned pixels of 2.4\,\AA) in the baseline analysis . 
         
    \item \textbf{nside $= 24$}:  we measure the correlations in HEALPix pixels defined by $\texttt{nside}=24$, rather than $\texttt{nside}=16$ as in the baseline analysis. Using smaller HEALPix regions reduces the number of quasars contributing to each subsample while increasing the total number of subsamples used to estimate the covariance matrix. This tests whether our results are sensitive to the choice of subsampling used in the covariance estimation.
     
     \item \textbf{$\Delta r = 5\,\hMpc$}: we bin the correlation function in 5 $\hMpc$ bins, rather than in 4 $\hMpc$ bins used in the baseline analysis.
    
\end{itemize}

All validation tests related to the measurement of the correlation function, distortion matrix, and covariance matrix do not exceed more than one third of the expected statistical uncertainty from the (blinded) baseline analysis, and are therefore not problematic.

\subsubsection{Variations in the Modeling of Correlations and Parameter Estimation}\label{Section: Model Variations}

We also consider variations of the modeling process in order to test the robustness of our baseline analysis. 

\begin{figure}
    \centering

    \begin{minipage}{\linewidth}
        \centering
        \includegraphics[width=\linewidth]{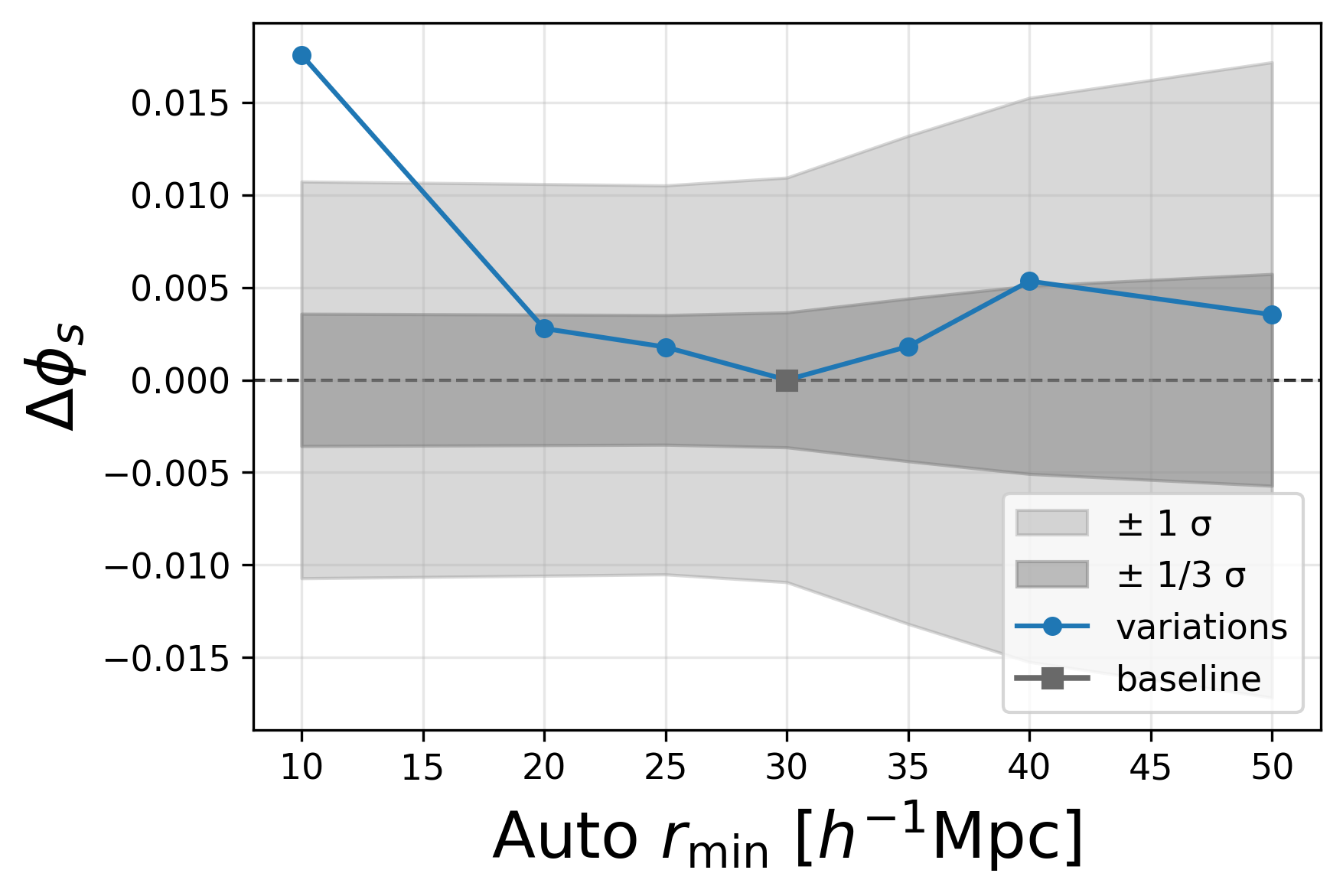}
    \end{minipage}

    \vspace{0.5em} 

    \begin{minipage}{\linewidth}
        \centering
        \includegraphics[width=\linewidth]{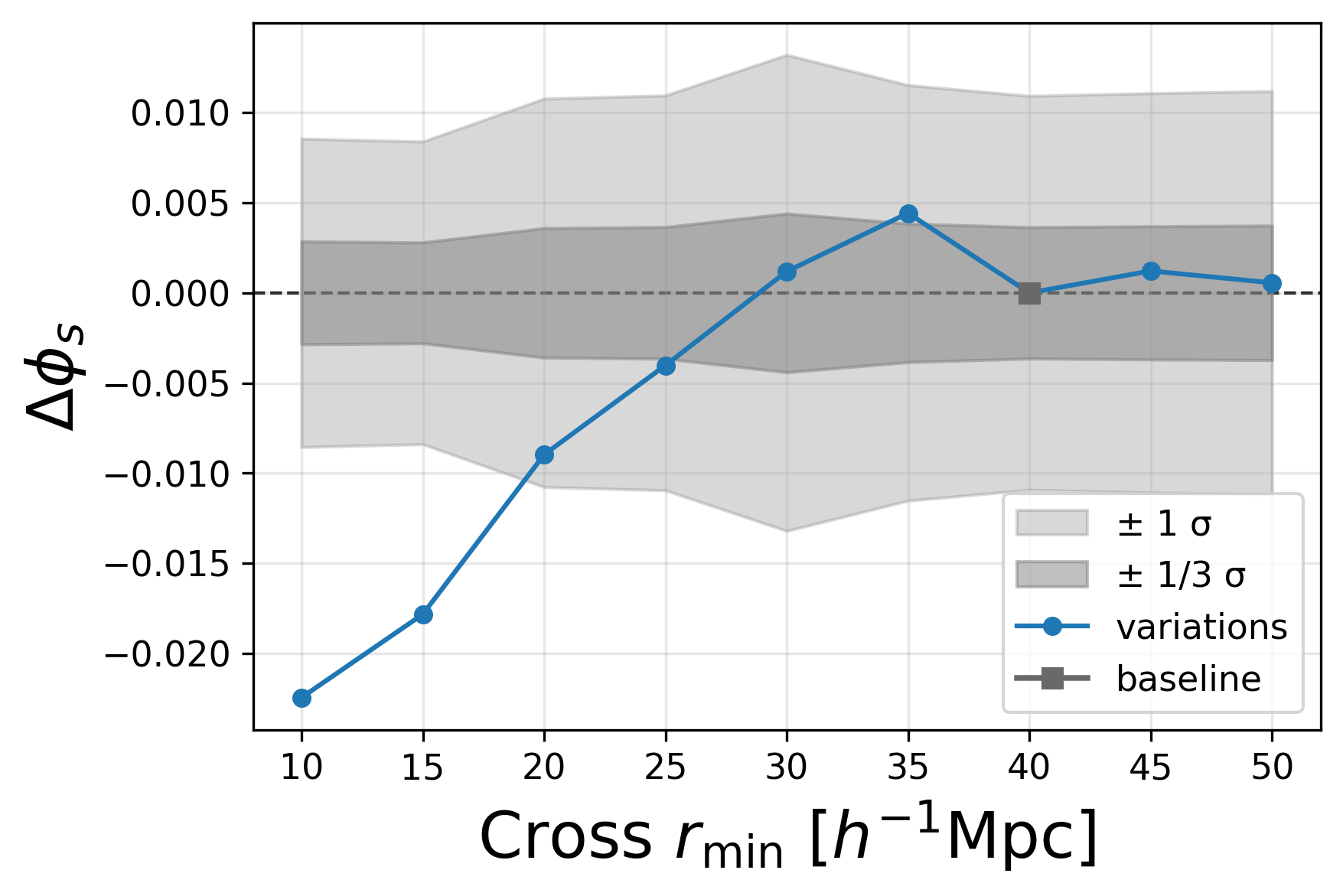}
    \end{minipage}
    
    \caption{Shifts in the AP measurement from a set of analysis variations that vary only the minimum scale cut in the Auto (top) and Cross (bottom) correlations. The baseline value (30 $\hMpc$ for Auto, and 40 $\hMpc$ for Cross) is centered at zero (gray) and variations (blue) are plotted with respect to the baseline value. The Auto (Cross) minimum scale cut is held constant while the Cross (Auto) value is varied. The shaded regions represent 1 (light gray) and 1/3 (dark gray) standard deviation of each variation with respect to the baseline analysis.}
    \label{fig:scale_cuts}
    
\end{figure}

The first subset of modeling variations are changes to scale cuts, the results of which are shown in Figure \ref{fig:scale_cuts}. The baseline analysis utilizes a single maximum fitted scale of $r_{\text{max}}<180\,\hMpc$ for both the auto- and cross-correlation functions, and a minimum fitted scale of $r_{\text{min}}>30\,\hMpc$ for the auto-correlation and $r_{\text{min}}>40\,\hMpc$  for the cross-correlation. We consider scale cut variations that are more or less conservative than the baseline analysis in both the auto- and cross-correlations, which can be seen by the top and bottom panels of Figure \ref{fig:scale_cuts}, respectively. Because these variations result in changes to which data, or what physical scales contribute to our results, some level of statistical fluctuations are expected. 

In Figure \ref{fig:scale_cuts}, we show results for different scale cuts. In the top plot, we vary the minimum separation included in the auto-correlation measurements, while keeping $r_{\rm min}$ fixed to $40\,\hMpc$ for the cross-correlations. In the bottom plot, we vary the minimum separation included in the cross-correlation measurements, while keeping $r_{\rm min}$ fixed to $30\,\hMpc$ for the auto-correlations. The dark and light gray shaded regions indicate the $1/3\,\sigma$ and  $1\,\sigma$  thresholds for each configuration of scale cuts, respectively.

These results show that \phis\ converges at roughly $r_{\rm min} \geq 20\,\hMpc$ in the auto-correlation and $r_{\rm min} \geq 30\,\hMpc$ in the cross-correlation. Note that we expect these results to be impacted by statistical fluctuations because the smallest scales in a fit contribute significantly to \phis, which means that the data driving these values can be quite different for large changes in $r_{\rm min}$. Therefore, we consider this test as a good validation of our results, but have not directly used it to choose our scale cuts, and instead relied on the same tests performed on stacks of mocks, which is detailed in \S \ref{Section: Mocks}.

\begin{figure}[htbp]
        \centering
        \includegraphics[width=1.0\linewidth]{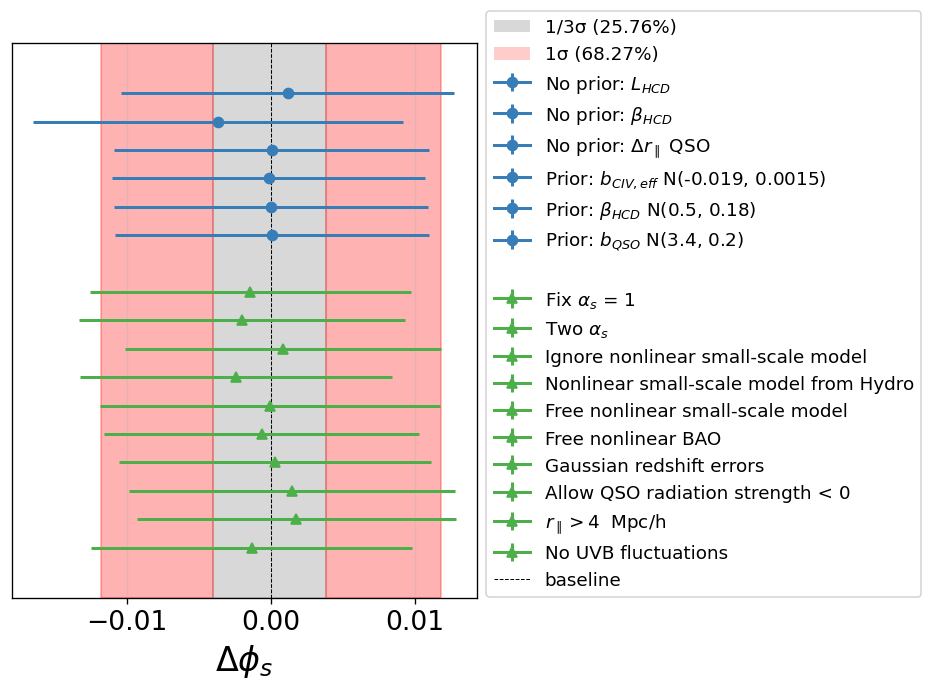}
        \caption{Shifts in the AP measurement from a set of analysis variations which change the model used to fit the correlation functions. This includes variations of the priors (blue, circle) and various modeling choices (green, triangle). The shaded regions represent the uncertainty in the baseline measurement, at 1 (red) and 1/3 (gray) standard deviation.
        }
        \label{fig: variations model}
    \end{figure}

The final selection of variations are variations in choices of priors and variations in the model used to fit the correlation functions. The shifts in  \phis\ as well as uncertainties compared to the baseline analysis are shown in Figure \ref{fig: variations model}. The blue (circle) points in Figure \ref{fig: variations model} demonstrate shifts from alternative prior choices for certain parameters, while the green (triangle) points demonstrate alternative modeling choices. 

The list of alternate prior variation tests are as follows: 

\begin{itemize}
     \item \textbf{no prior on $L_{\rm HCD}$}: we remove the informative Gaussian prior on the parameter $L_{\rm HCD}$ ($\mathcal{N}(5.0, 2.0)$) in the baseline analysis), in the model of HCD contamination.

     \item \textbf{no prior on $\beta_{\rm HCD}$}: we remove the informative Gaussian prior on the parameter $\beta_{\rm HCD}$ ($\mathcal{N}(0.5, 0.09)$) in the baseline analysis), in the model of HCD contamination. As this result is close to the threshold, we have also included an alternative test of the $\beta_{\rm HCD}$ prior below.

     \item \textbf{no prior on $\Delta r_\parallel$ QSO}: we remove the informative Gaussian prior on the parameter $\Delta r_\parallel$ ($\mathcal{N}(0.0, 1.0)$ in the baseline analysis) in the model of quasar redshift errors. 

    \item \textbf{prior on $b_{\rm CIV, eff}$}: we use an alternative informative prior on the parameter $b_{\rm CIV, eff}$ in the model of the contamination by \civ\ absorption. We use the 2024 measurement of  ($\mathcal{N}(-0.019, 0.0015)$ by \cite{KP6s5-Guy}, rather than the baseline prior,  ($\mathcal{N}(-0.019, 0.005)$.

    \item \textbf{prior on $\beta_{\rm HCD}$}: we double the width of the informative prior on the parameter $\beta_{\rm HCD}$ in the model of the contamination by HCDs. The prior in the baseline analysis is N(0.5, 0.09), and is widened to N(0.5, 0.18) in this variation. 

    \item \textbf{prior on $b_{QSO}$}: we add an informative prior on the quasar bias parameter N(3.4, 0.2), following the DESI DR2 BAO analysis \cite{DESI.DR2.BAO.lya}. The baseline analysis does not use a prior.  

\end{itemize}

The list of model variation tests that we conduct are as follows: 

\begin{itemize}

     \item \textbf{fix \alphas}: we do not fit for the isotropic scale of the broadband, \alphas\, as in the baseline analysis, and instead set it equal to one (i.e., fixed to the Planck 2018 \cite{Planck2018} fiducial cosmology).
    
     \item  \textbf{two \alphas}: we fit for the isotropic scale of the broadband separately in the auto- and cross-correlations, whereas the baseline analysis fits a single parameter in both correlations. 
    
     \item \textbf{ignore nonlinear small-scales correction}: we omit the small-scale correction from \cite{Arinyo2015} in the modeling of the \lya\ auto-correlation. The parameters used in this small-scale correction are fixed in the baseline analysis. 
    
    \item \textbf{nonlinear small-scales correction from Hydro}: we fix the small-scale correction parameters from \cite{Arinyo2015} in the modeling of the \lya\ auto-correlation to the best-fit values from hydro-simulations \cite{ACCEL2_2024}. The parameters used in this small scale correction are fixed in the baseline analysis to the best-fit values from simulations \cite{ForestFlow_2025} with tight priors from DESI DR1 P1D measurements \cite{Karacayli_2025, Ravoux_2025}.
    
    \item \textbf{free nonlinear small-scales correction}: we fit for the (six) small-scale correction parameters from \cite{Arinyo2015} in the modeling of the \lya\ auto-correlation, with Gaussian priors ($q_1: \mathcal{N}(1,2)$, $q_2: \mathcal{N}(0,1)$, $k_{\nu}: \mathcal{N}(1,2)$, $a_{\nu}: \mathcal{N}(0.3,0.5)$, $b_{\nu}: \mathcal{N}(1.6,0.5)$, $k_p: \mathcal{N}(14,10)$). The parameters used in this small-scale correction are fixed in the baseline analysis ($q_1: 0.303$, $q_2: 0.267$, $k_{\nu}: 0.576$, $a_{\nu}: 0.443$, $b_{\nu}: 1.66$, $k_p: 11.062$). 
    
    \item \textbf{free non-linear BAO}: we marginalize over the parameter that models the non-linear broadening of the BAO peak. This parameter is fixed to a value predicted by Lagrangian Perturbation Theory (LPT) in the baseline analysis. 
    
    \item \textbf{Gaussian redshift errors}: we use a Gaussian distribution to model quasar redshift errors and peculiar velocities. We use a Lorentzian model in the baseline analysis.
        
    \item \textbf{allow QSO radiation strength $< 0$}: we allow for the parameter characterizing quasar radiation strength to be negative. We require it to be positive in the baseline analysis. 
    
     \item \textbf{$r_{\parallel}$}\ > \ 4  $\hMpc$: we impose a cut below $r_{\parallel}=4 \,\hMpc$ for the auto-correlation only  (i.e., the first bin along the line-of-sight), and do not fit for DESI instrument systematics. This is a more realistic and conservative alternative to an older (now depreciated) model in which we removed the model for sky contamination while still fitting the bins that are impacted by this effect ($r_{\parallel}=0\,\hMpc$). See Appendix \ref{Appendix: Old Variations} for further discussion. 

    \item \textbf{no UVB fluctuations}: we do not model the impact of UV background fluctuations \cite{Pontzen2014,Gontcho2014} on the \lya\ forest auto-correlation, following the prescription of \cite{Bautista2017}.

\end{itemize}

Of all the validation tests related to choices in priors and modeling changes, none exceed more than one third of the expected statistical uncertainty from the (blinded) baseline analysis, and are therefore not problematic. However, the variation in which we remove the prior on  $\beta_{\rm HCD}$ did produce a notable shift relative to the baseline analysis, and is discussed further in \S \ref{Section: Discussion}.

\section{Discussion} \label{Section: Discussion}

We have presented numerous validation tests of our baseline analysis using synthetic spectra (\S \ref{Section: Mocks}), data splits (\S \ref{Section: data splits}), variations at the delta fluctuation level (\S \ref{Section: Fluctuations}), variations at the level of correlations (\S \ref{Section: Correlations}), and variations of the priors and modeling choices (\S \ref{Section: Model Variations}). Several of those tests produced significant shifts and in some cases exceeded our thresholds. We next discuss how we investigated and mitigated those cases in greater detail.

\subsection{Synthetic Spectra} \label{Section: Discussion-mocks}

Our full-shape analysis of 400 \clpt\ mocks demonstrated that both BAO parameters, \alphap\ and \phip, and the broadband AP parameter, \phis, are within our conservative threshold for unblinding. Additionally, we found \alphap\ and \phis\ are consistent with the truth, while \phip\ displays a $0.6\%$ bias at $\sim5\sigma$ significance. This is consistent in sign and magnitude with the results from the \textsc{AbacusSummit} simulations. Although this is still within the threshold for this analysis, the shift should be carefully handled in the systematic error budget, which is discussed in detail in \cite{Cuceu:2026}. The DR2 BAO analysis \cite{DESI.DR2.BAO.lya} accounted for a possible shift in the BAO peak due to nonlinear clustering with the inclusion of an isotropic systematic uncertainty to the total error budget. However, our full-shape tests indicate that this shift is primarily related to the anisotropic parameter \phip\, and therefore should be independently considered as an uncertainty, rather than using an equal, isotropic contribution to $\alpha_\perp$ and $\alpha_\parallel$. Additionally, though we report \phis\ for our validation tests in this work, as it was a blinded parameter, the final result will combine smooth and peak components (\phis\ and \phip\ ) into a single AP parameter, \phif.

Finally, we found the growth rate parameter \fsig\ to be outside of the threshold, as it displays a $10\%$ bias at high significance ($\gg5\sigma$). Despite numerous tests, and variations of scale cuts for the auto- and cross-correlations, we have found no reasonable change in our model that fully reconciles this measurement with the truth or significantly reduces the bias. Due to this large systematic bias, we do not report a measurement of \fsig\ in the main analysis, and therefore do not present the validation test results on blinded data with \phis\ in \S \ref{Section: Alternative Analyses}, as was done in the DR1 full-shape analysis \cite{Cuceu:2025}. Plots for \fsig\ are in Appendix \ref{Appendix: Growth Rate}.

\subsection{Blinded Data} \label{Section: Discussion-variations} 

Prior to unblinding the baseline analysis, we defined a threshold of one third of the expected statistical uncertainty from the (blinded) baseline analysis, corresponding to a tolerance of $\sim 0.0038$ for \phis. Of the many analysis variation tests that we performed, the only variation that exceeded our threshold is $\Delta\lambda=2.4$\AA. In this test, we rebin the \lya\ flux from the native DESI resolution of $0.8$\AA\ to $2.4$\AA\ before continuum fitting (i.e., at the level of the measured flux). This rebinning is done after continuum fitting (i.e., at the level of \lya\ fluctuations) in the baseline analysis.  

This test affects the data in several ways, which are expected to introduce small statistical fluctuations. The continuum fitting step includes a number of operations that would produce slightly different results when using a different binning scheme, including the masking of sky lines, the masking of DLAs and BALs, and the computation of the variance in the forest region, all of which affect the weights used for both continuum fitting and the measurement of the correlations. We consider that using the native DESI resolution for the continuum fitting step is the superior choice, which is the standard used for all previous DESI \lya\ analyses. 

To better understand and quantify the significance of the observed shift, we performed this variation on a set of 10 \clpt\  mocks prior to unblinding. We found that the distribution of shifts between the baseline analysis and this variation is consistent with zero, which indicates no significant systematic error, and has an RMS of $\sim0.25\%$ (indicating the level of statistical fluctuation). Comparing the RMS of $\sim0.25\%$ on a stack of mocks to the observed shift of $0.68\%$ indicates a significance of $\sim2.7\sigma$. We have also checked this using a larger set of bootstrap realizations generated from the data measurements of the correlation functions in Healpix pixels on the sky and found a shift of $0.0065 \pm 0.0023$. This corresponds to a significance of $\sim2.8\sigma$, which is consistent with the results on mocks. Given that we have performed 19 tests in the analysis variations category (\ref{Section: Alternative Analyses}), which are expected to exhibit some level of statistical fluctuation, the probability of observing a shift at this level of significance is $\sim10\%$. We have therefore concluded that this shift can be explained by a statistical fluctuation. In addition, we also tested the effect of using smaller wavelength bins of $1.6$\AA\ instead of $2.4$\AA\ when computing the correlation functions. That choice gives minimal deviations from the baseline analysis.
    
Another variation test that produces a large change, though it does not exceed our threshold, is when we remove the prior on the $\beta_{\rm HCD}$ parameter. The choice of this prior is based on measurements from BOSS \lya-DLA cross-correlations \cite{PerezRafols2018}. The fact that the \phis\ shift is close to the threshold when removing this prior indicates that our measurement retains some sensitivity to $\beta_{\rm HCD}$, making an informative prior necessary. Indeed, while this test passes, it may fail in future analyses with tighter statistical constraints. We therefore perform an additional test in which the width of the Gaussian prior is doubled. This produces no measurable change in \phis, indicating that while the parameter should be constrained by a physically motivated prior, the results are not sensitive to the exact width of a reasonable prior at this time. Future measurements from DESI \lya-DLA cross-correlations could provide a more precise constraint on $\beta_{\rm HCD}$, but the current prior is sufficient for the present analysis.

\section{Conclusion}\label{Section: Conclusion}

In this work, we have performed a comprehensive validation of the DESI DR2 Lyman-$\alpha$ forest full-shape analysis presented in \cite{Cuceu:2026}. Using both synthetic or mock datasets and blinded observational data, we evaluated the accuracy, robustness, and reliability of the analysis pipeline and cosmological parameter measurements, and the reliability of the resulting cosmological constraints. We described the DR2 datasets and catalogs used in this work in \S\ref{section: data}, the analysis methodology in \S\ref{section: analysis}, and the principal methodological developments relative to the DR2 BAO analysis \cite{DESI.DR2.BAO.lya} in \S\ref{section: changes since dr2 bao}. We then presented validation tests on mock datasets in \S\ref{Section: Mocks}, robustness tests on blinded observational data--including data splits and alternative analysis configurations--in \S\ref{Section: data splits} and \S\ref{Section: Alternative Analyses}, and discussed the investigation of potentially concerning results and their resolution in \S\ref{Section: Discussion}. 

Compared to the DR1 full-shape analysis, the DR2 analysis benefits from substantial improvements in both the datasets and the analysis methodology, which motivated a correspondingly more extensive validation effort. In addition to mock datasets that are both larger in number and more realistic than those available for DR1,  the DR2 analysis employs updated catalogs and significant methodological developments relative to the DR2 \lya\ BAO analysis \cite{DESI.DR2.BAO.lya}. These developments improve the fidelity of both the data and modeling framework, but also introduce additional complexity and potential sources of systematic uncertainty. The validation program presented in this work was therefore designed to verify that these choices do not introduce significant biases in the recovered cosmological parameters, and that the resulting measurements remain robust to reasonable variations in the data, modeling assumptions, and analysis configuration.

Our tests on mock datasets have shown that the modeling choices and scale cuts recover the expected cosmology within the predefined validation thresholds for the BAO and AP measurements. These conclusions are further supported by population tests on individual mock realizations, which recover parameter uncertainties that are consistent with the data, and recover BAO and AP parameter values that are consistent with the input cosmology. Additional fiducial cosmology tests, presented in Appendix \ref{Appendix: Fiducial Cosmology}, demonstrate that the measurements remain stable under reasonable changes to the assumed cosmological model, with any observed shifts consistent with statistical fluctuations. While a small bias was observed for \phip\ and a more significant discrepancy was found for \fsig, the latter was consistently identified across multiple mock-based tests and ultimately led to our decision to de-scope \fsig\ from the primary cosmological results.

The blinded-data validation tests similarly demonstrate a high degree of robustness. All catalog and correlation-function data splits satisfy the predefined validation criteria and are consistent with the baseline analysis. Likewise, nearly all alternative analyses—including variations in the treatment of fluctuations, correlations, covariances, distortion matrices, modeling assumptions, and priors—produced shifts substantially below the adopted tolerance threshold. A single analysis variation involving rebinning the flux prior to continuum fitting exceeded the nominal threshold, was investigated in detail with mock datasets and alternative binning schemes, and no evidence for systematic bias introduced by baseline analysis choices was found. 

Furthermore, the baseline model provides a satisfactory description of the DR2 dataset, yielding a goodness-of-fit consistent with statistical expectations with a reduced $\chi^2 = 1.011$, and PTE = $0.20$. Additionally, the covariance-normalized residuals (Figure \ref{fig:residuals}), which are consistent with the expected standard normal distribution, provide an independent validation of both the covariance model and the assumed Gaussian likelihood. Together, these results indicate that the adopted modeling framework accurately captures the statistical properties of the observed correlations over the scales used in the analysis. 

The agreement between mock-based validation tests, blinded-data robustness studies, and the goodness of fit of the baseline model provides strong evidence that the DR2 \lya\ full-shape measurements are stable against reasonable variations in the data and analysis methodology, and that the quoted uncertainties accurately capture the dominant sources of statistical error. These validation results provide the foundation for the cosmological interpretation presented in the DESI DR2 \lya\ full-shape measurement and cosmological analysis paper \cite{Cuceu:2026}. More broadly, the validation framework developed in this work establishes a template for future DESI \lya\ analyses, including upcoming DR3 BAO and future full-shape measurements, which will benefit from larger datasets, improved simulations, and continued refinements to the modeling of astrophysical and instrumental systematics. As the statistical precision of DESI continues to improve, similarly rigorous validation efforts will remain essential to ensuring the robustness and reliability of cosmological constraints derived from the \lya\ forest.

\begin{acknowledgments}

MH and PM acknowledge support from the United States Department of Energy, Office of High Energy Physics under Award Number DE-SC0011726. AC acknowledges support provided by NASA through the NASA Hubble Fellowship grant HST-HF2-51526.001-A awarded by the Space Telescope Science Institute, which is operated by the Association of Universities for Research in Astronomy, Incorporated, under NASA contract NAS5-26555.

This material is based upon work supported by the U.S. Department of Energy (DOE), Office of Science, Office of High-Energy Physics, under Contract No. DE–AC02–05CH11231, and by the National Energy Research Scientific Computing Center, a DOE Office of Science User Facility under the same contract. Additional support for DESI was provided by the U.S. National Science Foundation (NSF), Division of Astronomical Sciences under Contract No. AST-0950945 to the NSF’s National Optical-Infrared Astronomy Research Laboratory; the Science and Technology Facilities Council of the United Kingdom; the Gordon and Betty Moore Foundation; the Heising-Simons Foundation; the French Alternative Energies and Atomic Energy Commission (CEA); the National Council of Humanities, Science and Technology of Mexico (CONAHCYT); the Ministry of Science, Innovation and Universities of Spain (MICIU/AEI/10.13039/501100011033), and by the DESI Member Institutions: \url{https://www.desi.lbl.gov/collaborating-institutions}. Any opinions, findings, and conclusions or recommendations expressed in this material are those of the author(s) and do not necessarily reflect the views of the U. S. National Science Foundation, the U. S. Department of Energy, or any of the listed funding agencies.

The authors are honored to be permitted to conduct scientific research on I'oligam Du'ag (Kitt Peak), a mountain with particular significance to the Tohono O’odham Nation.

\end{acknowledgments}

\appendix

\section{Data Availability} \label{Appendix: Data Availability}
Information and data associated with current and future DESI data releases, including the Early Data Release (EDR) and Data Release 1 (DR1), are publicly available at \url{https://data.desi.lbl.gov/doc/releases/}. Data Release 2 (DR2), referenced in this work, will be available on the same webpage. 

The data points corresponding to the figures from this paper will be available in a Zenodo repository (\url{https://zenodo.org}); the exact link will be provided upon publication.

\section{Fiducial Cosmology on Mocks} \label{Appendix: Fiducial Cosmology}

Our mocks were generated with a fixed set of cosmological parameters and in this section we assess the robustness of our analysis to this fiducial cosmology choice. Specifically, we performed validation tests that vary the fiducial cosmology that is used to measure the correlation functions and perform the fits. In all of these variations, we generate a new template to describe the new cosmology, recompute correlations for several mocks with the new cosmology, and then run the full baseline analysis on the stack of those mocks. 

Since the cosmological inference uses both the AP signal from the BAO and broadband, the fits are performed to the full-shape AP signal, and therefore we show \phif\ rather than \phis.  We investigate two alternative cosmologies in this section: a flat \lcdm\ with a fixed sound horizon and dynamic dark energy model ($w_0-w_a$).

\begin{figure}
    \centering
    \begin{minipage}{0.9\linewidth}
        \centering
        \includegraphics[width=\linewidth]{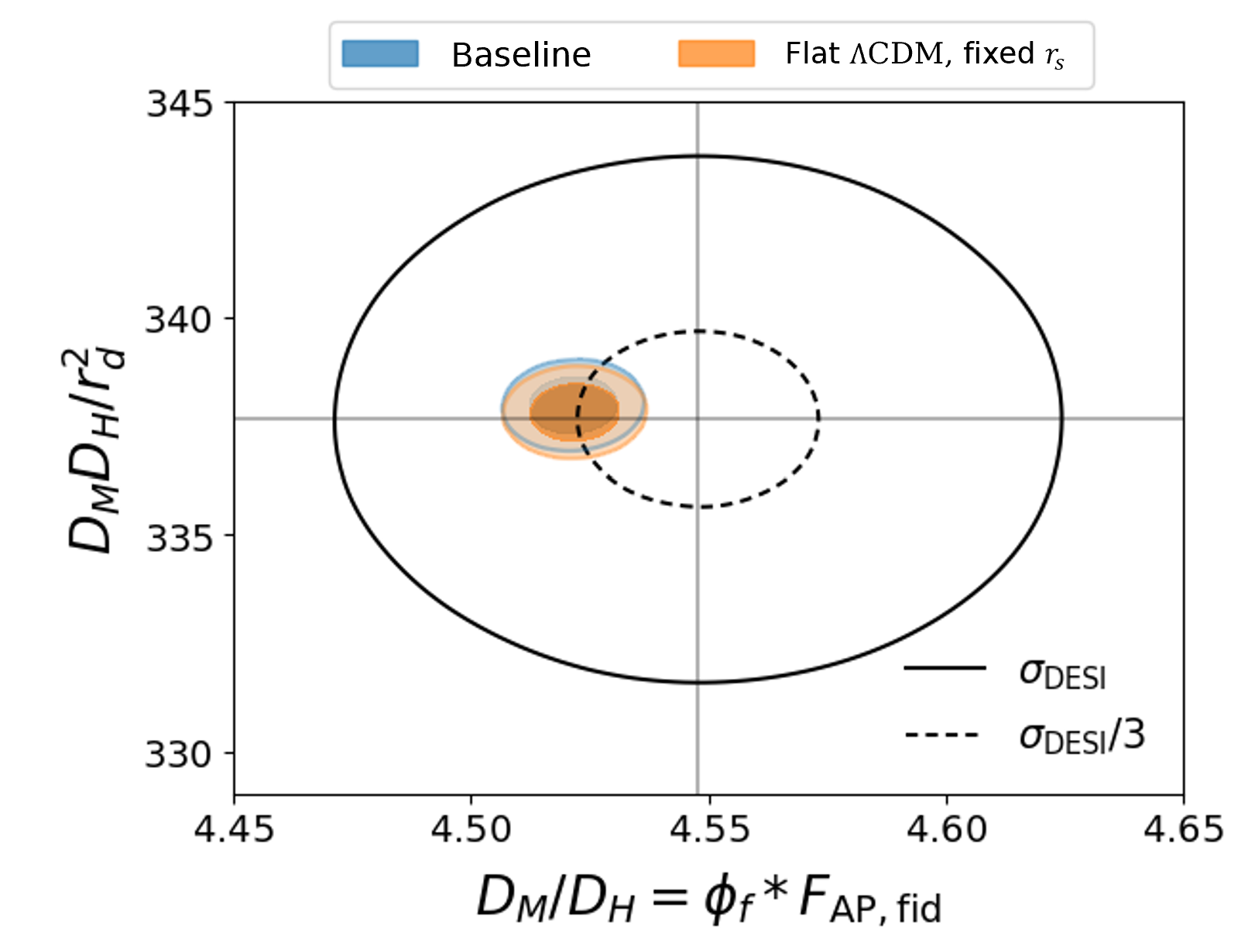}
    \end{minipage}
    
    \vspace{0.5em} 

    \begin{minipage}{0.9\linewidth}
      \includegraphics[width=\linewidth]{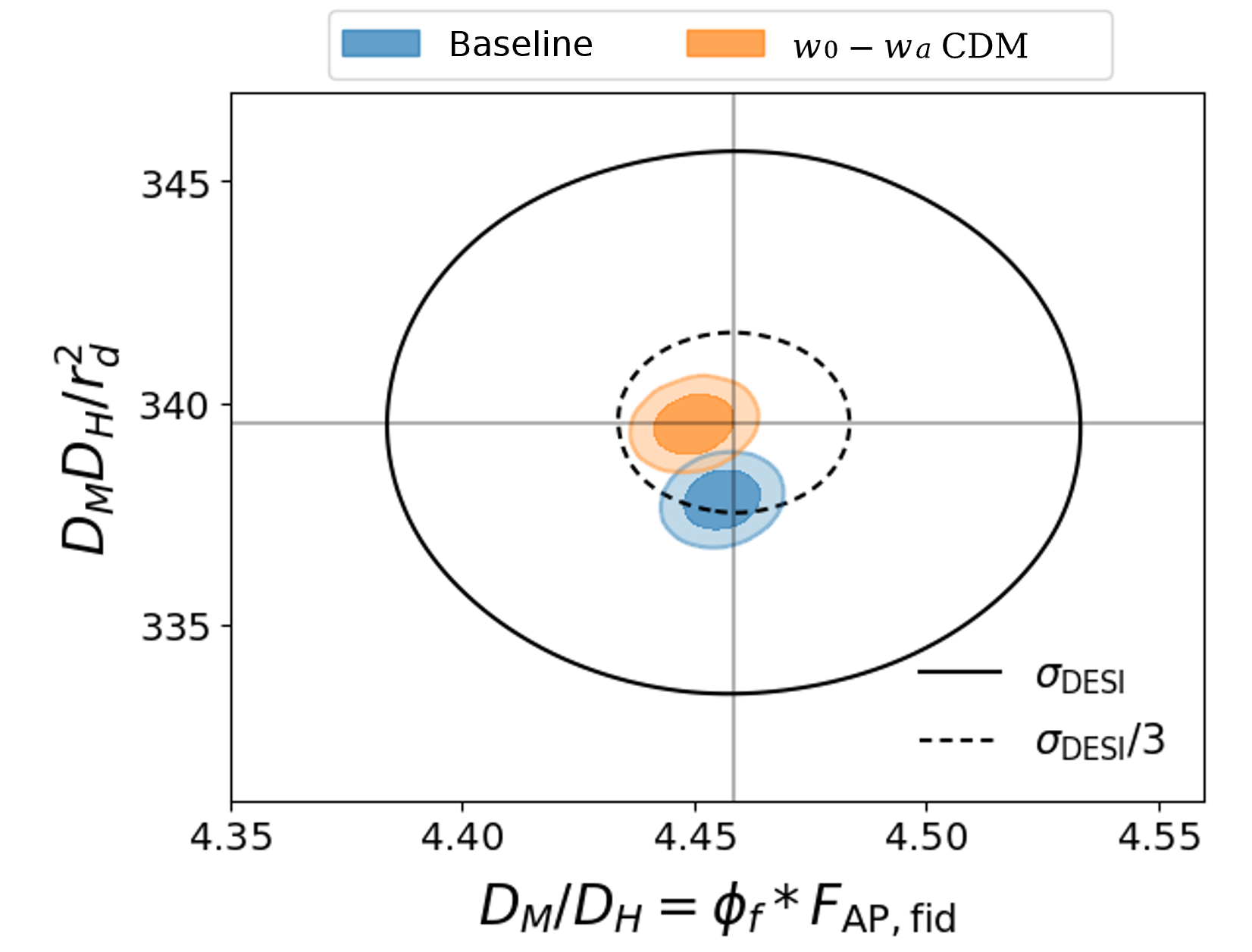}
    \end{minipage}
    
    \caption{
        Full-shape fit results from a stack of 100 mock correlations computed with different fiducial cosmologies.  (Top) We vary cosmological parameters within $\Lambda$CDM using a set of 100 \clpt\ mocks. (Bottom) We vary cosmological parameters within $w_0w_a$CDM using a set of 100 \texttt{IFAE-QL} mocks. The derived distance ratio from full-shape AP constraints (\phif), which includes AP from both BAO and broadband, is shown on the x-axes. The derived distances from the isotropic BAO constraint (\alphap), is shown on the y-axes. Both alternative fiducial cosmologies demonstrate excellent agreement in the derived distance ratios from \phif. 
        }
    \label{fig:fid_cosmo}
\end{figure}

\subsubsection{Flat $\Lambda$CDM with fixed sound horizon}\label{section:fid_cosmo_1}

The first fiducial cosmology test is a flat \lcdm\ universe with a fixed sound horizon. The purpose of this test is to verify that the recovered full-shape AP measurement is insensitive to the assumed fiducial cosmology when converting observed redshifts and angular positions into comoving distances. By holding the sound horizon $r_d$ fixed, the isotropic component is unchanged, allowing this test to isolate the impact of the geometric distortion introduced by changing the fiducial cosmology. We select a subset of 100 \clpt\ mocks  with true cosmology defined by Planck 2018 \cite{Planck2018}. We then modify $\Omega_\mathrm{m}$ such that $F_\mathrm{AP, new}/F_\mathrm{AP, truth} = \frac{[D_M/D_H]_\mathrm{new}}{[D_M/D_H]_\mathrm{truth}} \approx 1.032$. This corresponds to a $\sim3\sigma$ shift in the AP parameter. Additionally, we tune the Hubble parameter $H_0$ and density parameters to hold all physical densities (e.g., $\Omega_\mathrm{b}, h^2$) fixed relative to the truth. However, we leave the sound horizon $r_d$ unchanged. This results in a new template power spectrum $\mathrm{P}(k)$ that remains unchanged in units of $\mathrm{Mpc}^{-1}$, but varies in units of $\hMpc$. The new fiducial cosmology is $\Omega_\mathrm{m} = 0.375$, $\Omega_\mathrm{b} = 0.0587$, and $H_0 = 61.755 \, \mathrm{km\,s^{-1}\,Mpc^{-1}}$.

The top panel of Figure \ref{fig:fid_cosmo} shows the results of the full-shape fits for both the baseline (blue) and different fiducial cosmology (orange). We find that the alternative fiducial cosmology is in good agreement with the baseline analysis, demonstrating that the analysis and results are robust to changes in the assumed cosmology. For this subset of mocks, the isotropic BAO result (y-axis) aligns well with the true cosmology, however, the AP results (x-axis) are slightly shifted by $\sim 1/3\ \sigma$. Given that the full set of 400 mocks is consistent with the truth, this small shift can be explained by a statistical fluctuation. 

\subsubsection{$w_0-w_a$ model}\label{section:fid_cosmo_w0wa}

The other alternative fiducial cosmology that we test is the best-fit $w_0w_a$CDM cosmology from the DESI-DR2 + CMB analysis \cite{DESI.DR2.BAO.cosmo} where the dark energy equation of state is parameterized by $w(a)=w_0+w_a(1-a)$. For this test, we use a stack of 100 contaminated \texttt{IFAE-QL} mocks, which are based on the Planck 2015 cosmology (see \cite{Casas2025}). We perform a joint fit of the A-region correlations, Ly$\alpha$(A)×Ly$\alpha$(A) and Ly$\alpha$(A)×QSO. Here, the new fiducial cosmology is defined by $\Omega_\mathrm{m} = 0.353$, $H_0 = 63.6 \, \mathrm{km\,s^{-1}\,Mpc^{-1}}$, $w_0 = -0.42$, and $w_a = -1.75$.

The bottom panel of Figure \ref{fig:fid_cosmo} shows the results of fits performed using the baseline analysis (blue) and the alternative $w_0-w_a$ model. We find that the AP measurements (x-axis) are in good agreement with the baseline analysis. The two isotropic BAO constraints (y-axis) displays a small shift, however the shift is within the threshold of 1/3 $\sigma$ of the expected statistical uncertainty. Therefore, we find that the results are robust to alternative fiducial cosmology choices at the precision relevant for this analysis.  

\section{Supplementary Validation Results}\label{Appendix: supplementary validation}

This appendix collects several supplementary validation results that complement the primary analysis presented in the main text. These include the behavior of the growth-rate parameter \fsig\ under the analysis variations considered in this work; the corresponding BAO peak-component measurements, which are required to construct the final full-shape constraints; and a set of deprecated validation tests retained for completeness and comparison with previous analyses.

\subsection{Growth Rate} \label{Appendix: Growth Rate}

\begin{figure}
        \centering
        \includegraphics[width=0.9\linewidth]{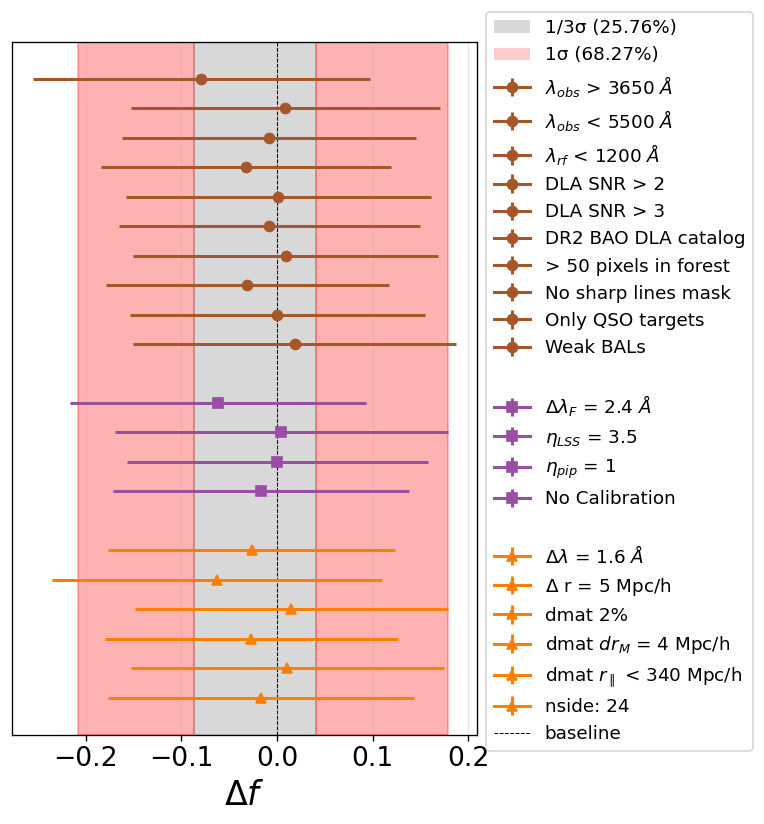}
        \caption{Shifts in the growth rate parameter ($f$) measurement from a set of analysis variations. Analysis variations shown here are those that result in changes to the dataset (brown, circle), changes in how we estimate \lya\ fluctuations (purple, square), and variations that change the method of computing correlations, covariances, and distortion matrices (orange, triangle). The shaded regions represent the uncertainty in the baseline measurement, at 1 (red) and 1/3 (gray) standard deviation.}
        \label{fig:variations_growth_rate}
    \end{figure}

\begin{figure}
        \centering
        \includegraphics[width=0.9\linewidth]{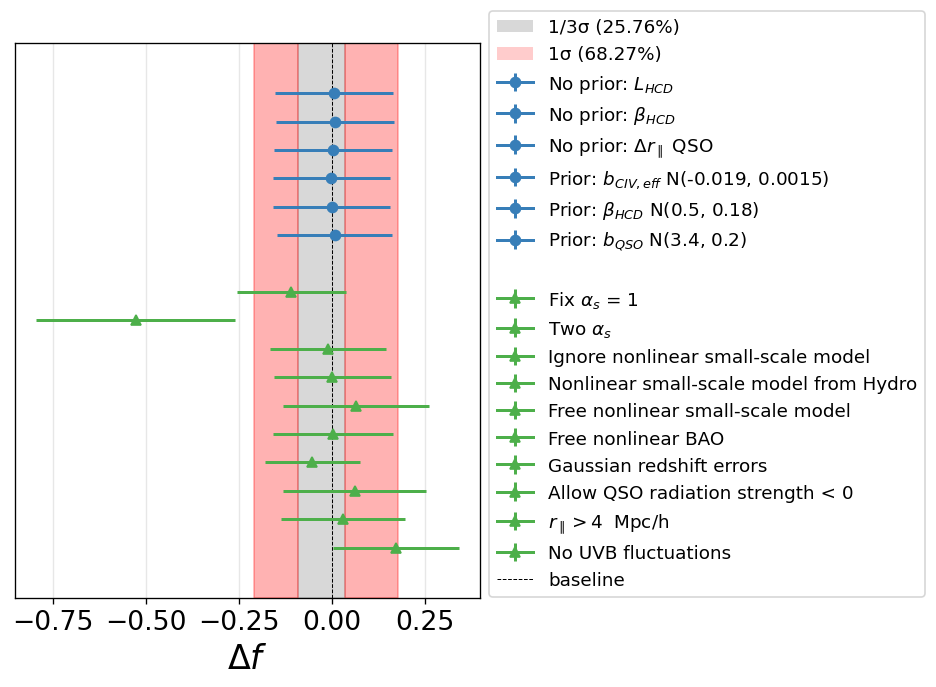}
        \caption{Shifts in the growth rate parameter ($f$) measurement from a set of analysis variations which change the model used to fit the correlation functions. This includes variations of the priors (blue, circle) and various modeling choices (green, triangle). The shaded regions represent the uncertainty in the baseline measurement, at 1 (red) and 1/3 (gray) standard deviation.}
        \label{fig:variations_model_growth_rate}
    \end{figure}

As discussed in \S \ref{Section: Discussion}, we chose to de-scope \fsig\ from the primary analysis because of a significant and unresolved bias observed in the mock data, as well as a lack of robustness under several blinded analysis variations. While \fsig\ remains stable within the one-third statistical uncertainty threshold for variations affecting the measured fluctuations and correlations, as shown in Figure \ref{fig:variations_growth_rate}, and for prior variations, its behavior is substantially less stable under modeling variations, as illustrated in Figure \ref{fig:variations_model_growth_rate}. 

In particular, we find that \fsig\ is sensitive to changes in the modeling of the smooth anisotropic parameter, \alphas\, the quasar radiation strength (proximity effect), and UVB fluctuations. We also find that \fsig\ exhibits significant degeneracies with several model parameters, most notably \alphas. As discussed previously, \alphas\ is largely treated as a nuisance parameter because it is difficult to disentangle from other effects, including the \lya\ flux bias. Combined, these results indicate that the current model lacks sufficient sensitivity to robustly constrain \fsig. We therefore conclude that \fsig\ cannot be reliably interpreted as a cosmological measurement in this analysis and do not report it among our primary results.

\subsection{BAO Parameters} \label{Appendix: BAO params}

The primary focus of this work is the validation of the smooth anisotropic component of the full-shape model, characterized by the parameters \alphas\ and \phis. Consequently, the analysis variations presented throughout the main text are reported in terms of \phis, which was treated as the blinded parameter during the validation process. The BAO component of the model has already undergone extensive validation as part of the DESI DR2 BAO analysis \cite{DESI.DR2.BAO.lya} and is therefore not re-examined in detail here.

Nevertheless, the final full-shape cosmological measurement is obtained by combining the smooth and peak components into the full-shape parameters, \alphaf\ and \phif. The BAO component parameters, \alphap\ and \phip, therefore remain an essential ingredient of the final result. Additionally, as discussed in \S \ref{section: changes since dr2 bao}, this work introduces modeling and catalog changes relative to the DR2 BAO analysis. While we have shown the full-shape measurement to be robust under these modifications, and we do not expect these changes to affect the BAO measurement, we include for completeness the corresponding measurements of these parameters for the same set of analysis variations considered throughout this work. These figures provide a complimentary comparison of the BAO peak component across the validation suite and demonstrate that the variations affecting the smooth component similarly leave the BAO measurements stable, consistent with the conclusions of the dedicated DR2 BAO validation analysis, which is discussed in Appendix B of \cite{DESI.DR2.BAO.lya}.

    \begin{figure}
        \centering
        \includegraphics[width=1\linewidth]{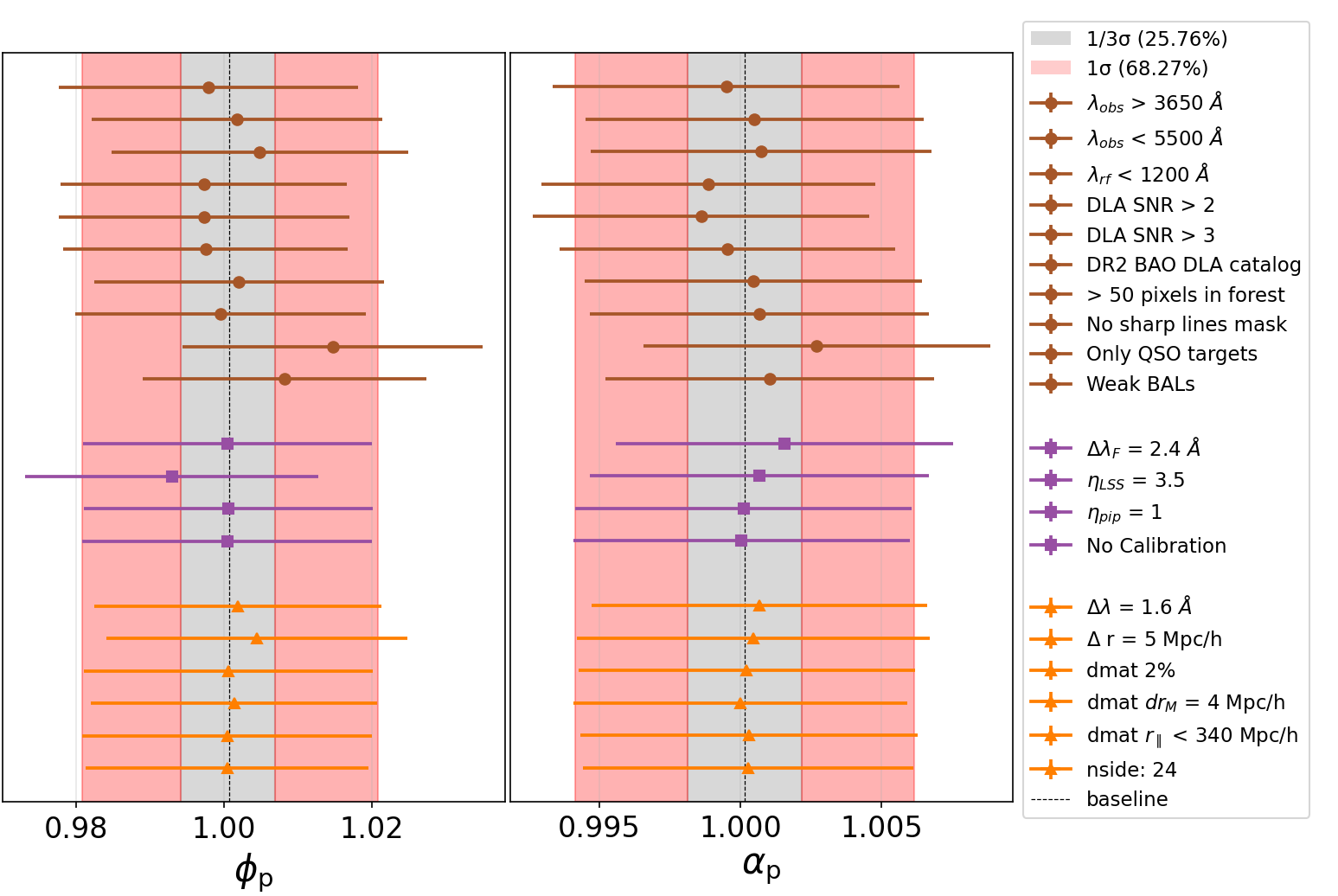}
        \caption{
        Shifts in the measurement of BAO peak parameters (\phip, \alphap) from a set of analysis variations. Analysis variations shown here are those that result in changes to the dataset (brown, circle), changes in how we estimate \lya\ fluctuations (purple, square), and variations that change the method of computing correlations, covariances, and distortion matrices (orange, triangle). The shaded regions represent the uncertainty in the baseline measurement, at 1 (red) and 1/3 (gray) standard deviation.
        }
        \label{fig:BAO_params_data}
    \end{figure}
    
    \begin{figure}
        \centering
        \includegraphics[width=1\linewidth]{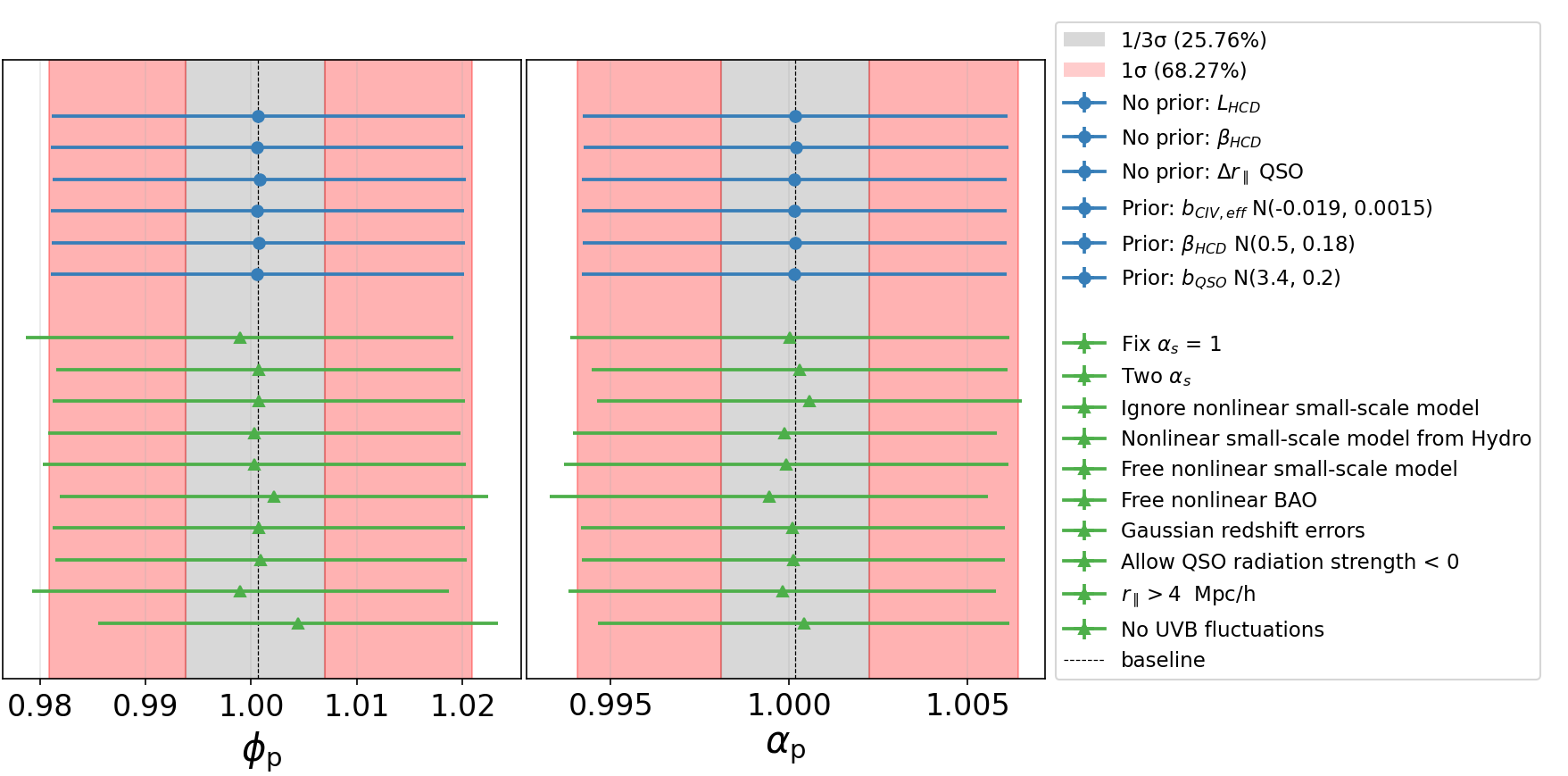}
        \caption{
        Shifts in the measurement of BAO parameters (\phip, \alphap) from a set of analysis variations which change the model used to fit the correlation functions. This includes variations of the priors (blue, circle) and various modeling choices (green, triangle). The shaded regions represent the uncertainty in the baseline measurement, at 1 (red) and 1/3 (gray) standard deviation.
        }
        \label{fig:BAO_params_model}
    \end{figure}

Figures \ref{fig:BAO_params_data} and \ref{fig:BAO_params_model} summarize the shifts in the recovered BAO peak parameters, \alphap\ and \phip, for the same analysis and model variations considered throughout this work. Overall, the BAO measurements are stable across the validation suite, with the vast majority of variations producing shifts well within the adopted $1/3\,\sigma$ threshold. This demonstrates that the modeling and pipeline modifications introduced for the full-shape analysis do not significantly alter the recovered BAO component.

Among the analysis variations, three produce shifts in \phip\ exceeding the $1/3\,\sigma$ threshold: \textbf{$\eta_{\rm LSS}=3.5$}, \textbf{Only QSO Targets}, and \textbf{Weak BALs}. The \textbf{Only QSO Targets} variation also produces a shift larger than $1/3\,\sigma$ in \alphap. Of these, only \textbf{Only QSO Targets} exceeded the corresponding threshold in the DR2 BAO validation, where it produced a significant shift in the transverse BAO parameter. This behavior is unsurprising, as restricting the sample to quasars originally targeted as quasars removes approximately $12\%$ of the quasar catalog used in the baseline analysis, making it one of the largest changes to the dataset considered. The remaining two variations were also among the largest shifts observed in the DR2 BAO analysis, with \textbf{$\eta_{\rm LSS} = 3.5$} lying close to the threshold for the line-of-sight BAO parameter and \textbf{Weak BALs} lying close to the threshold for the transverse BAO parameter. Additionally, we have investigated these three variations, and verified that they pass the threshold in the case of 2D covariance contours. These slightly larger shifts in the full-shape parameterization are therefore consistent with the behavior already observed in the dedicated BAO validation and do not indicate any new sensitivity introduced by the updated full-shape analysis.

\subsection{Deprecated Tests} \label{Appendix: Old Variations}

This appendix contains validation tests that we have depreciated, and therefore their consistency with the baseline analysis was not a requirement for unblinding. These include tests that are inferior or redundant relative to the baseline analysis, or are included for comparison with prior conventions that are no longer relevant.

The first two depreciated validation tests use alternative distortion matrix models that are notably inferior to the baseline analysis; we include these variations here for historical context, as they were used in previous analyses (e.g. \cite{DESI2024.IV.KP6, Cuceu:2025}). The remainder of the tests in this section demonstrate important effects that cannot be ignored, and are therefore not valid alternatives to the baseline analysis. For instance, the test in which the effect of sky residuals is ignored removes the model for a well-known systematic, and produces a significant but understood impact on the AP measurement. The variation "$\Delta \lambda = 3.2$\,\AA", in which we rebin pixels, results in an understood loss of information due to oversmoothing, which was not detected in previous analyses.

\begin{figure}
    \centering
    \includegraphics[width=0.9\columnwidth,keepaspectratio]{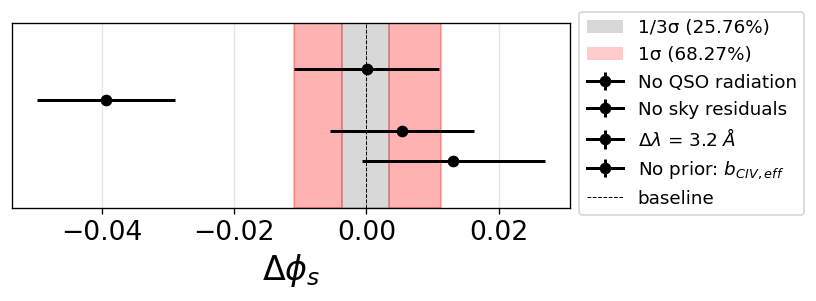}
    \caption{
    Shifts in the AP measurement from a set of analysis variations that have been retired, and therefore were not considered in the path to unblinding. These include variations that were included in previous analyses and inferior models.
    }
    \label{fig:variations_depreciated}
\end{figure}

The list of depreciated variation tests are as follows: 

\begin{itemize}
    
     \item \textbf{dmat $r_\parallel < 200 \,\hMpc$}: we model the distortion matrix up to $r_\parallel = 200 \,\hMpc$, whereas the baseline analysis extends the model to $r_\parallel = 300 \,\hMpc$.  This analysis variation is inferior to the baseline and has been included only for comparison with previous approaches (see \cite{Busca2025}), not as a validation test. In Figure \ref{fig: variations} we introduced a new variation in which we model the distortion matrix up to $r_\parallel = 340 \,\hMpc$.

     \item \textbf{dmat model z-indep}: we use an older method for the computation of the distortion matrix in which we do not model redshift evolution. This is consistent with previous methods, as done in eBOSS and DESI DR1. This method is now considered inferior in the current baseline analysis, but has been included for historical comparison.

     \item \textbf{ignore quasar radiation}: we ignore the impact of quasar radiation in the cross-correlation, known as the transverse proximity effect. While this does not have a significant impact on our baseline \phis\ measurement, as seen in Figure \ref{fig:variations_depreciated}, we decided to retire it because it is an inferior variation. Instead, we added the test "allow negative quasar radiation" as a better alternative to gauge the impact of quasar radiation. 

     \item \textbf{ignore modeling of sky residuals}: we ignore contamination from correlated sky residuals in the \lya\ auto-correlation, as discussed in \cite{KP6s5-Guy}. This test has a very large impact on AP, because it removes the model for a known systematic that directly changes the anisotropy of the \lya\ auto-correlation, showing that this effect cannot be ignored. Therefore, this does not constitute a relevant validation test and has been retired. To test the robustness of our model for sky residuals, we have instead added the test "$r_{\parallel}> 4 \, \hMpc$", which is more relevant because sky residuals only affect $r_{\parallel}=0\,\hMpc$. This alternative test is consistent with the baseline analysis as seen in Figure~\ref{fig: variations model}.
     
     \item \textbf{$\Delta \lambda = 3.2$\,\AA}: we rebin the continuum-fitted \lya\ pixels into groups of four (rebinned pixels of 3.2\,\AA), rather than groups of three (rebinned pixels of 2.4\,\AA), as is done in the baseline analysis. This validation test results in a loss of information along the line-of-sight due to oversmoothing, which directly impacts the anisotropy of the correlation function and therefore AP. This test is therefore not a valid alternative to our baseline analysis.

    \item \textbf{no prior on $b_{\rm CIV, eff}$}: we remove the informative Gaussian prior on the parameter $b_{\rm CIV, eff}$ ($\mathcal{N}(-0.019, 0.005)$ in the baseline analysis) in the model of \civ\ absorption contamination. This test demonstrates that the prior on this parameter cannot be entirely ignored, as this effect is degenerate in our model. The test, "[a]dd prior on $b_{\rm CIV, eff}$," is included as a better alternative in our validation and is consistent with the baseline analysis as seen in Figure~\ref{fig: variations model}.

\end{itemize}

\bibliographystyle{apsrev4-2-5authors}
\bibliography{apssamp}

@article{Ennesser2022,
   title ={The impact and mitigation of broad-absorption-line quasars in Lyman-$\alpha$ forest correlations},
   volume = {511},
   ISSN ={1365-2966},
   url={http://dx.doi.org/10.1093/mnras/stac301},
   DOI={10.1093/mnras/stac301},
   number={3},
   journal={Monthly Notices of the Royal Astronomical Society},
   publisher={Oxford University Press (OUP)},
   author={Ennesser, Lauren and Martini, Paul and Font-Ribera, Andreu and Pérez-Ràfols, Ignasi},
   year={2022},
   month=feb, pages={3514–3523} }

@article{Bautista2017,
   title={Measurement of baryon acoustic oscillation correlations at z = 2.3 with SDSS DR12 Ly$\alpha$-Forests},
   volume={603},
   ISSN={1432-0746},
   url={http://dx.doi.org/10.1051/0004-6361/201730533},
   DOI={10.1051/0004-6361/201730533},
   journal={Astronomy \& Astrophysics},
   publisher={EDP Sciences},
   author={Bautista, Julian E. and Busca, Nicolás G. and Guy, Julien and Rich, James and Blomqvist, Michael and du Mas des Bourboux, Hélion and Pieri, Matthew M. and Font-Ribera, Andreu and Bailey, Stephen and Delubac, Timothée and Kirkby, David and Le Goff, Jean-Marc and Margala, Daniel and Slosar, Anže and Vazquez, Jose Alberto and Brownstein, Joel R. and Dawson, Kyle S. and Eisenstein, Daniel J. and Miralda-Escudé, Jordi and Noterdaeme, Pasquier and Palanque-Delabrouille, Nathalie and Pâris, Isabelle and Petitjean, Patrick and Ross, Nicholas P. and Schneider, Donald P. and Weinberg, David H. and Yèche, Christophe},
   year={2017},
   month=jun, pages={A12} }

@article{Pontzen2014,
   title={Scale-dependent bias in the baryonic-acoustic-oscillation-scale intergalactic neutral hydrogen},
   volume={89},
   ISSN={1550-2368},
   url={http://dx.doi.org/10.1103/PhysRevD.89.083010},
   DOI={10.1103/physrevd.89.083010},
   number={8},
   journal={Physical Review D},
   publisher={American Physical Society (APS)},
   author={Pontzen, Andrew},
   year={2014},
   month=apr }

@article{Gontcho2014,
   title={On the effect of the ionizing background on the Lyα forest autocorrelation function},
   volume={442},
   ISSN={0035-8711},
   url={http://dx.doi.org/10.1093/mnras/stu860},
   DOI={10.1093/mnras/stu860},
   number={1},
   journal={Monthly Notices of the Royal Astronomical Society},
   publisher={Oxford University Press (OUP)},
   author={Gontcho A Gontcho, Satya and Miralda-Escudé, Jordi and Busca, Nicolás G.},
   year={2014},
   month=jun, pages={187–195} }

@article{Arinyo2015,
   title={The non-linear power spectrum of the Lyman alpha forest},
   volume={2015},
   ISSN={1475-7516},
   url={http://dx.doi.org/10.1088/1475-7516/2015/12/017},
   DOI={10.1088/1475-7516/2015/12/017},
   number={12},
   journal={Journal of Cosmology and Astroparticle Physics},
   publisher={IOP Publishing},
   author={Arinyo-i-Prats, Andreu and Miralda-Escudé, Jordi and Viel, Matteo and Cen, Renyue},
   year={2015},
   month=dec, pages={017–017} }

@article{Baldwin1977,
       author = {{Baldwin}, Jack A.},
        title = "{Luminosity Indicators in the Spectra of Quasi-Stellar Objects}",
      journal = {\apj},
         year = 1977,
        month = jun,
       volume = {214},
        pages = {679-684},
          doi = {10.1086/155294},
       adsurl = {https://ui.adsabs.harvard.edu/abs/1977ApJ...214..679B}
}

@article{Cuceu2021,
    author = {Cuceu, Andrei and Font-Ribera, Andreu and Joachimi, Benjamin and Nadathur, Seshadri},
    title = {Cosmology beyond BAO from the 3D distribution of the Lyman-α forest},
    journal = {Monthly Notices of the Royal Astronomical Society},
    volume = {506},
    number = {4},
    pages = {5439-5450},
    year = {2021},
    month = {07},
    issn = {0035-8711},
    doi = {10.1093/mnras/stab1999},
    url = {https://doi.org/10.1093/mnras/stab1999},
    }

@ARTICLE{Ho21,
       author = {{Ho}, Ming-Feng and {Bird}, Simeon and {Garnett}, Roman},
        title = "{Damped Lyman-{\ensuremath{\alpha}} absorbers from Sloan digital sky survey DR16Q with Gaussian processes}",
      journal = {\mnras},
         year = 2021,
        month = oct,
       volume = {507},
       number = {1},
        pages = {704-719},
          doi = {10.1093/mnras/stab2169},
archivePrefix = {arXiv},
       eprint = {2103.10964},
 primaryClass = {astro-ph.GA},
       adsurl = {https://ui.adsabs.harvard.edu/abs/2021MNRAS.507..704H}
}

@ARTICLE{FiberSystem.Poppett.2024,
	       author = {{Poppett}, Claire and {Tyas}, Luke and {Aguilar}, J. and {Bebek}, Christopher and {Bramall}, D. and {Claybaugh}, T. and {Edelstein}, J. and {Fagrelius}, P. and {Heetderks}, H. and {Jelinsky}, P. and {Jelinsky}, S. and {Lafever}, Robin and {Lambert}, A. and {Lampton}, M. and {Levi}, Michael E. and {Martini}, P. and {Rockosi}, C. and {Schmoll}, J. and {Sharples}, Ray M. and {Sirk}, Martin and {Wishnow}, Edward and {Yu}, Jiaxi and {Ahlen}, S. and {Bault}, A. and {BenZvi}, S. and {Brooks}, D. and {Cole}, S. and {de la Macorra}, A. and {Dey}, Arjun and {Doel}, P. and {Fanning}, K. and {Font-Ribera}, A. and {Forero-Romero}, J.~E. and {Gazta{\~n}aga}, E. and {Gontcho A Gontcho}, S. and {Gonzalez-Morales}, A.~X. and {Hahn}, C. and {Honscheid}, K. and {Jimenez}, J. and {Juneau}, S. and {Kirkby}, D. and {Kremin}, A. and {Landriau}, M. and {Le Guillou}, L. and {Manera}, M. and {Meisner}, A. and {Miquel}, R. and {Moustakas}, J. and {Mueller}, E. and {Mu{\~n}oz-Guti{\'e}rrez}, A. and {Myers}, A.~D. and {Nie}, J. and {Niz}, G. and {Palanque-Delabrouille}, N. and {Percival}, W.~J. and {Prada}, F. and {Rabinowitz}, D. and {Rezaie}, M. and {Rossi}, G. and {Sanchez}, E. and {Schlafly}, Edward F. and {Schlegel}, D. and {Schubnell}, M. and {Seo}, H. and {Sprayberry}, D. and {Tarl{\'e}}, G. and {Vargas-Maga{\~n}a}, M. and {Weaver}, B.~A. and {Zhou}, R.},
	        title = "{Overview of the Fiber System for the Dark Energy Spectroscopic Instrument}",
	      journal = {\aj},
	         year = 2024,
	        month = dec,
	       volume = {168},
	       number = {6},
	          eid = {245},
	        pages = {245},
	          doi = {10.3847/1538-3881/ad76a4},
	       adsurl = {https://ui.adsabs.harvard.edu/abs/2024AJ....168..245P}
	}

@ARTICLE{DESI2016a.Science,
       author = {{DESI Collaboration} and {Aghamousa}, Amir and {Aguilar}, Jessica and {Ahlen}, Steve and {Alam}, Shadab and {Allen}, Lori E. and {Allende Prieto}, Carlos and {Annis}, James and {Bailey}, Stephen and {Balland}, Christophe and {Ballester}, Otger and {Baltay}, Charles and {Beaufore}, Lucas and {Bebek}, Chris and {Beers}, Timothy C. and {Bell}, Eric F. and {Bernal}, Jos{\'e} Luis and {Besuner}, Robert and {Beutler}, Florian and {Blake}, Chris and {Bleuler}, Hannes and {Blomqvist}, Michael and {Blum}, Robert and {Bolton}, Adam S. and {Briceno}, Cesar and {Brooks}, David and {Brownstein}, Joel R. and {Buckley-Geer}, Elizabeth and {Burden}, Angela and {Burtin}, Etienne and {Busca}, Nicolas G. and {Cahn}, Robert N. and {Cai}, Yan-Chuan and {Cardiel-Sas}, Laia and {Carlberg}, Raymond G. and {Carton}, Pierre-Henri and {Casas}, Ricard and {Castander}, Francisco J. and {Cervantes-Cota}, Jorge L. and {Claybaugh}, Todd M. and {Close}, Madeline and {Coker}, Carl T. and {Cole}, Shaun and {Comparat}, Johan and {Cooper}, Andrew P. and {Cousinou}, M. -C. and {Crocce}, Martin and {Cuby}, Jean-Gabriel and {Cunningham}, Daniel P. and {Davis}, Tamara M. and {Dawson}, Kyle S. and {de la Macorra}, Axel and {De Vicente}, Juan and {Delubac}, Timoth{\'e}e and {Derwent}, Mark and {Dey}, Arjun and {Dhungana}, Govinda and {Ding}, Zhejie and {Doel}, Peter and {Duan}, Yutong T. and {Ealet}, Anne and {Edelstein}, Jerry and {Eftekharzadeh}, Sarah and {Eisenstein}, Daniel J. and {Elliott}, Ann and {Escoffier}, St{\'e}phanie and {Evatt}, Matthew and {Fagrelius}, Parker and {Fan}, Xiaohui and {Fanning}, Kevin and {Farahi}, Arya and {Farihi}, Jay and {Favole}, Ginevra and {Feng}, Yu and {Fernandez}, Enrique and {Findlay}, Joseph R. and {Finkbeiner}, Douglas P. and {Fitzpatrick}, Michael J. and {Flaugher}, Brenna and {Flender}, Samuel and {Font-Ribera}, Andreu and {Forero-Romero}, Jaime E. and {Fosalba}, Pablo and {Frenk}, Carlos S. and {Fumagalli}, Michele and {Gaensicke}, Boris T. and {Gallo}, Giuseppe and {Garcia-Bellido}, Juan and {Gaztanaga}, Enrique and {Pietro Gentile Fusillo}, Nicola and {Gerard}, Terry and {Gershkovich}, Irena and {Giannantonio}, Tommaso and {Gillet}, Denis and {Gonzalez-de-Rivera}, Guillermo and {Gonzalez-Perez}, Violeta and {Gott}, Shelby and {Graur}, Or and {Gutierrez}, Gaston and {Guy}, Julien and {Habib}, Salman and {Heetderks}, Henry and {Heetderks}, Ian and {Heitmann}, Katrin and {Hellwing}, Wojciech A. and {Herrera}, David A. and {Ho}, Shirley and {Holland}, Stephen and {Honscheid}, Klaus and {Huff}, Eric and {Hutchinson}, Timothy A. and {Huterer}, Dragan and {Hwang}, Ho Seong and {Illa Laguna}, Joseph Maria and {Ishikawa}, Yuzo and {Jacobs}, Dianna and {Jeffrey}, Niall and {Jelinsky}, Patrick and {Jennings}, Elise and {Jiang}, Linhua and {Jimenez}, Jorge and {Johnson}, Jennifer and {Joyce}, Richard and {Jullo}, Eric and {Juneau}, St{\'e}phanie and {Kama}, Sami and {Karcher}, Armin and {Karkar}, Sonia and {Kehoe}, Robert and {Kennamer}, Noble and {Kent}, Stephen and {Kilbinger}, Martin and {Kim}, Alex G. and {Kirkby}, David and {Kisner}, Theodore and {Kitanidis}, Ellie and {Kneib}, Jean-Paul and {Koposov}, Sergey and {Kovacs}, Eve and {Koyama}, Kazuya and {Kremin}, Anthony and {Kron}, Richard and {Kronig}, Luzius and {Kueter-Young}, Andrea and {Lacey}, Cedric G. and {Lafever}, Robin and {Lahav}, Ofer and {Lambert}, Andrew and {Lampton}, Michael and {Landriau}, Martin and {Lang}, Dustin and {Lauer}, Tod R. and {Le Goff}, Jean-Marc and {Le Guillou}, Laurent and {Le Van Suu}, Auguste and {Lee}, Jae Hyeon and {Lee}, Su-Jeong and {Leitner}, Daniela and {Lesser}, Michael and {Levi}, Michael E. and {L'Huillier}, Benjamin and {Li}, Baojiu and {Liang}, Ming and {Lin}, Huan and {Linder}, Eric and {Loebman}, Sarah R. and Luki{\'c}, Zarija and {Ma}, Jun and {MacCrann}, Niall and {Magneville}, Christophe and {Makarem}, Laleh and {Manera}, Marc and {Manser}, Christopher J. and {Marshall}, Robert and {Martini}, Paul and {Massey}, Richard and {Matheson}, Thomas and {McCauley}, Jeremy and {McDonald}, Patrick and {McGreer}, Ian D. and {Meisner}, Aaron and {Metcalfe}, Nigel and {Miller}, Timothy N. and {Miquel}, Ramon and {Moustakas}, John and {Myers}, Adam and {Naik}, Milind and {Newman}, Jeffrey A. and {Nichol}, Robert C. and {Nicola}, Andrina and {Nicolati da Costa}, Luiz and {Nie}, Jundan and {Niz}, Gustavo and {Norberg}, Peder and {Nord}, Brian and {Norman}, Dara and {Nugent}, Peter and {O'Brien}, Thomas and {Oh}, Minji and {Olsen}, Knut A.~G. and {Padilla}, Cristobal and {Padmanabhan}, Hamsa and {Padmanabhan}, Nikhil and {Palanque-Delabrouille}, Nathalie and {Palmese}, Antonella and {Pappalardo}, Daniel and {P{\^a}ris}, Isabelle and {Park}, Changbom and {Patej}, Anna and {Peacock}, John A. and {Peiris}, Hiranya V. and {Peng}, Xiyan and {Percival}, Will J. and {Perruchot}, Sandrine and {Pieri}, Matthew M. and {Pogge}, Richard and {Pollack}, Jennifer E. and {Poppett}, Claire and {Prada}, Francisco and {Prakash}, Abhishek and {Probst}, Ronald G. and {Rabinowitz}, David and {Raichoor}, Anand and {Ree}, Chang Hee and {Refregier}, Alexandre and {Regal}, Xavier and {Reid}, Beth and {Reil}, Kevin and {Rezaie}, Mehdi and {Rockosi}, Constance M. and {Roe}, Natalie and {Ronayette}, Samuel and {Roodman}, Aaron and {Ross}, Ashley J. and {Ross}, Nicholas P. and {Rossi}, Graziano and {Rozo}, Eduardo and {Ruhlmann-Kleider}, Vanina and {Rykoff}, Eli S. and {Sabiu}, Cristiano and {Samushia}, Lado and {Sanchez}, Eusebio and {Sanchez}, Javier and {Schlegel}, David J. and {Schneider}, Michael and {Schubnell}, Michael and {Secroun}, Aur{\'e}lia and {Seljak}, Uros and {Seo}, Hee-Jong and {Serrano}, Santiago and {Shafieloo}, Arman and {Shan}, Huanyuan and {Sharples}, Ray and {Sholl}, Michael J. and {Shourt}, William V. and {Silber}, Joseph H. and {Silva}, David R. and {Sirk}, Martin M. and {Slosar}, Anze and {Smith}, Alex and {Smoot}, George F. and {Som}, Debopam and {Song}, Yong-Seon and {Sprayberry}, David and {Staten}, Ryan and {Stefanik}, Andy and {Tarle}, Gregory and {Sien Tie}, Suk and {Tinker}, Jeremy L. and {Tojeiro}, Rita and {Valdes}, Francisco and {Valenzuela}, Octavio and {Valluri}, Monica and {Vargas-Magana}, Mariana and {Verde}, Licia and {Walker}, Alistair R. and {Wang}, Jiali and {Wang}, Yuting and {Weaver}, Benjamin A. and {Weaverdyck}, Curtis and {Wechsler}, Risa H. and {Weinberg}, David H. and {White}, Martin and {Yang}, Qian and {Yeche}, Christophe and {Zhang}, Tianmeng and {Zhao}, Gong-Bo and {Zheng}, Yi and {Zhou}, Xu and {Zhou}, Zhimin and {Zhu}, Yaling and {Zou}, Hu and {Zu}, Ying},
    title = "{The DESI Experiment Part I: Science,Targeting, and Survey Design}",
      journal = {arXiv e-prints},
         year = 2016,
        month = oct,
          eid = {arXiv:1611.00036},
        pages = {arXiv:1611.00036},
          doi = {10.48550/arXiv.1611.00036},
archivePrefix = {arXiv},
       eprint = {1611.00036},
 primaryClass = {astro-ph.IM},
       adsurl = {https://ui.adsabs.harvard.edu/abs/2016arXiv161100036D}
}

@ARTICLE{DESI2023b.KP1.EDR,
       author = {{DESI Collaboration} and {Adame}, A.~G. and {Aguilar}, J. and {Ahlen}, S. and {Alam}, S. and {Aldering}, G. and {Alexander}, D.~M. and {Alfarsy}, R. and {Allende Prieto}, C. and {Alvarez}, M. and {Alves}, O. and {Anand}, A. and {Andrade-Oliveira}, F. and {Armengaud}, E. and {Asorey}, J. and {Avila}, S. and {Aviles}, A. and {Bailey}, S. and {Balaguera-Antol{\'\i}nez}, A. and {Ballester}, O. and {Baltay}, C. and {Bault}, A. and {Bautista}, J. and {Behera}, J. and {Beltran}, S.~F. and {BenZvi}, S. and {Beraldo e Silva}, L. and {Bermejo-Climent}, J.~R. and {Berti}, A. and {Besuner}, R. and {Beutler}, F. and {Bianchi}, D. and {Blake}, C. and {Blum}, R. and {Bolton}, A.~S. and {Brieden}, S. and {Brodzeller}, A. and {Brooks}, D. and {Brown}, Z. and {Buckley-Geer}, E. and {Burtin}, E. and {Cabayol-Garcia}, L. and {Cai}, Z. and {Canning}, R. and {Cardiel-Sas}, L. and {Carnero Rosell}, A. and {Castander}, F.~J. and {Cervantes-Cota}, J.~L. and {Chabanier}, S. and {Chaussidon}, E. and {Chaves-Montero}, J. and {Chen}, S. and {Chuang}, C. and {Claybaugh}, T. and {Cole}, S. and {Cooper}, A.~P. and {Cuceu}, A. and {Davis}, T.~M. and {Dawson}, K. and {de Belsunce}, R. and {de la Cruz}, R. and {de la Macorra}, A. and {de Mattia}, A. and {Demina}, R. and {Demirbozan}, U. and {DeRose}, J. and {Dey}, A. and {Dey}, B. and {Dhungana}, G. and {Ding}, J. and {Ding}, Z. and {Doel}, P. and {Doshi}, R. and {Douglass}, K. and {Edge}, A. and {Eftekharzadeh}, S. and {Eisenstein}, D.~J. and {Elliott}, A. and {Escoffier}, S. and {Fagrelius}, P. and {Fan}, X. and {Fanning}, K. and {Fawcett}, V.~A. and {Ferraro}, S. and {Ereza}, J. and {Flaugher}, B. and {Font-Ribera}, A. and {Forero-S{\'a}nchez}, D. and {Forero-Romero}, J.~E. and {Frenk}, C.~S. and {G{\"a}nsicke}, B.~T. and {Garc{\'\i}a}, L. {\'A}. and {Garc{\'\i}a-Bellido}, J. and {Garcia-Quintero}, C. and {Garrison}, L.~H. and {Gil-Mar{\'\i}n}, H. and {Golden-Marx}, J. and {Gontcho}, S. Gontcho A and {Gonzalez-Morales}, A.~X. and {Gonzalez-Perez}, V. and {Gordon}, C. and {Graur}, O. and {Green}, D. and {Gruen}, D. and {Guy}, J. and {Hadzhiyska}, B. and {Hahn}, C. and {Han}, J.~J. and {Hanif}, M.~M. S and {Herrera-Alcantar}, H.~K. and {Honscheid}, K. and {Hou}, J. and {Howlett}, C. and {Huterer}, D. and {Ir{\v{s}}i{\v{c}}}, V. and {Ishak}, M. and {Jacques}, A. and {Jana}, A. and {Jiang}, L. and {Jimenez}, J. and {Jing}, Y.~P. and {Joudaki}, S. and {Jullo}, E. and {Juneau}, S. and {Kizhuprakkat}, N. and {Kara{\c{c}}ayl{\i}}, N.~G. and {Karim}, T. and {Kehoe}, R. and {Kent}, S. and {Khederlarian}, A. and {Kim}, S. and {Kirkby}, D. and {Kisner}, T. and {Kitaura}, F. and {Kneib}, J. and {Koposov}, S.~E. and {Kov{\'a}cs}, A. and {Kremin}, A. and {Krolewski}, A. and {L'Huillier}, B. and {Lambert}, A. and {Lamman}, C. and {Lan}, T. -W. and {Landriau}, M. and {Lang}, D. and {Lange}, J.~U. and {Lasker}, J. and {Le Guillou}, L. and {Leauthaud}, A. and {Levi}, M.~E. and {Li}, T.~S. and {Linder}, E. and {Lyons}, A. and {Magneville}, C. and {Manera}, M. and {Manser}, C.~J. and {Margala}, D. and {Martini}, P. and {McDonald}, P. and {Medina}, G.~E. and {Medina-Varela}, L. and {Meisner}, A. and {Mena-Fern{\'a}ndez}, J. and {Meneses-Rizo}, J. and {Mezcua}, M. and {Miquel}, R. and {Montero-Camacho}, P. and {Moon}, J. and {Moore}, S. and {Moustakas}, J. and {Mueller}, E. and {Mundet}, J. and {Mu{\~n}oz-Guti{\'e}rrez}, A. and {Myers}, A.~D. and {Nadathur}, S. and {Napolitano}, L. and {Neveux}, R. and {Newman}, J.~A. and {Nie}, J. and {Nikutta}, R. and {Niz}, G. and {Norberg}, P. and {Noriega}, H.~E. and {Paillas}, E. and {Palanque-Delabrouille}, N. and {Palmese}, A. and {Zhiwei}, P. and {Parkinson}, D. and {Penmetsa}, S. and {Percival}, W.~J. and {P{\'e}rez-Fern{\'a}ndez}, A. and {P{\'e}rez-R{\`a}fols}, I. and {Pieri}, M. and {Poppett}, C. and {Porredon}, A. and {Pothier}, S. and {Prada}, F. and {Pucha}, R. and {Raichoor}, A. and {Ram{\'\i}rez-P{\'e}rez}, C. and {Ramirez-Solano}, S. and {Rashkovetskyi}, M. and {Ravoux}, C. and {Rocher}, A. and {Rockosi}, C. and {Ross}, A.~J. and {Rossi}, G. and {Ruggeri}, R. and {Ruhlmann-Kleider}, V. and {Sabiu}, C.~G. and {Said}, K. and {Saintonge}, A. and {Samushia}, L. and {Sanchez}, E. and {Saulder}, C. and {Schaan}, E. and {Schlafly}, E.~F. and {Schlegel}, D. and {Scholte}, D. and {Schubnell}, M. and {Seo}, H. and {Shafieloo}, A. and {Sharples}, R. and {Sheu}, W. and {Silber}, J. and {Sinigaglia}, F. and {Siudek}, M. and {Slepian}, Z. and {Smith}, A. and {Sprayberry}, D. and {Stephey}, L. and {Su{\'a}rez-P{\'e}rez}, J. and {Sun}, Z. and {Tan}, T. and {Tarl{\'e}}, G. and {Tojeiro}, R. and {Ure{\~n}a-L{\'o}pez}, L.~A. and {Vaisakh}, R. and {Valcin}, D. and {Valdes}, F. and {Valluri}, M. and {Vargas-Maga{\~n}a}, M. and {Variu}, A. and {Verde}, L. and {Walther}, M. and {Wang}, B. and {Wang}, M.~S. and {Weaver}, B.~A. and {Weaverdyck}, N. and {Wechsler}, R.~H. and {White}, M. and {Xie}, Y. and {Yang}, J. and {Y{\`e}che}, C. and {Yu}, J. and {Yuan}, S. and {Zhang}, H. and {Zhang}, Z. and {Zhao}, C. and {Zheng}, Z. and {Zhou}, R. and {Zhou}, Z. and {Zou}, H. and {Zou}, S. and {Zu}, Y.},
        title = "{The Early Data Release of the Dark Energy Spectroscopic Instrument}",
      journal = {\aj},
         year = 2024,
        month = aug,
       volume = {168},
       number = {2},
          eid = {58},
        pages = {58},
          doi = {10.3847/1538-3881/ad3217},
archivePrefix = {arXiv},
       eprint = {2306.06308},
 primaryClass = {astro-ph.CO},
       adsurl = {https://ui.adsabs.harvard.edu/abs/2024AJ....168...58D}
}

@ARTICLE{DESI2024.I.DR1,
doi = {10.3847/1538-3881/ae4c43},
url = {https://doi.org/10.3847/1538-3881/ae4c43},
year = {2026},
month = {apr},
publisher = {The American Astronomical Society},
volume = {171},
number = {5},
pages = {285},
author = {DESI Collaboration and Abdul Karim, M. and Adame, A. G. and Aguado, D. and Aguilar, J. and Ahlen, S. and Alam, S. and Aldering, G. and Alexander, D. M. and Alfarsy, R. and Allen, L. and Allende Prieto, C. and Alves, O. and Anand, A. and Andrade, U. and Armengaud, E. and Avila, S. and Aviles, A. and Awan, H. and Bailey, S. and Baleato Lizancos, A. and Ballester, O. and Bault, A. and Bautista, J. and Bean, R. and Behera, J. and BenZvi, S. and Beraldo e Silva, L. and Bermejo-Climent, J. R. and Beutler, F. and Bianchi, D. and Blake, C. and Blum, R. and Bolton, A. S. and Bonici, M. and Brieden, S. and Brodzeller, A. and Brooks, D. and Buckley-Geer, E. and Burtin, E. and Byström, A. and Canning, R. and Carnero Rosell, A. and Carr, A. and Carrilho, P. and Casas, L. and Castander, F. J. and Cereskaite, R. and Cervantes-Cota, J. L. and Chaussidon, E. and Chaves-Montero, J. and Chen, S. and Chen, X. and Circosta, C. and Claybaugh, T. and Cole, S. and Cooper, A. P. and Cousinou, M.-C. and Cuceu, A. and Davis, T. M. and Dawson, K. S. and de Belsunce, R. and de la Cruz, R. and de la Macorra, A. and de Mattia, A. and Deiosso, N. and Della Costa, J. and Demina, R. and Demirbozan, U. and DeRose, J. and Dey, A. and Dey, B. and Ding, J. and Ding, Z. and Doel, P. and Douglass, K. and Dowicz, M. and Ebina, H. and Edelstein, J. and Eisenstein, D. J. and Elbers, W. and Emas, N. and Escoffier, S. and Fagrelius, P. and Fan, X. and Fanning, K. and Favole, G. and Fawcett, V. A. and Fernández-García, E. and Ferraro, S. and Findlay, N. and Font-Ribera, A. and Forero-Romero, J. E. and Forero-Sánchez, D. and Frenk, C. S. and Gänsicke, B. T. and Galbany, L. and García-Bellido, J. and Garcia-Quintero, C. and Garrison, L. H. and Gaztañaga, E. and Gil-Marín, H. and Gloudemans, A. and Gnedin, O. Y. and Gontcho A Gontcho, S. and Gonzalez, D. and Gonzalez-Morales, A. X. and Gonzalez-Perez, V. and Gordon, C. and Graur, O. and Green, D. and Gruen, D. and Gsponer, R. and Guandalin, C. and Gutierrez, G. and Guy, J. and Hahn, C. and Han, J. J. and Han, J. and He, S. and Herrera-Alcantar, H. K. and Heydenreich, S. and Honscheid, K. and Hou, J. and Howlett, C. and Huterer, D. and Iršič, V. and Ishak, M. and Jacques, A. and Jiang, L. and Jimenez, J. and Jing, Y. P. and Joachimi, B. and Joudaki, S. and Joyce, R. and Jullo, E. and Juneau, S. and Karaçaylı, N. G. and Karim, T. and Kehoe, R. and Kent, S. and Khederlarian, A. and Kirkby, D. and Kisner, T. and Kitaura, F.-S. and Kizhuprakkat, N. and Kong, H. and Koposov, S. E. and Kremin, A. and Krolewski, A. and Lahav, O. and Lai, Y. and Lamman, C. and Lan, T.-W. and Landriau, M. and Lang, D. and Lange, J. U. and Lasker, J. and Le Goff, J.M. and Le Guillou, L. and Leauthaud, A. and Levi, M. E. and Li, S. and Li, T. S. and Liu, W. and Lodha, K. and Lokken, M. and Luo, Y. and Luo, Y. and Magneville, C. and Manera, M. and Manser, C. J. and Margala, D. and Martini, P. and Maus, M. and McCullough, J. and McDonald, P. and Medina, G. E. and Medina-Varela, L. and Meisner, A. and Mena-Fernández, J. and Menegas, A. and Meneses-Rizo, J. and Mezcua, M. and Miquel, R. and Montero-Camacho, P. and Moon, J. and Moustakas, J. and Muñoz-Gutiérrez, A. and Mu noz-Santos, D. and Myers, A. D. and Myles, J. and Nadathur, S. and Najita, J. and Napolitano, L. and Newman, J. A. and Nikakhtar, F. and Nikutta, R. and Niz, G. and Noriega, H. E. and Nugent, P. and Padmanabhan, N. and Paillas, E. and Palanque-Delabrouille, N. and Palmese, A. and Pan, J. and Pan, Z. and Parkinson, D. and Peacock, J. A. and Ibanez, M. P. and Percival, W. J. and Pérez-Fernández, A. and Pérez-Ràfols, I. and Peterson, P. and Piat, J. and Pieri, M. M. and Pinon, M. and Poppett, C. and Porredon, A. and Prada, F. and Pucha, R. and Qin, F. and Rabinowitz, D. and Raichoor, A. and Ramírez-Pérez, C. and Ramirez-Solano, S. and Rashkovetskyi, M. and Ravoux, C. and Ried Guachalla, B. and Riley, A. H. and Rocher, A. and Rockosi, C. and Rohlf, J. and Rosado-Marín, A. J. and Ross, A. J. and Ross, C. and Rossi, G. and Ruggeri, R. and Ruhlmann-Kleider, V. and Sabiu, C. G. and Said, K. and Sailer, N. and Saintonge, A. and Salcedo Hernandez, Y. and Samushia, L. and Sanchez, E. and Sanders, N. and Sandford, N. and Satyavolu, S. and Saulder, C. and Saydjari, A. K. and Schlafly, E. F. and Schlegel, D. and Scholte, D. and Schubnell, M. and Semenaite, A. and Seo, H. and Shafieloo, A. and Sharples, R. and Silber, J. and Sinigaglia, F. and Siudek, M. and Slepian, Z. and Smith, A. and Soumagnac, M. and Sprayberry, D. and Suárez-Pérez, J. and Swanson, J. and Tan, T. and Tarlé, G. and Taylor, P. and Thomas, G. and Tojeiro, R. and Turner, R. J. and Turner, W. and Ureña-López, L. A. and Vaisakh, R. and Valluri, M. and Valogiannis, G. and Vargas-Magaña, M. and Verde, L. and Vielzeuf, P. and Walther, M. and Wang, B. and Wang, M. S. and Wang, W. and Weaver, B. A. and Weaverdyck, N. and Wechsler, R. H. and Weinberg, D. H. and White, M. and Whitford, A. and Wolfson, M. and Yang, J. and Yèche, C. and Youles, S. and Yu, J. and Yuan, S. and Zaborowski, E. A. and Zarrouk, P. and Zhang, H. and Zhao, C. and Zhao, R. and Zheng, Z. and Zhou, C. and Zhou, R. and Zhou, Y. and Zou, H. and Zou, S. and Zu, Y.},
title = {Data Release 1 of the Dark Energy Spectroscopic Instrument},
journal = {The Astronomical Journal}
}

@ARTICLE{DESI2024.IV.KP6,
       author = {{DESI Collaboration} and {Adame}, A.~G. and {Aguilar}, J. and {Ahlen}, S. and {Alam}, S. and {Alexander}, D.~M. and {Alvarez}, M. and {Alves}, O. and {Anand}, A. and {Andrade}, U. and {Armengaud}, E. and {Avila}, S. and {Aviles}, A. and {Awan}, H. and {Bailey}, S. and {Baltay}, C. and {Bault}, A. and {Bautista}, J. and {Behera}, J. and {BenZvi}, S. and {Beutler}, F. and {Bianchi}, D. and {Blake}, C. and {Blum}, R. and {Brieden}, S. and {Brodzeller}, A. and {Brooks}, D. and {Buckley-Geer}, E. and {Burtin}, E. and {Calderon}, R. and {Canning}, R. and {Carnero Rosell}, A. and {Cereskaite}, R. and {Cervantes-Cota}, J.~L. and {Chabanier}, S. and {Chaussidon}, E. and {Chaves-Montero}, J. and {Chen}, S. and {Chen}, X. and {Claybaugh}, T. and {Cole}, S. and {Cuceu}, A. and {Davis}, T.~M. and {Dawson}, K. and {de la Cruz}, R. and {de la Macorra}, A. and {de Mattia}, A. and {Deiosso}, N. and {Dey}, A. and {Dey}, B. and {Ding}, J. and {Ding}, Z. and {Doel}, P. and {Edelstein}, J. and {Eftekharzadeh}, S. and {Eisenstein}, D.~J. and {Elliott}, A. and {Fagrelius}, P. and {Fanning}, K. and {Ferraro}, S. and {Ereza}, J. and {Findlay}, N. and {Flaugher}, B. and {Font-Ribera}, A. and {Forero-S{\'a}nchez}, D. and {Forero-Romero}, J.~E. and {Garcia-Quintero}, C. and {Gazta{\~n}aga}, E. and {Gil-Mar{\'\i}n}, H. and {Gontcho}, S. Gontcho A and {Gonzalez-Morales}, A.~X. and {Gonzalez-Perez}, V. and {Gordon}, C. and {Green}, D. and {Gruen}, D. and {Gsponer}, R. and {Gutierrez}, G. and {Guy}, J. and {Hadzhiyska}, B. and {Hahn}, C. and {Hanif}, M.~M. S and {Herrera-Alcantar}, H.~K. and {Honscheid}, K. and {Howlett}, C. and {Huterer}, D. and {Ir{\v{s}}i{\v{c}}}, V. and {Ishak}, M. and {Juneau}, S. and {Kara{\c{c}}ayli}, N.~G. and {Kehoe}, R. and {Kent}, S. and {Kirkby}, D. and {Kremin}, A. and {Krolewski}, A. and {Lai}, Y. and {Lan}, T. -W. and {Landriau}, M. and {Lang}, D. and {Lasker}, J. and {Le Goff}, J.~M. and {Le Guillou}, L. and {Leauthaud}, A. and {Levi}, M.~E. and {Li}, T.~S. and {Linder}, E. and {Lodha}, K. and {Magneville}, C. and {Manera}, M. and {Margala}, D. and {Martini}, P. and {Maus}, M. and {McDonald}, P. and {Medina-Varela}, L. and {Meisner}, A. and {Mena-Fern{\'a}ndez}, J. and {Miquel}, R. and {Moon}, J. and {Moore}, S. and {Moustakas}, J. and {Mueller}, E. and {Mu{\~n}oz-Guti{\'e}rrez}, A. and {Myers}, A.~D. and {Nadathur}, S. and {Napolitano}, L. and {Neveux}, R. and {Newman}, J.~A. and {Nguyen}, N.~M. and {Nie}, J. and {Niz}, G. and {Noriega}, H.~E. and {Padmanabhan}, N. and {Paillas}, E. and {Palanque-Delabrouille}, N. and {Pan}, J. and {Penmetsa}, S. and {Percival}, W.~J. and {Pieri}, M.~M. and {Pinon}, M. and {Poppett}, C. and {Porredon}, A. and {Prada}, F. and {P{\'e}rez-Fern{\'a}ndez}, A. and {P{\'e}rez-R{\`a}fols}, I. and {Rabinowitz}, D. and {Raichoor}, A. and {Ram{\'\i}rez-P{\'e}rez}, C. and {Ramirez-Solano}, S. and {Rashkovetskyi}, M. and {Ravoux}, C. and {Rezaie}, M. and {Rich}, J. and {Rocher}, A. and {Rockosi}, C. and {Roe}, N.~A. and {Rosado-Marin}, A. and {Ross}, A.~J. and {Rossi}, G. and {Ruggeri}, R. and {Ruhlmann-Kleider}, V. and {Samushia}, L. and {Sanchez}, E. and {Saulder}, C. and {Schlafly}, E.~F. and {Schlegel}, D. and {Schubnell}, M. and {Seo}, H. and {Sharples}, R. and {Silber}, J. and {Sinigaglia}, F. and {Slosar}, A. and {Smith}, A. and {Sprayberry}, D. and {Tan}, T. and {Tarl{\'e}}, G. and {Trusov}, S. and {Vaisakh}, R. and {Valcin}, D. and {Valdes}, F. and {Vargas-Maga{\~n}a}, M. and {Verde}, L. and {Walther}, M. and {Wang}, B. and {Wang}, M.~S. and {Weaver}, B.~A. and {Weaverdyck}, N. and {Wechsler}, R.~H. and {Weinberg}, D.~H. and {White}, M. and {Yu}, J. and {Yu}, Y. and {Yuan}, S. and {Y{\`e}che}, C. and {Zaborowski}, E.~A. and {Zarrouk}, P. and {Zhang}, H. and {Zhao}, C. and {Zhao}, R. and {Zhou}, R. and {Zou}, H.},
    title = "{DESI 2024 IV: Baryon Acoustic Oscillations from the Lyman alpha forest}",
    journal = {\jcap},
    year = 2025,
    month = jan,
    volume = {2025},
    number = {1},
    eid = {124},
    pages = {124},
    doi = {10.1088/1475-7516/2025/01/124},
    archivePrefix = {arXiv},
    eprint = {2404.03001},
    primaryClass = {astro-ph.CO},
    adsurl = {https://ui.adsabs.harvard.edu/abs/2025JCAP...01..124A}
}

@ARTICLE{DESI2024.VII.KP7B,
       author = {{DESI Collaboration} and {Adame}, A.~G. and {Aguilar}, J. and {Ahlen}, S. and {Alam}, S. and {Alexander}, D.~M. and {Allende Prieto}, C. and {Alvarez}, M. and {Alves}, O. and {Anand}, A. and {Andrade}, U. and {Armengaud}, E. and {Avila}, S. and {Aviles}, A. and {Awan}, H. and {Bahr-Kalus}, B. and {Bailey}, S. and {Baltay}, C. and {Bault}, A. and {Behera}, J. and {BenZvi}, S. and {Beutler}, F. and {Bianchi}, D. and {Blake}, C. and {Blum}, R. and {Bonici}, M. and {Brieden}, S. and {Brodzeller}, A. and {Brooks}, D. and {Buckley-Geer}, E. and {Burtin}, E. and {Calderon}, R. and {Canning}, R. and {Carnero Rosell}, A. and {Cereskaite}, R. and {Cervantes-Cota}, J.~L. and {Chabanier}, S. and {Chaussidon}, E. and {Chaves-Montero}, J. and {Chebat}, D. and {Chen}, S. and {Chen}, X. and {Claybaugh}, T. and {Cole}, S. and {Cuceu}, A. and {Davis}, T.~M. and {Dawson}, K. and {de la Macorra}, A. and {de Mattia}, A. and {Deiosso}, N. and {Dey}, A. and {Dey}, B. and {Ding}, Z. and {Doel}, P. and {Edelstein}, J. and {Eftekharzadeh}, S. and {Eisenstein}, D.~J. and {Elbers}, W. and {Elliott}, A. and {Fagrelius}, P. and {Fanning}, K. and {Ferraro}, S. and {Ereza}, J. and {Findlay}, N. and {Flaugher}, B. and {Font-Ribera}, A. and {Forero-S{\'a}nchez}, D. and {Forero-Romero}, J.~E. and {Frenk}, C.~S. and {Garcia-Quintero}, C. and {Garrison}, L.~H. and {Gazta{\~n}aga}, E. and {Gil-Mar{\'\i}n}, H. and {Gontcho}, S. Gontcho A and {Gonzalez-Morales}, A.~X. and {Gonzalez-Perez}, V. and {Gordon}, C. and {Green}, D. and {Gruen}, D. and {Gsponer}, R. and {Gutierrez}, G. and {Guy}, J. and {Hadzhiyska}, B. and {Hahn}, C. and {Hanif}, M.~M. S and {Herrera-Alcantar}, H.~K. and {Honscheid}, K. and {Howlett}, C. and {Huterer}, D. and {Ir{\v{s}}i{\v{c}}}, V. and {Ishak}, M. and {Joyce}, R. and {Juneau}, S. and {Kara{\c{c}}ayl{\i}}, N.~G. and {Kehoe}, R. and {Kent}, S. and {Kirkby}, D. and {Kong}, H. and {Koposov}, S.~E. and {Kremin}, A. and {Krolewski}, A. and {Lahav}, O. and {Lai}, Y. and {Lan}, T. -W. and {Landriau}, M. and {Lang}, D. and {Lasker}, J. and {Le Goff}, J.~M. and {Le Guillou}, L. and {Leauthaud}, A. and {Levi}, M.~E. and {Li}, T.~S. and {Lodha}, K. and {Magneville}, C. and {Manera}, M. and {Margala}, D. and {Martini}, P. and {Matthewson}, W. and {Maus}, M. and {McDonald}, P. and {Medina-Varela}, L. and {Meisner}, A. and {Mena-Fern{\'a}ndez}, J. and {Miquel}, R. and {Moon}, J. and {Moore}, S. and {Moustakas}, J. and {Mudur}, N. and {Mueller}, E. and {Mu{\~n}oz-Guti{\'e}rrez}, A. and {Myers}, A.~D. and {Nadathur}, S. and {Napolitano}, L. and {Neveux}, R. and {Newman}, J.~A. and {Nguyen}, N.~M. and {Nie}, J. and {Niz}, G. and {Noriega}, H.~E. and {Padmanabhan}, N. and {Paillas}, E. and {Palanque-Delabrouille}, N. and {Pan}, J. and {Penmetsa}, S. and {Percival}, W.~J. and {Pieri}, M.~M. and {Pinon}, M. and {Poppett}, C. and {Porredon}, A. and {Prada}, F. and {P{\'e}rez-Fern{\'a}ndez}, A. and {P{\'e}rez-R{\`a}fols}, I. and {Rabinowitz}, D. and {Raichoor}, A. and {Ram{\'\i}rez-P{\'e}rez}, C. and {Ramirez-Solano}, S. and {Rashkovetskyi}, M. and {Ravoux}, C. and {Rezaie}, M. and {Rich}, J. and {Rocher}, A. and {Rockosi}, C. and {Roe}, N.~A. and {Rosado-Marin}, A. and {Ross}, A.~J. and {Rossi}, G. and {Ruggeri}, R. and {Ruhlmann-Kleider}, V. and {Samushia}, L. and {Sanchez}, E. and {Saulder}, C. and {Schlafly}, E.~F. and {Schlegel}, D. and {Schubnell}, M. and {Seo}, H. and {Shafieloo}, A. and {Sharples}, R. and {Silber}, J. and {Slosar}, A. and {Smith}, A. and {Sprayberry}, D. and {Tan}, T. and {Tarl{\'e}}, G. and {Taylor}, P. and {Trusov}, S. and {Vaisakh}, R. and {Valcin}, D. and {Valdes}, F. and {Valogiannis}, G. and {Vargas-Maga{\~n}a}, M. and {Verde}, L. and {Walther}, M. and {Wang}, B. and {Wang}, M.~S. and {Weaver}, B.~A. and {Weaverdyck}, N. and {Wechsler}, R.~H. and {Weinberg}, D.~H. and {White}, M. and {Wilson}, M.~J. and {Yi}, L. and {Yu}, J. and {Yu}, Y. and {Yuan}, S. and {Y{\`e}che}, C. and {Zaborowski}, E.~A. and {Zarrouk}, P. and {Zhang}, H. and {Zhao}, C. and {Zhao}, R. and {Zhou}, R. and {Zhuang}, T. and {Zou}, H.},
        title = "{DESI 2024 VII: cosmological constraints from the full-shape modeling of clustering measurements}",
      journal = {\jcap},
         year = 2025,
        month = jul,
       volume = {2025},
       number = {7},
          eid = {028},
        pages = {028},
          doi = {10.1088/1475-7516/2025/07/028},
archivePrefix = {arXiv},
       eprint = {2411.12022},
 primaryClass = {astro-ph.CO},
       adsurl = {https://ui.adsabs.harvard.edu/abs/2025JCAP...07..028A}
}

@ARTICLE{Ramirez2024,
       author = {{Ram{\'\i}rez-P{\'e}rez}, C{\'e}sar and {P{\'e}rez-R{\`a}fols}, Ignasi and {Font-Ribera}, Andreu and {Karim}, M. Abdul and {Armengaud}, E. and {Bautista}, J. and {Beltran}, S.~F. and {Cabayol-Garcia}, L. and {Cai}, Z. and {Chabanier}, S. and {Chaussidon}, E. and {Chaves-Montero}, J. and {Cuceu}, A. and {de la Cruz}, R. and {Garc{\'\i}a-Bellido}, J. and {Gonzalez-Morales}, A.~X. and {Gordon}, C. and {Herrera-Alcantar}, H.~K. and {Ir{\v{s}}i{\v{c}}}, V. and {Ishak}, M. and {Kara{\c{c}}ayl{\i}}, N.~G. and Luki{\'c}, Zarija and {Manser}, C.~J. and {Montero-Camacho}, P. and {Napolitano}, L. and {Niz}, G. and {Pieri}, M.~M. and {Ravoux}, C. and {Sinigaglia}, F. and {Tan}, T. and {Walther}, M. and {Wang}, B. and {Aguilar}, J. and {Ahlen}, S. and {Bailey}, S. and {Brooks}, D. and {Claybaugh}, T. and {Dawson}, K. and {de la Macorra}, A. and {Dhungana}, G. and {Doel}, P. and {Fanning}, K. and {Forero-Romero}, J.~E. and {Gontcho}, S. Gontcho A. and {Guy}, J. and {Honscheid}, K. and {Kehoe}, R. and {Kisner}, T. and {Landriau}, M. and {Le Guillou}, L. and {Levi}, M.~E. and {Magneville}, C. and {Martini}, P. and {Meisner}, A. and {Miquel}, R. and {Moustakas}, J. and {Mueller}, E. and {Mu{\~n}oz-Guti{\'e}rrez}, A. and {Nie}, J. and {Palanque-Delabrouille}, N. and {Percival}, W.~J. and {Rossi}, G. and {Sanchez}, E. and {Schlafly}, E.~F. and {Schlegel}, D. and {Seo}, H. and {Tarl{\'e}}, G. and {Weaver}, B.~A. and {Y{\'e}che}, C. and {Zhou}, Z.},
        title = "{The Lyman-{\ensuremath{\alpha}} forest catalogue from the Dark Energy Spectroscopic Instrument Early Data Release}",
      journal = {\mnras},
         year = 2024,
        month = mar,
       volume = {528},
       number = {4},
        pages = {6666-6679},
          doi = {10.1093/mnras/stad3781},
archivePrefix = {arXiv},
       eprint = {2306.06312},
 primaryClass = {astro-ph.CO},
       adsurl = {https://ui.adsabs.harvard.edu/abs/2024MNRAS.528.6666R}
}

@ARTICLE{Gordon2023,
       author = {{Gordon}, C. and {Cuceu}, A. and {Chaves-Montero}, J. and {Font-Ribera}, A. and {Gonz{\'a}lez-Morales}, A.~X. and {Aguilar}, J. and {Ahlen}, S. and {Armengaud}, E. and {Bailey}, S. and {Bault}, A. and {Brodzeller}, A. and {Brooks}, D. and {Claybaugh}, T. and {de la Cruz}, R. and {Dawson}, K. and {Doel}, P. and {Forero-Romero}, J.~E. and {Gontcho}, S. Gontcho A. and {Guy}, J. and {Herrera-Alcantar}, H.~K. and {Ir{\v{s}}i{\v{c}}}, V. and {Kara{\c{c}}ayl{\i}}, N.~G. and {Kirkby}, D. and {Landriau}, M. and {Le Guillou}, L. and {Levi}, M.~E. and {de la Macorra}, A. and {Manera}, M. and {Martini}, P. and {Meisner}, A. and {Miquel}, R. and {Montero-Camacho}, P. and {Mu{\~n}oz-Guti{\'e}rrez}, A. and {Napolitano}, L. and {Nie}, J. and {Niz}, G. and {Palanque-Delabrouille}, N. and {Percival}, W.~J. and {Pieri}, M. and {Poppett}, C. and {Prada}, F. and {P{\'e}rez-R{\`a}fols}, I. and {Ram{\'\i}rez-P{\'e}rez}, C. and {Ravoux}, C. and {Rezaie}, M. and {Ross}, A.~J. and {Rossi}, G. and {Sanchez}, E. and {Schlegel}, D. and {Schubnell}, M. and {Seo}, H. and {Sinigaglia}, F. and {Tan}, T. and {Tarl{\'e}}, G. and {Walther}, M. and {Weaver}, B.~A. and {Y{\`e}che}, C. and {Zhou}, Z. and {Zou}, H.},
        title = "{3D correlations in the Lyman-{\ensuremath{\alpha}} forest from early DESI data}",
      journal = {\jcap},
         year = 2023,
        month = nov,
       volume = {2023},
       number = {11},
          eid = {045},
        pages = {045},
          doi = {10.1088/1475-7516/2023/11/045},
archivePrefix = {arXiv},
       eprint = {2308.10950},
 primaryClass = {astro-ph.CO},
       adsurl = {https://ui.adsabs.harvard.edu/abs/2023JCAP...11..045G}
}

@ARTICLE{2024arXiv240100303H,
       author = {{Herrera-Alcantar}, Hiram K. and {Mu{\~n}oz-Guti{\'e}rrez}, Andrea and {Tan}, Ting and {Gonz{\'a}lez-Morales}, Alma X. and {Font-Ribera}, Andreu and {Guy}, Julien and {Moustakas}, John and {Kirkby}, David and {Armengaud}, E. and {Bault}, A. and {Cabayol-Garcia}, L. and {Chaves-Montero}, J. and {Cuceu}, A. and {de la Cruz}, R. and {Garc{\'\i}a}, L. {\'A}. and {Gordon}, C. and {Ir{\v{s}}i{\v{c}}}, V. and {Kara{\c{c}}ayl{\i}}, N.~G. and {Le Goff}, J.~M. and {Montero-Camacho}, P. and {Niz}, G. and {P{\'e}rez-R{\`a}fols}, I. and {Ram{\'\i}rez-P{\'e}rez}, C. and {Ravoux}, C. and {Walther}, M. and {Aguilar}, J. and {Ahlen}, S. and {Brooks}, D. and {Claybaugh}, T. and {Dawson}, K. and {de la Macorra}, A. and {Doel}, P. and {Forero-Romero}, J.~E. and {Gazta{\~n}aga}, E. and {Gontcho}, S. Gontcho A. and {Honscheid}, K. and {Kehoe}, R. and {Kisner}, T. and {Landriau}, M. and {Levi}, Michael E. and {Manera}, M. and {Martini}, P. and {Meisner}, A. and {Miquel}, R. and {Nie}, J. and {Palanque-Delabrouille}, N. and {Poppett}, C. and {Rezaie}, M. and {Rossi}, G. and {Sanchez}, E. and {Seo}, H. and {Tarl{\'e}}, G. and {Weamver}, B.~A. and {Zhou}, Z.},
        title = "{Synthetic spectra for Lyman-{\ensuremath{\alpha}} forest analysis in the Dark Energy Spectroscopic Instrument}",
      journal = {\jcap},
         year = 2025,
        month = jan,
       volume = {2025},
       number = {1},
          eid = {141},
        pages = {141},
          doi = {10.1088/1475-7516/2025/01/141},
archivePrefix = {arXiv},
       eprint = {2401.00303},
 primaryClass = {astro-ph.CO},
       adsurl = {https://ui.adsabs.harvard.edu/abs/2025JCAP...01..141H}
}

@ARTICLE{KP6s6-Cuceu,
       author = {{Cuceu}, Andrei and {Herrera-Alcantar}, Hiram K. and {Gordon}, Calum and {Martini}, Paul and {Guy}, Julien and {Font-Ribera}, Andreu and {Gonzalez-Morales}, Alma X. and {Abdul Karim}, M. and {Aguilar}, J. and {Ahlen}, S. and {Armengaud}, E. and {Bault}, A. and {Brooks}, D. and {Claybaugh}, T. and {de la Macorra}, A. and {Doel}, P. and {Fanning}, K. and {Ferraro}, S. and {Forero-Romero}, J.~E. and {Gazta{\~n}aga}, E. and {Gontcho}, S. Gontcho A. and {Gutierrez}, G. and {Honscheid}, K. and {Howlett}, C. and {Kara{\c{c}}ayl{\i}}, N.~G. and {Kirkby}, D. and {Kremin}, A. and {Landriau}, M. and {Le Goff}, J.~M. and {Le Guillou}, L. and {Levi}, M.~E. and {Manera}, M. and {Meisner}, A. and {Miquel}, R. and {Moustakas}, J. and {Mu{\~n}oz-Guti{\'e}rrez}, A. and {Myers}, A.~D. and {Niz}, G. and {Palanque-Delabrouille}, N. and {Percival}, W.~J. and {Poppett}, C. and {Prada}, F. and {P{\'e}rez-R{\`a}fols}, I. and {Ram{\'\i}rez-P{\'e}rez}, C. and {Ravoux}, C. and {Rezaie}, M. and {Rossi}, G. and {Sanchez}, E. and {Schlegel}, D. and {Schubnell}, M. and {Seo}, H. and {Sprayberry}, D. and {Tan}, T. and {Tarl{\'e}}, G. and {Vargas-Maga{\~n}a}, M. and {Walther}, M. and {Weaver}, B.~A. and {Zhou}, R. and {Zou}, H.},
        title = "{Validation of the DESI 2024 Ly{\ensuremath{\alpha}} forest BAO analysis using synthetic datasets}",
      journal = {\jcap},
         year = 2025,
        month = jan,
       volume = {2025},
       number = {1},
          eid = {148},
        pages = {148},
          doi = {10.1088/1475-7516/2025/01/148},
archivePrefix = {arXiv},
       eprint = {2404.03004},
 primaryClass = {astro-ph.CO},
       adsurl = {https://ui.adsabs.harvard.edu/abs/2025JCAP...01..148C}
}

@ARTICLE{2023arXiv230903434F,
       author = {{Filbert}, S. and {Martini}, P. and {Seebaluck}, K. and {Ennesser}, L. and {Alexander}, D.~M. and {Bault}, A. and {Brodzeller}, A. and {Herrera-Alcantar}, H.~K. and {Montero-Camacho}, P. and {P{\'e}rez-R{\`a}fols}, I. and {Ram{\'\i}rez-P{\'e}rez}, C. and {Ravoux}, C. and {Tan}, T. and {Aguilar}, J. and {Ahlen}, S. and {Bailey}, S. and {Brooks}, D. and {Claybaugh}, T. and {Dawson}, K. and {de la Macorra}, A. and {Doel}, P. and {Fanning}, K. and {Font-Ribera}, A. and {Forero-Romero}, J.~E. and {Gontcho A Gontcho}, S. and {Guy}, J. and {Kirkby}, D. and {Kremin}, A. and {Magneville}, C. and {Manera}, M. and {Meisner}, A. and {Miquel}, R. and {Moustakas}, J. and {Nie}, J. and {Percival}, W.~J. and {Prada}, F. and {Rezaie}, M. and {Rossi}, G. and {Sanchez}, E. and {Schubnell}, M. and {Seo}, H. and {Tarl{\'e}}, G. and {Weaver}, B.~A. and {Zhou}, Z.},
        title = "{Broad absorption line quasars in the Dark Energy Spectroscopic Instrument Early Data Release}",
      journal = {\mnras},
         year = 2024,
        month = aug,
       volume = {532},
       number = {4},
        pages = {3669-3681},
          doi = {10.1093/mnras/stae1610},
archivePrefix = {arXiv},
       eprint = {2309.03434},
 primaryClass = {astro-ph.CO},
       adsurl = {https://ui.adsabs.harvard.edu/abs/2024MNRAS.532.3669F}
}

@ARTICLE{DESI2016b.Instr,
       author = {{DESI Collaboration} and {Aghamousa}, Amir and {Aguilar}, Jessica and {Ahlen}, Steve and {Alam}, Shadab and {Allen}, Lori E. and {Allende Prieto}, Carlos and {Annis}, James and {Bailey}, Stephen and {Balland}, Christophe and {Ballester}, Otger and {Baltay}, Charles and {Beaufore}, Lucas and {Bebek}, Chris and {Beers}, Timothy C. and {Bell}, Eric F. and {Bernal}, Jos{\'e} Luis and {Besuner}, Robert and {Beutler}, Florian and {Blake}, Chris and {Bleuler}, Hannes and {Blomqvist}, Michael and {Blum}, Robert and {Bolton}, Adam S. and {Briceno}, Cesar and {Brooks}, David and {Brownstein}, Joel R. and {Buckley-Geer}, Elizabeth and {Burden}, Angela and {Burtin}, Etienne and {Busca}, Nicolas G. and {Cahn}, Robert N. and {Cai}, Yan-Chuan and {Cardiel-Sas}, Laia and {Carlberg}, Raymond G. and {Carton}, Pierre-Henri and {Casas}, Ricard and {Castander}, Francisco J. and {Cervantes-Cota}, Jorge L. and {Claybaugh}, Todd M. and {Close}, Madeline and {Coker}, Carl T. and {Cole}, Shaun and {Comparat}, Johan and {Cooper}, Andrew P. and {Cousinou}, M.-C. and {Crocce}, Martin and {Cuby}, Jean-Gabriel and {Cunningham}, Daniel P. and {Davis}, Tamara M. and {Dawson}, Kyle S. and {de la Macorra}, Axel and {De Vicente}, Juan and {Delubac}, Timoth{\'e}e and {Derwent}, Mark and {Dey}, Arjun and {Dhungana}, Govinda and {Ding}, Zhejie and {Doel}, Peter and {Duan}, Yutong T. and {Ealet}, Anne and {Edelstein}, Jerry and {Eftekharzadeh}, Sarah and {Eisenstein}, Daniel J. and {Elliott}, Ann and {Escoffier}, St{\'e}phanie and {Evatt}, Matthew and {Fagrelius}, Parker and {Fan}, Xiaohui and {Fanning}, Kevin and {Farahi}, Arya and {Farihi}, Jay and {Favole}, Ginevra and {Feng}, Yu and {Fernandez}, Enrique and {Findlay}, Joseph R. and {Finkbeiner}, Douglas P. and {Fitzpatrick}, Michael J. and {Flaugher}, Brenna and {Flender}, Samuel and {Font-Ribera}, Andreu and {Forero-Romero}, Jaime E. and {Fosalba}, Pablo and {Frenk}, Carlos S. and {Fumagalli}, Michele and {Gaensicke}, Boris T. and {Gallo}, Giuseppe and {Garcia-Bellido}, Juan and {Gaztanaga}, Enrique and {Pietro Gentile Fusillo}, Nicola and {Gerard}, Terry and {Gershkovich}, Irena and {Giannantonio}, Tommaso and {Gillet}, Denis and {Gonzalez-de-Rivera}, Guillermo and {Gonzalez-Perez}, Violeta and {Gott}, Shelby and {Graur}, Or and {Gutierrez}, Gaston and {Guy}, Julien and {Habib}, Salman and {Heetderks}, Henry and {Heetderks}, Ian and {Heitmann}, Katrin and {Hellwing}, Wojciech A. and {Herrera}, David A. and {Ho}, Shirley and {Holland}, Stephen and {Honscheid}, Klaus and {Huff}, Eric and {Hutchinson}, Timothy A. and {Huterer}, Dragan and {Hwang}, Ho Seong and {Illa Laguna}, Joseph Maria and {Ishikawa}, Yuzo and {Jacobs}, Dianna and {Jeffrey}, Niall and {Jelinsky}, Patrick and {Jennings}, Elise and {Jiang}, Linhua and {Jimenez}, Jorge and {Johnson}, Jennifer and {Joyce}, Richard and {Jullo}, Eric and {Juneau}, St{\'e}phanie and {Kama}, Sami and {Karcher}, Armin and {Karkar}, Sonia and {Kehoe}, Robert and {Kennamer}, Noble and {Kent}, Stephen and {Kilbinger}, Martin and {Kim}, Alex G. and {Kirkby}, David and {Kisner}, Theodore and {Kitanidis}, Ellie and {Kneib}, Jean-Paul and {Koposov}, Sergey and {Kovacs}, Eve and {Koyama}, Kazuya and {Kremin}, Anthony and {Kron}, Richard and {Kronig}, Luzius and {Kueter-Young}, Andrea and {Lacey}, Cedric G. and {Lafever}, Robin and {Lahav}, Ofer and {Lambert}, Andrew and {Lampton}, Michael and {Landriau}, Martin and {Lang}, Dustin and {Lauer}, Tod R. and {Le Goff}, Jean-Marc and {Le Guillou}, Laurent and {Le Van Suu}, Auguste and {Lee}, Jae Hyeon and {Lee}, Su-Jeong and {Leitner}, Daniela and {Lesser}, Michael and {Levi}, Michael E. and {L'Huillier}, Benjamin and {Li}, Baojiu and {Liang}, Ming and {Lin}, Huan and {Linder}, Eric and {Loebman}, Sarah R. and {Luki{\'c}}, Zarija and {Ma}, Jun and {MacCrann}, Niall and {Magneville}, Christophe and {Makarem}, Laleh and {Manera}, Marc and {Manser}, Christopher J. and {Marshall}, Robert and {Martini}, Paul and {Massey}, Richard and {Matheson}, Thomas and {McCauley}, Jeremy and {McDonald}, Patrick and {McGreer}, Ian D. and {Meisner}, Aaron and {Metcalfe}, Nigel and {Miller}, Timothy N. and {Miquel}, Ramon and {Moustakas}, John and {Myers}, Adam and {Naik}, Milind and {Newman}, Jeffrey A. and {Nichol}, Robert C. and {Nicola}, Andrina and {Nicolati da Costa}, Luiz and {Nie}, Jundan and {Niz}, Gustavo and {Norberg}, Peder and {Nord}, Brian and {Norman}, Dara and {Nugent}, Peter and {O'Brien}, Thomas and {Oh}, Minji and {Olsen}, Knut A.~G.},
        title = "{The DESI Experiment Part II: Instrument Design}",
      journal = {arXiv e-prints},
         year = 2016,
        month = oct,
          eid = {arXiv:1611.00037},
        pages = {arXiv:1611.00037},
          doi = {10.48550/arXiv.1611.00037},
archivePrefix = {arXiv},
       eprint = {1611.00037},
 primaryClass = {astro-ph.IM},
       adsurl = {https://ui.adsabs.harvard.edu/abs/2016arXiv161100037D}
}

@ARTICLE{FocalPlane.Silber.2023,
       author = {{Silber}, Joseph Harry and {Fagrelius}, Parker and {Fanning}, Kevin and {Schubnell}, Michael and {Aguilar}, Jessica Nicole and {Ahlen}, Steven and {Ameel}, Jon and {Ballester}, Otger and {Baltay}, Charles and {Bebek}, Chris and {Benton Beard}, Dominic and {Besuner}, Robert and {Cardiel-Sas}, Laia and {Casas}, Ricard and {Castander}, Francisco Javier and {Claybaugh}, Todd and {Dobson}, Carl and {Duan}, Yutong and {Dunlop}, Patrick and {Edelstein}, Jerry and {Emmet}, William T. and {Elliott}, Ann and {Evatt}, Matthew and {Gershkovich}, Irena and {Guy}, Julien and {Harris}, Stu and {Heetderks}, Henry and {Heetderks}, Ian and {Honscheid}, Klaus and {Illa}, Jose Maria and {Jelinsky}, Patrick and {Jelinsky}, Sharon R. and {Jimenez}, Jorge and {Karcher}, Armin and {Kent}, Stephen and {Kirkby}, David and {Kneib}, Jean-Paul and {Lambert}, Andrew and {Lampton}, Mike and {Leitner}, Daniela and {Levi}, Michael and {McCauley}, Jeremy and {Meisner}, Aaron and {Miller}, Timothy N. and {Miquel}, Ramon and {Mundet}, Juli{\'a} and {Poppett}, Claire and {Rabinowitz}, David and {Reil}, Kevin and {Roman}, David and {Schlegel}, David and {Serrano}, Santiago and {Van Shourt}, William and {Sprayberry}, David and {Tarl{\'e}}, Gregory and {Tie}, Suk Sien and {Weaverdyck}, Curtis and {Zhang}, Kai and {Azzaro}, Marco and {Bailey}, Stephen and {Becerril}, Santiago and {Blackwell}, Tami and {Bouri}, Mohamed and {Brooks}, David and {Buckley-Geer}, Elizabeth and {Castro}, Jose Pe{\~n}ate and {Derwent}, Mark and {Dey}, Arjun and {Dhungana}, Govinda and {Doel}, Peter and {Eisenstein}, Daniel J. and {Fahim}, Nasib and {Garcia-Bellido}, Juan and {Gazta{\~n}aga}, Enrique and {A Gontcho}, Satya Gontcho and {Gutierrez}, Gaston and {H{\"o}rler}, Philipp and {Kehoe}, Robert and {Kisner}, Theodore and {Kremin}, Anthony and {Kronig}, Luzius and {Landriau}, Martin and {Le Guillou}, Laurent and {Martini}, Paul and {Moustakas}, John and {Palanque-Delabrouille}, Nathalie and {Peng}, Xiyan and {Percival}, Will and {Prada}, Francisco and {Allende Prieto}, Carlos and {de Rivera}, Guillermo Gonzalez and {Sanchez}, Eusebio and {Sanchez}, Justo and {Sharples}, Ray and {Soares-Santos}, Marcelle and {Schlafly}, Edward and {Weaver}, Benjamin Alan and {Zhou}, Zhimin and {Zhu}, Yaling and {Zou}, Hu and {DESI Collaboration}},
        title = "{The Robotic Multiobject Focal Plane System of the Dark Energy Spectroscopic Instrument (DESI)}",
      journal = {\aj},
         year = 2023,
        month = jan,
       volume = {165},
       number = {1},
          eid = {9},
        pages = {9},
          doi = {10.3847/1538-3881/ac9ab1},
archivePrefix = {arXiv},
       eprint = {2205.09014},
 primaryClass = {astro-ph.IM},
       adsurl = {https://ui.adsabs.harvard.edu/abs/2023AJ....165....9S}
}

@ARTICLE{BASS.Zou.2017,
       author = {{Zou}, Hu and {Zhou}, Xu and {Fan}, Xiaohui and {Zhang}, Tianmeng and {Zhou}, Zhimin and {Nie}, Jundan and {Peng}, Xiyan and {McGreer}, Ian and {Jiang}, Linhua and {Dey}, Arjun and {Fan}, Dongwei and {He}, Boliang and {Jiang}, Zhaoji and {Lang}, Dustin and {Lesser}, Michael and {Ma}, Jun and {Mao}, Shude and {Schlegel}, David and {Wang}, Jiali},
        title = "{Project Overview of the Beijing-Arizona Sky Survey}",
      journal = {\pasp},
         year = 2017,
        month = jun,
       volume = {129},
       number = {976},
        pages = {064101},
          doi = {10.1088/1538-3873/aa65ba},
archivePrefix = {arXiv},
       eprint = {1702.03653},
 primaryClass = {astro-ph.GA},
       adsurl = {https://ui.adsabs.harvard.edu/abs/2017PASP..129f4101Z}
}

@ARTICLE{LS.Overview.Dey.2019,
       author = {{Dey}, Arjun and {Schlegel}, David J. and {Lang}, Dustin and {Blum}, Robert and {Burleigh}, Kaylan and {Fan}, Xiaohui and {Findlay}, Joseph R. and {Finkbeiner}, Doug and {Herrera}, David and {Juneau}, St{\'e}phanie and {Landriau}, Martin and {Levi}, Michael and {McGreer}, Ian and {Meisner}, Aaron and {Myers}, Adam D. and {Moustakas}, John and {Nugent}, Peter and {Patej}, Anna and {Schlafly}, Edward F. and {Walker}, Alistair R. and {Valdes}, Francisco and {Weaver}, Benjamin A. and {Y{\`e}che}, Christophe and {Zou}, Hu and {Zhou}, Xu and {Abareshi}, Behzad and {Abbott}, T.~M.~C. and {Abolfathi}, Bela and {Aguilera}, C. and {Alam}, Shadab and {Allen}, Lori and {Alvarez}, A. and {Annis}, James and {Ansarinejad}, Behzad and {Aubert}, Marie and {Beechert}, Jacqueline and {Bell}, Eric F. and {BenZvi}, Segev Y. and {Beutler}, Florian and {Bielby}, Richard M. and {Bolton}, Adam S. and {Brice{\~n}o}, C{\'e}sar and {Buckley-Geer}, Elizabeth J. and {Butler}, Karen and {Calamida}, Annalisa and {Carlberg}, Raymond G. and {Carter}, Paul and {Casas}, Ricard and {Castander}, Francisco J. and {Choi}, Yumi and {Comparat}, Johan and {Cukanovaite}, Elena and {Delubac}, Timoth{\'e}e and {DeVries}, Kaitlin and {Dey}, Sharmila and {Dhungana}, Govinda and {Dickinson}, Mark and {Ding}, Zhejie and {Donaldson}, John B. and {Duan}, Yutong and {Duckworth}, Christopher J. and {Eftekharzadeh}, Sarah and {Eisenstein}, Daniel J. and {Etourneau}, Thomas and {Fagrelius}, Parker A. and {Farihi}, Jay and {Fitzpatrick}, Mike and {Font-Ribera}, Andreu and {Fulmer}, Leah and {G{\"a}nsicke}, Boris T. and {Gaztanaga}, Enrique and {George}, Koshy and {Gerdes}, David W. and {Gontcho}, Satya Gontcho A. and {Gorgoni}, Claudio and {Green}, Gregory and {Guy}, Julien and {Harmer}, Diane and {Hernandez}, M. and {Honscheid}, Klaus and {Huang}, Lijuan Wendy and {James}, David J. and {Jannuzi}, Buell T. and {Jiang}, Linhua and {Joyce}, Richard and {Karcher}, Armin and {Karkar}, Sonia and {Kehoe}, Robert and {Kneib}, Jean-Paul and {Kueter-Young}, Andrea and {Lan}, Ting-Wen and {Lauer}, Tod R. and {Le Guillou}, Laurent and {Le Van Suu}, Auguste and {Lee}, Jae Hyeon and {Lesser}, Michael and {Perreault Levasseur}, Laurence and {Li}, Ting S. and {Mann}, Justin L. and {Marshall}, Robert and {Mart{\'\i}nez-V{\'a}zquez}, C.~E. and {Martini}, Paul and {du Mas des Bourboux}, H{\'e}lion and {McManus}, Sean and {Meier}, Tobias Gabriel and {M{\'e}nard}, Brice and {Metcalfe}, Nigel and {Mu{\~n}oz-Guti{\'e}rrez}, Andrea and {Najita}, Joan and {Napier}, Kevin and {Narayan}, Gautham and {Newman}, Jeffrey A. and {Nie}, Jundan and {Nord}, Brian and {Norman}, Dara J. and {Olsen}, Knut A.~G. and {Paat}, Anthony and {Palanque-Delabrouille}, Nathalie and {Peng}, Xiyan and {Poppett}, Claire L. and {Poremba}, Megan R. and {Prakash}, Abhishek and {Rabinowitz}, David and {Raichoor}, Anand and {Rezaie}, Mehdi and {Robertson}, A.~N. and {Roe}, Natalie A. and {Ross}, Ashley J. and {Ross}, Nicholas P. and {Rudnick}, Gregory and {Safonova}, Sasha and {Saha}, Abhijit and {S{\'a}nchez}, F. Javier and {Savary}, Elodie and {Schweiker}, Heidi and {Scott}, Adam and {Seo}, Hee-Jong and {Shan}, Huanyuan and {Silva}, David R. and {Slepian}, Zachary and {Soto}, Christian and {Sprayberry}, David and {Staten}, Ryan and {Stillman}, Coley M. and {Stupak}, Robert J. and {Summers}, David L. and {Sien Tie}, Suk and {Tirado}, H. and {Vargas-Maga{\~n}a}, Mariana and {Vivas}, A. Katherina and {Wechsler}, Risa H. and {Williams}, Doug and {Yang}, Jinyi and {Yang}, Qian and {Yapici}, Tolga and {Zaritsky}, Dennis and {Zenteno}, A. and {Zhang}, Kai and {Zhang}, Tianmeng and {Zhou}, Rongpu and {Zhou}, Zhimin},
        title = "{Overview of the DESI Legacy Imaging Surveys}",
      journal = {\aj},
         year = 2019,
        month = may,
       volume = {157},
       number = {5},
          eid = {168},
        pages = {168},
          doi = {10.3847/1538-3881/ab089d},
archivePrefix = {arXiv},
       eprint = {1804.08657},
 primaryClass = {astro-ph.IM},
       adsurl = {https://ui.adsabs.harvard.edu/abs/2019AJ....157..168D}
}

@ARTICLE{Redrock.Bailey.2024,
   author = {{Bailey et al.}},
    title = "Redrock: Spectroscopic Classification and Redshift Fitting for the Dark Energy Spectroscopic Instrument",
    year = 2024,
  journal = {in preparation}
}

@ARTICLE{SurveyOps.Schlafly.2023,
       author = {{Schlafly}, Edward F. and {Kirkby}, David and {Schlegel}, David J. and {Myers}, Adam D. and {Raichoor}, Anand and {Dawson}, Kyle and {Aguilar}, Jessica and {Allende Prieto}, Carlos and {Bailey}, Stephen and {BenZvi}, Segev and {Bermejo-Climent}, Jose and {Brooks}, David and {de la Macorra}, Axel and {Dey}, Arjun and {Doel}, Peter and {Fanning}, Kevin and {Font-Ribera}, Andreu and {Forero-Romero}, Jaime E. and {Garc{\'\i}a-Bellido}, Juan and {Gontcho A Gontcho}, Satya and {Guy}, Julien and {Hahn}, ChangHoon and {Honscheid}, Klaus and {Ishak}, Mustapha and {Juneau}, St{\'e}phanie and {Kehoe}, Robert and {Kisner}, Theodore and {Kremin}, Anthony and {Landriau}, Martin and {Lang}, Dustin A. and {Lasker}, James and {Levi}, Michael E. and {Magneville}, Christophe and {Manser}, Christopher J. and {Martini}, Paul and {Meisner}, Aaron M. and {Miquel}, Ramon and {Moustakas}, John and {Newman}, Jeffrey A. and {Nie}, Jundan and {Palanque-Delabrouille}, Nathalie. and {Percival}, Will J. and {Poppett}, Claire and {Rockosi}, Constance and {Ross}, Ashley J. and {Rossi}, Graziano and {Tarl{\'e}}, Gregory and {Weaver}, Benjamin A. and {Y{\`e}che}, Christophe and {Zhou}, Rongpu and {DESI Collaboration}},
        title = "{Survey Operations for the Dark Energy Spectroscopic Instrument}",
      journal = {\aj},
         year = 2023,
        month = dec,
       volume = {166},
       number = {6},
          eid = {259},
        pages = {259},
          doi = {10.3847/1538-3881/ad0832},
archivePrefix = {arXiv},
       eprint = {2306.06309},
 primaryClass = {astro-ph.CO},
       adsurl = {https://ui.adsabs.harvard.edu/abs/2023AJ....166..259S}
}

@ARTICLE{QSOPrelim.Yeche.2020,
       author = {{Y{\`e}che}, Christophe and {Palanque-Delabrouille}, Nathalie and {Claveau}, Charles-Antoine and {Brooks}, David D. and {Chaussidon}, Edmond and {Davis}, Tamara M. and {Dawson}, Kyle S. and {Dey}, Arjun and {Duan}, Yutong and {Eftekharzadeh}, Sarah and {Eisenstein}, Daniel J. and {Gazta{\~n}aga}, Enrique and {Kehoe}, Robert and {Landriau}, Martin and {Lang}, Dustin and {Levi}, Michael E. and {Meisner}, Aaron M. and {Myers}, Adam D. and {Newman}, Jeffrey A. and {Poppett}, Claire and {Prada}, Francisco and {Raichoor}, Anand and {Schlegel}, David J. and {Schubnell}, Michael and {Staten}, Ryan and {Tarl{\'e}}, Gregory and {Zhou}, Rongpu},
        title = "{Preliminary Target Selection for the DESI Quasar (QSO) Sample}",
      journal = {Research Notes of the American Astronomical Society},
         year = 2020,
        month = oct,
       volume = {4},
       number = {10},
          eid = {179},
        pages = {179},
          doi = {10.3847/2515-5172/abc01a},
archivePrefix = {arXiv},
       eprint = {2010.11280},
 primaryClass = {astro-ph.CO},
       adsurl = {https://ui.adsabs.harvard.edu/abs/2020RNAAS...4..179Y}
}

@ARTICLE{VIQSO.Alexander.2023,
       author = {{Alexander}, David M. and {Davis}, Tamara M. and {Chaussidon}, E. and {Fawcett}, V.~A. and {X. Gonzalez-Morales}, Alma and {Lan}, Ting-Wen and {Y{\`e}che}, Christophe and {Ahlen}, S. and {Aguilar}, J.~N. and {Armengaud}, E. and {Bailey}, S. and {Brooks}, D. and {Cai}, Z. and {Canning}, R. and {Carr}, A. and {Chabanier}, S. and {Cousinou}, Marie-Claude and {Dawson}, K. and {de la Macorra}, A. and {Dey}, A. and {Dey}, Biprateep and {Dhungana}, G. and {Edge}, A.~C. and {Eftekharzadeh}, S. and {Fanning}, K. and {Farr}, James and {Font-Ribera}, A. and {Garcia-Bellido}, J. and {Garrison}, Lehman and {Gazta{\~n}aga}, E. and {A Gontcho}, Satya Gontcho and {Gordon}, C. and {Medellin Gonzalez}, Stefany Guadalupe and {Guy}, J. and {Herrera-Alcantar}, Hiram K. and {Jiang}, L. and {Juneau}, S. and {Kara{\c{c}}ayl{\i}}, N.~G. and {Kehoe}, R. and {Kisner}, T. and {Kov{\'a}cs}, A. and {Landriau}, M. and {Levi}, Michael E. and {Magneville}, C. and {Martini}, P. and {Meisner}, Aaron M. and {Mezcua}, M. and {Miquel}, R. and {Camacho}, P. Montero and {Moustakas}, J. and {Mu{\~n}oz-Guti{\'e}rrez}, Andrea and {Myers}, Adam D. and {Nadathur}, S. and {Napolitano}, L. and {Nie}, J.~D. and {Palanque-Delabrouille}, N. and {Pan}, Z. and {Percival}, W.~J. and {P{\'e}rez-R{\`a}fols}, I. and {Poppett}, C. and {Prada}, F. and {Ram{\'\i}rez-P{\'e}rez}, C{\'e}sar and {Ravoux}, C. and {Rosario}, D.~J. and {Schubnell}, M. and {Tarl{\'e}}, Gregory and {Walther}, M. and {Weiner}, B. and {Youles}, S. and {Zhou}, Zhimin and {Zou}, H. and {Zou}, Siwei},
        title = "{The DESI Survey Validation: Results from Visual Inspection of the Quasar Survey Spectra}",
      journal = {\aj},
         year = 2023,
        month = mar,
       volume = {165},
       number = {3},
          eid = {124},
        pages = {124},
          doi = {10.3847/1538-3881/acacfc},
archivePrefix = {arXiv},
       eprint = {2208.08517},
 primaryClass = {astro-ph.GA},
       adsurl = {https://ui.adsabs.harvard.edu/abs/2023AJ....165..124A}
}

@ARTICLE{QSO.TS.Chaussidon.2023,
       author = {{Chaussidon}, Edmond and {Y{\`e}che}, Christophe and {Palanque-Delabrouille}, Nathalie and {Alexander}, David M. and {Yang}, Jinyi and {Ahlen}, Steven and {Bailey}, Stephen and {Brooks}, David and {Cai}, Zheng and {Chabanier}, Sol{\`e}ne and {Davis}, Tamara M. and {Dawson}, Kyle and {de laMacorra}, Axel and {Dey}, Arjun and {Dey}, Biprateep and {Eftekharzadeh}, Sarah and {Eisenstein}, Daniel J. and {Fanning}, Kevin and {Font-Ribera}, Andreu and {Gazta{\~n}aga}, Enrique and {A Gontcho}, Satya Gontcho and {Gonzalez-Morales}, Alma X. and {Guy}, Julien and {Herrera-Alcantar}, Hiram K. and {Honscheid}, Klaus and {Ishak}, Mustapha and {Jiang}, Linhua and {Juneau}, Stephanie and {Kehoe}, Robert and {Kisner}, Theodore and {Kov{\'a}cs}, Andras and {Kremin}, Anthony and {Lan}, Ting-Wen and {Landriau}, Martin and {Le Guillou}, Laurent and {Levi}, Michael E. and {Magneville}, Christophe and {Martini}, Paul and {Meisner}, Aaron M. and {Moustakas}, John and {Mu{\~n}oz-Guti{\'e}rrez}, Andrea and {Myers}, Adam D. and {Newman}, Jeffrey A. and {Nie}, Jundan and {Percival}, Will J. and {Poppett}, Claire and {Prada}, Francisco and {Raichoor}, Anand and {Ravoux}, Corentin and {Ross}, Ashley J. and {Schlafly}, Edward and {Schlegel}, David and {Tan}, Ting and {Tarl{\'e}}, Gregory and {Zhou}, Rongpu and {Zhou}, Zhimin and {Zou}, Hu},
        title = "{Target Selection and Validation of DESI Quasars}",
      journal = {\apj},
         year = 2023,
        month = feb,
       volume = {944},
       number = {1},
          eid = {107},
        pages = {107},
          doi = {10.3847/1538-4357/acb3c2},
archivePrefix = {arXiv},
       eprint = {2208.08511},
 primaryClass = {astro-ph.CO},
       adsurl = {https://ui.adsabs.harvard.edu/abs/2023ApJ...944..107C}
}

@article{ACCEL2_2024,
   title={The ACCEL2 project: Simulating Lyman-$\alpha$ forest in large-volume hydrodynamical simulations},
   ISSN={1365-2966},
   url={http://dx.doi.org/10.1093/mnras/stae2255},
   DOI={10.1093/mnras/stae2255},
   journal={Monthly Notices of the Royal Astronomical Society},
   publisher={Oxford University Press (OUP)},
   author={Chabanier, Sol{\`e}ne and Ravoux, Corentin and Latrille, Lucas and Sexton, Jean and Armengaud, {\'E}ric and Bautista, Julian and Dumerchat, Tyann and Luki{\'c}, Zarija},
   year={2024},
   month=oct }

@article{ForestFlow_2025,
title = {ForestFlow: predicting the Lyman-α forest clustering from linear to nonlinear scales},
author = {Chaves-Montero, J. and Cabayol-Garcia, L. and Lokken, M. and Font-Ribera, A. and Aguilar, J. and Ahlen, S. and Bianchi, D. and Brooks, D. and Claybaugh, T. and Cole, S. and de la Macorra, A. and Ferraro, S. and Forero-Romero, J. E. and Gaztañaga, E. and A.Gontcho, S. Gontcho and Gutierrez, G. and Honscheid, K. and Kehoe, R. and Kirkby, D. and Kremin, A. and Lambert, A. and Landriau, M. and Manera, M. and Martini, P. and Miquel, R. and Muñoz-Gutiérrez, A. and Niz, G. and Pérez-Ràfols, I. and Rossi, G. and Sanchez, E. and Schubnell, M. and Sprayberry, D. and Tarlé, G. and Weaver, B. A.},
abstractNote = {On large scales, the Lyman-α forest provides insights into the expansion history of the Universe, while on small scales, it imposes strict constraints on the growth history, the nature of dark matter, and the sum of neutrino masses. This work introduces ForestFlow, a novel framework that bridges the gap between large- and small-scale analyses, which have traditionally relied on distinct modeling approaches. Using conditional normalizing flows, ForestFlow predicts the two Lyman-α linear biases (bδ and bη) and six parameters describing small-scale deviations of the three-dimensional flux power spectrum (P3D) from linear theory as a function of cosmology and intergalactic medium physics. These are then combined with a Boltzmann solver to make consistent predictions, from arbitrarily large scales down to the nonlinear regime, for P3D and any other statistics derived from it. Trained on a suite of 30 fixed-and-paired cosmological hydrodynamical simulations spanning redshifts from z = 2 to 4.5, ForestFlow achieves 3 and 1.5% precision in describing P3D and the one-dimensional flux power spectrum (P1D) from linear scales to k = 5 Mpc−1 and k∥ = 4 Mpc−1, respectively. Thanks to its conditional parameterization, ForestFlow shows similar performance for ionization histories and two ΛCDM model extensions – massive neutrinos and curvature – even though none of these are included in the training set. This framework will enable full-scale cosmological analyses of Lyman-α forest measurements from the DESI survey.Key words: cosmological parameters / cosmology: theory / large-scale structure of Universe},
doi = {10.1051/0004-6361/202452039},
journal = {Astron.Astrophys.},
volume = 694,
place = {United States},
year = {2025},
month = {2}
}

@ARTICLE{Karacayli_2025,
       author = {{Kara{\c{c}}ayl{\i}}, Naim G{\"o}ksel and {Martini}, Paul and {Aguilar}, J. and {Ahlen}, S. and {Armengaud}, E. and {Bailey}, S. and {Bault}, A. and {Bianchi}, D. and {Brodzeller}, A. and {Brooks}, D. and {Chaves-Montero}, J. and {Claybaugh}, T. and {Cuceu}, A. and {de la Macorra}, A. and {Dey}, A. and {Dey}, B. and {Doel}, P. and {Ferraro}, S. and {Font-Ribera}, A. and {Forero-Romero}, J.~E. and {Gazta{\~n}aga}, E. and {Gontcho}, S. Gontcho A. and {Gutierrez}, G. and {Guy}, J. and {Hahn}, C. and {Herrera-Alcantar}, H.~K. and {Honscheid}, K. and {Ishak}, M. and {Kehoe}, R. and {Kirkby}, D. and {Kremin}, A. and {Landriau}, M. and {Le Goff}, J.~M. and {Le Guillou}, L. and {Levi}, M.~E. and {Manera}, M. and {Meisner}, A. and {Miquel}, R. and {Montero-Camacho}, P. and {Nadathur}, S. and {Niz}, G. and {Palanque-Delabrouille}, N. and {Pan}, Z. and {Percival}, W.~J. and {Pieri}, Matthew M. and {Prada}, F. and {P{\'e}rez-R{\`a}fols}, I. and {Ravoux}, C. and {Rossi}, G. and {Sanchez}, E. and {Saulder}, C. and {Schlegel}, D. and {Schubnell}, M. and {Seo}, H. and {Siudek}, M. and {Sprayberry}, D. and {Tan}, T. and {Tang}, Ji-Jia and {Tarl{\'e}}, G. and {Walther}, M. and {Weaver}, B.~A. and {Yu}, J. and {Zhou}, R. and {Zou}, H.},
        title = "{DESI DR1 Ly{\ensuremath{\alpha}} 1D power spectrum: the optimal estimator measurement}",
      journal = {\jcap},
         year = 2025,
        month = oct,
       volume = {2025},
       number = {10},
          eid = {004},
        pages = {004},
          doi = {10.1088/1475-7516/2025/10/004},
archivePrefix = {arXiv},
       eprint = {2505.07974},
 primaryClass = {astro-ph.CO},
       adsurl = {https://ui.adsabs.harvard.edu/abs/2025JCAP...10..004K}
}

@ARTICLE{Ravoux_2025,
       author = {{Ravoux}, Corentin and {Abdul-Karim}, Marie-Lynn and {Le Goff}, Jean-Marc and {Armengaud}, Eric and {Aguilar}, Jessica N. and {Ahlen}, Steven and {Bailey}, Stephen and {Bianchi}, Davide and {Brodzeller}, Allyson and {Brooks}, David and {Chaves-Montero}, Jon{\'a}s and {Claybaugh}, Todd and {Cuceu}, Andrei and {de Belsunce}, Roger and {de la Macorra}, Axel and {Dey}, Arjun and {Ding}, Zhejie and {Doel}, Peter and {Ferraro}, Simone and {Font-Ribera}, Andreu and {Forero-Romero}, Jaime E. and {Gazta{\~n}aga}, Enrique and {Kara{\c{c}}ayl{\i}}, Naim G{\"o}ksel and {Gontcho}, Satya Gontcho A. and {Gutierrez}, Gaston and {Guy}, Julien and {Herrera-Alcantar}, Hiram K. and {Ishak}, Mustapha and {Kehoe}, Robert and {Kirkby}, David and {Kisner}, Theodore and {Kremin}, Anthony and {Landriau}, Martin and {Le Guillou}, Laurent and {Levi}, Michael E. and {Manera}, Marc and {Martini}, Paul and {Meisner}, Aaron and {Miquel}, Ramon and {Montero-Camacho}, Paulo and {Mu{\~n}oz-Guti{\'e}rrez}, Andrea and {Nadathur}, Seshadri and {Niz}, Gustavo and {Palanque-Delabrouille}, Nathalie and {Pan}, Zhiwei and {Percival}, Will J. and {P{\'e}rez-R{\`a}fols}, Ignasi and {Pieri}, Matthew M. and {Prada}, Francisco and {Rossi}, Graziano and {Sanchez}, Eusebio and {Saulder}, Christoph and {Schlegel}, David and {Schubnell}, Michael and {Seo}, Hee-Jong and {Silber}, Joseph H. and {Siudek}, Ma{\l}gorzata and {Sprayberry}, David and {Tan}, Ting and {Tang}, Ji-Jia and {Tarl{\'e}}, Gregory and {Walther}, Michael and {Weaver}, Benjamin A. and {Y{\`e}che}, Christophe and {Yu}, Jiaxi and {Zhou}, Rongpu and {Zou}, Hu},
        title = "{DESI DR1 Ly{\ensuremath{\alpha}} 1D power spectrum: the Fast Fourier Transform estimator measurement}",
      journal = {\jcap},
         year = 2025,
        month = nov,
       volume = {2025},
       number = {11},
          eid = {079},
        pages = {079},
          doi = {10.1088/1475-7516/2025/11/079},
archivePrefix = {arXiv},
       eprint = {2505.09493},
 primaryClass = {astro-ph.CO},
       adsurl = {https://ui.adsabs.harvard.edu/abs/2025JCAP...11..079R}
}

@ARTICLE{Planck2018,
       author = {{Planck Collaboration} and {Aghanim}, N. and {Akrami}, Y. and
         {Ashdown}, M. and {Aumont}, J. and {Baccigalupi}, C. and
         {Ballardini}, M. and {Banday}, A.~J. and {Barreiro}, R.~B. and
         {Bartolo}, N. and {Basak}, S. and {Battye}, R. and {Benabed}, K. and
         {Bernard}, J. -P. and {Bersanelli}, M. and {Bielewicz}, P. and
         {Bock}, J.~J. and {Bond}, J.~R. and {Borrill}, J. and {Bouchet}, F.~R. and
         {Boulanger}, F. and {Bucher}, M. and {Burigana}, C. and
         {Butler}, R.~C. and {Calabrese}, E. and {Cardoso}, J. -F. and
         {Carron}, J. and {Challinor}, A. and {Chiang}, H.~C. and {Chluba}, J. and
         {Colombo}, L.~P.~L. and {Combet}, C. and {Contreras}, D. and
         {Crill}, B.~P. and {Cuttaia}, F. and {de Bernardis}, P. and
         {de Zotti}, G. and {Delabrouille}, J. and {Delouis}, J. -M. and
         {Di Valentino}, E. and {Diego}, J.~M. and {Dor{\'e}}, O. and
         {Douspis}, M. and {Ducout}, A. and {Dupac}, X. and {Dusini}, S. and
         {Efstathiou}, G. and {Elsner}, F. and {En{\ss}lin}, T.~A. and
         {Eriksen}, H.~K. and {Fantaye}, Y. and {Farhang}, M. and
         {Fergusson}, J. and {Fernandez-Cobos}, R. and {Finelli}, F. and
         {Forastieri}, F. and {Frailis}, M. and {Fraisse}, A.~A. and
         {Franceschi}, E. and {Frolov}, A. and {Galeotta}, S. and {Galli}, S. and
         {Ganga}, K. and {G{\'e}nova-Santos}, R.~T. and {Gerbino}, M. and
         {Ghosh}, T. and {Gonz{\'a}lez-Nuevo}, J. and {G{\'o}rski}, K.~M. and
         {Gratton}, S. and {Gruppuso}, A. and {Gudmundsson}, J.~E. and
         {Hamann}, J. and {Handley}, W. and {Hansen}, F.~K. and {Herranz}, D. and
         {Hildebrandt}, S.~R. and {Hivon}, E. and {Huang}, Z. and
         {Jaffe}, A.~H. and {Jones}, W.~C. and {Karakci}, A. and
         {Keih{\"a}nen}, E. and {Keskitalo}, R. and {Kiiveri}, K. and {Kim}, J. and
         {Kisner}, T.~S. and {Knox}, L. and {Krachmalnicoff}, N. and {Kunz}, M. and
         {Kurki-Suonio}, H. and {Lagache}, G. and {Lamarre}, J. -M. and
         {Lasenby}, A. and {Lattanzi}, M. and {Lawrence}, C.~R. and
         {Le Jeune}, M. and {Lemos}, P. and {Lesgourgues}, J. and {Levrier}, F. and
         {Lewis}, A. and {Liguori}, M. and {Lilje}, P.~B. and {Lilley}, M. and
         {Lindholm}, V. and {L{\'o}pez-Caniego}, M. and {Lubin}, P.~M. and
         {Ma}, Y. -Z. and {Mac{\'\i}as-P{\'e}rez}, J.~F. and {Maggio}, G. and
         {Maino}, D. and {Mandolesi}, N. and {Mangilli}, A. and
         {Marcos-Caballero}, A. and {Maris}, M. and {Martin}, P.~G. and
         {Martinelli}, M. and {Mart{\'\i}nez-Gonz{\'a}lez}, E. and
         {Matarrese}, S. and {Mauri}, N. and {McEwen}, J.~D. and
         {Meinhold}, P.~R. and {Melchiorri}, A. and {Mennella}, A. and
         {Migliaccio}, M. and {Millea}, M. and {Mitra}, S. and
         {Miville-Desch{\^e}nes}, M. -A. and {Molinari}, D. and {Montier}, L. and
         {Morgante}, G. and {Moss}, A. and {Natoli}, P. and
         {N{\o}rgaard-Nielsen}, H.~U. and {Pagano}, L. and {Paoletti}, D. and
         {Partridge}, B. and {Patanchon}, G. and {Peiris}, H.~V. and
         {Perrotta}, F. and {Pettorino}, V. and {Piacentini}, F. and
         {Polastri}, L. and {Polenta}, G. and {Puget}, J. -L. and
         {Rachen}, J.~P. and {Reinecke}, M. and {Remazeilles}, M. and
         {Renzi}, A. and {Rocha}, G. and {Rosset}, C. and {Roudier}, G. and
         {Rubi{\~n}o-Mart{\'\i}n}, J.~A. and {Ruiz-Granados}, B. and
         {Salvati}, L. and {Sandri}, M. and {Savelainen}, M. and {Scott}, D. and
         {Shellard}, E.~P.~S. and {Sirignano}, C. and {Sirri}, G. and
         {Spencer}, L.~D. and {Sunyaev}, R. and {Suur-Uski}, A. -S. and
         {Tauber}, J.~A. and {Tavagnacco}, D. and {Tenti}, M. and
         {Toffolatti}, L. and {Tomasi}, M. and {Trombetti}, T. and
         {Valenziano}, L. and {Valiviita}, J. and {Van Tent}, B. and
         {Vibert}, L. and {Vielva}, P. and {Villa}, F. and {Vittorio}, N. and {Wand
        elt}, B.~D. and {Wehus}, I.~K. and {White}, M. and {White}, S.~D.~M. and
         {Zacchei}, A. and {Zonca}, A.},
        title = "{Planck 2018 results. VI. Cosmological parameters}",
      journal = {\aap},
         year = 2020,
        month = sep,
       volume = {641},
          eid = {A6},
        pages = {A6},
          doi = {10.1051/0004-6361/201833910},
archivePrefix = {arXiv},
       eprint = {1807.06209},
 primaryClass = {astro-ph.CO},
       adsurl = {https://ui.adsabs.harvard.edu/abs/2020A&A...641A...6P}
      }

@ARTICLE{PerezRafols2018,
       author = {{P{\'e}rez-R{\`a}fols}, Ignasi and {Font-Ribera}, Andreu and {Miralda-Escud{\'e}}, Jordi and {Blomqvist}, Michael and {Bird}, Simeon and {Busca}, Nicol{\'a}s and {du Mas des Bourboux}, H{\'e}lion and {Mas-Ribas}, Llu{\'\i}s and {Noterdaeme}, Pasquier and {Petitjean}, Patrick and {Rich}, James and {Schneider}, Donald P.},
        title = "{The SDSS-DR12 large-scale cross-correlation of damped Lyman alpha systems with the Lyman alpha forest}",
      journal = {\mnras},
         year = 2018,
        month = jan,
       volume = {473},
       number = {3},
        pages = {3019-3038},
          doi = {10.1093/mnras/stx2525},
archivePrefix = {arXiv},
       eprint = {1709.00889},
 primaryClass = {astro-ph.CO},
       adsurl = {https://ui.adsabs.harvard.edu/abs/2018MNRAS.473.3019P}
}

@ARTICLE{DESI.DR2.DR2,
       author = {{DESI Collaboration}},
        title = "{DESI DR2: Data Release 2 of the Dark Energy Spectroscopic Instrument}",
         year = 2026,
      journal = {in preparation}
}

@ARTICLE{DESI.DR2.BAO.cosmo,
       author = {{Abdul Karim}, M. and {Aguilar}, J. and {Ahlen}, S. and {Alam}, S. and {Allen}, L. and {Allende Prieto}, C. and {Alves}, O. and {Anand}, A. and {Andrade}, U. and {Armengaud}, E. and {Aviles}, A. and {Bailey}, S. and {Baltay}, C. and {Bansal}, P. and {Bault}, A. and {Behera}, J. and {BenZvi}, S. and {Bianchi}, D. and {Blake}, C. and {Brieden}, S. and {Brodzeller}, A. and {Brooks}, D. and {Buckley-Geer}, E. and {Burtin}, E. and {Calderon}, R. and {Canning}, R. and {Rosell}, A. Carnero and {Carrilho}, P. and {Casas}, L. and {Castander}, F.~J. and {Charles}, M. and {Chaussidon}, E. and {Chaves-Montero}, J. and {Chebat}, D. and {Chen}, X. and {Claybaugh}, T. and {Cole}, S. and {Cooper}, A.~P. and {Cuceu}, A. and {Dawson}, K.~S. and {de la Macorra}, A. and {de Mattia}, A. and {Deiosso}, N. and {Della Costa}, J. and {Demina}, R. and {Dey}, A. and {Dey}, B. and {Ding}, Z. and {Doel}, P. and {Edelstein}, J. and {Eisenstein}, D.~J. and {Elbers}, W. and {Fagrelius}, P. and {Fanning}, K. and {Fern{\'a}ndez-Garc{\'\i}a}, E. and {Ferraro}, S. and {Font-Ribera}, A. and {Forero-Romero}, J.~E. and {Frenk}, C.~S. and {Garcia-Quintero}, C. and {Garrison}, L.~H. and {Gazta{\~n}aga}, E. and {Gil-Mar{\'\i}n}, H. and {Gontcho A Gontcho}, S. and {Gonzalez}, D. and {Gonzalez-Morales}, A.~X. and {Gordon}, C. and {Green}, D. and {Gutierrez}, G. and {Guy}, J. and {Hadzhiyska}, B. and {Hahn}, C. and {He}, S. and {Herbold}, M. and {Herrera-Alcantar}, H.~K. and {Ho}, M.-F. and {Honscheid}, K. and {Howlett}, C. and {Huterer}, D. and {Ishak}, M. and {Juneau}, S. and {Kamble}, N.~V. and {Kara{\c{c}}ayl{\i}}, N.~G. and {Kehoe}, R. and {Kent}, S. and {Kim}, A.~G. and {Kirkby}, D. and {Kisner}, T. and {Koposov}, S.~E. and {Kremin}, A. and {Krolewski}, A. and {Lahav}, O. and {Lamman}, C. and {Landriau}, M. and {Lang}, D. and {Lasker}, J. and {Le Goff}, J.~M. and {Le Guillou}, L. and {Leauthaud}, A. and {Levi}, M.~E. and {Li}, Q. and {Li}, T.~S. and {Lodha}, K. and {Lokken}, M. and {Lozano-Rodr{\'\i}guez}, F. and {Magneville}, C. and {Manera}, M. and {Martini}, P. and {Matthewson}, W.~L. and {Meisner}, A. and {Mena-Fern{\'a}ndez}, J. and {Menegas}, A. and {Mergulh{\~a}o}, T. and {Miquel}, R. and {Moustakas}, J. and {Mu{\~n}oz-Guti{\'e}rrez}, A. and {Mu{\~n}oz-Santos}, D. and {Myers}, A.~D. and {Nadathur}, S. and {Naidoo}, K. and {Napolitano}, L. and {Newman}, J.~A. and {Niz}, G. and {Noriega}, H.~E. and {Paillas}, E. and {Palanque-Delabrouille}, N. and {Pan}, J. and {Peacock}, J.~A. and {Pellejero Ibanez}, M. and {Percival}, W.~J. and {P{\'e}rez-Fern{\'a}ndez}, A. and {P{\'e}rez-R{\`a}fols}, I. and {Pieri}, M.~M. and {Poppett}, C. and {Prada}, F. and {Rabinowitz}, D. and {Raichoor}, A. and {Ram{\'\i}rez-P{\'e}rez}, C. and {Rashkovetskyi}, M. and {Ravoux}, C. and {Rich}, J. and {Rocher}, A. and {Rockosi}, C. and {Rohlf}, J. and {Rom{\'a}n-Herrera}, J.~O. and {Ross}, A.~J. and {Rossi}, G. and {Ruggeri}, R. and {Ruhlmann-Kleider}, V. and {Samushia}, L. and {Sanchez}, E. and {Sanders}, N. and {Schlegel}, D. and {Schubnell}, M. and {Seo}, H. and {Shafieloo}, A. and {Sharples}, R. and {Silber}, J. and {Sinigaglia}, F. and {Sprayberry}, D. and {Tan}, T. and {Tarl{\'e}}, G. and {Taylor}, P. and {Turner}, W. and {Ure{\~n}a-L{\'o}pez}, L.~A. and {Vaisakh}, R. and {Valdes}, F. and {Valogiannis}, G. and {Vargas-Maga{\~n}a}, M. and {Verde}, L. and {Walther}, M. and {Weaver}, B.~A. and {Weinberg}, D.~H. and {White}, M. and {Wolfson}, M. and {Y{\`e}che}, C. and {Yu}, J. and {Zaborowski}, E.~A. and {Zarrouk}, P. and {Zhai}, Z. and {Zhang}, H. and {Zhao}, C. and {Zhao}, G.~B. and {Zhou}, R. and {Zou}, H. and {DESI Collaboration}},
        title = "{DESI DR2 results. II. Measurements of baryon acoustic oscillations and cosmological constraints}",
      journal = {\prd},
         year = 2025,
        month = oct,
       volume = {112},
       number = {8},
          eid = {083515},
        pages = {083515},
          doi = {10.1103/tr6y-kpc6},
archivePrefix = {arXiv},
       eprint = {2503.14738},
 primaryClass = {astro-ph.CO},
       adsurl = {https://ui.adsabs.harvard.edu/abs/2025PhRvD.112h3515A}
}

@ARTICLE{DESI.DR2.BAO.lya,
       author = {{DESI Collaboration} and {Karim}, M. Abdul and {Aguilar}, J. and {Ahlen}, S. and {Allende Prieto}, C. and {Alves}, O. and {Anand}, A. and {Andrade}, U. and {Armengaud}, E. and {Aviles}, A. and {Bailey}, S. and {Bault}, A. and {BenZvi}, S. and {Bianchi}, D. and {Blake}, C. and {Brodzeller}, A. and {Brooks}, D. and {Buckley-Geer}, E. and {Burtin}, E. and {Calderon}, R. and {Canning}, R. and {Carnero Rosell}, A. and {Carrilho}, P. and {Casas}, L. and {Castander}, F.~J. and {Cereskaite}, R. and {Charles}, M. and {Chaussidon}, E. and {Chaves-Montero}, J. and {Chebat}, D. and {Claybaugh}, T. and {Cole}, S. and {Cooper}, A.~P. and {Cuceu}, A. and {Dawson}, K.~S. and {de Belsunce}, R. and {de la Macorra}, A. and {de Mattia}, A. and {Deiosso}, N. and {Della Costa}, J. and {Dey}, A. and {Dey}, B. and {Ding}, Z. and {Doel}, P. and {Edelstein}, J. and {Eisenstein}, D.~J. and {Elbers}, W. and {Fagrelius}, P. and {Fanning}, K. and {Ferraro}, S. and {Font-Ribera}, A. and {Forero-Romero}, J.~E. and {Garcia-Quintero}, C. and {Garrison}, L.~H. and {Gazta{\~n}aga}, E. and {Gil-Mar{\'\i}n}, H. and {Gontcho}, S. Gontcho A and {Gonzalez-Morales}, A.~X. and {Gordon}, C. and {Green}, D. and {Gutierrez}, G. and {Guy}, J. and {Hahn}, C. and {Herbold}, M. and {Herrera-Alcantar}, H.~K. and {Ho}, M. and {Honscheid}, K. and {Howlett}, C. and {Huterer}, D. and {Ishak}, M. and {Juneau}, S. and {Kara{\c{c}}ayl{\i}}, N.~G. and {Kehoe}, R. and {Kent}, S. and {Kirkby}, D. and {Kisner}, T. and {Kitaura}, F. -S. and {Koposov}, S.~E. and {Kremin}, A. and {Lahav}, O. and {Lamman}, C. and {Landriau}, M. and {Lang}, D. and {Lasker}, J. and {Le Goff}, J.~M. and {Le Guillou}, L. and {Leauthaud}, A. and {Levi}, M.~E. and {Li}, Q. and {Li}, T.~S. and {Lodha}, K. and {Lokken}, M. and {Magneville}, C. and {Manera}, M. and {Martini}, P. and {Matthewson}, W. and {McDonald}, P. and {Meisner}, A. and {Mena-Fern{\'a}ndez}, J. and {Miquel}, R. and {Moustakas}, J. and {Mu{\~n}oz-Guti{\'e}rrez}, A. and {Mu{\~n}oz-Santos}, D. and {Myers}, A.~D. and {Newman}, J.~A. and {Niz}, G. and {Noriega}, H.~E. and {Paillas}, E. and {Palanque-Delabrouille}, N. and {Pan}, J. and {Percival}, W.~J. and {P{\'e}rez-R{\`a}fols}, I. and {Pieri}, M.~M. and {Poppett}, C. and {Prada}, F. and {Rabinowitz}, D. and {Raichoor}, A. and {Ram{\'\i}rez-P{\'e}rez}, C. and {Rashkovetskyi}, M. and {Ravoux}, C. and {Rich}, J. and {Rockosi}, C. and {Ross}, A.~J. and {Rossi}, G. and {Ruhlmann-Kleider}, V. and {Sanchez}, E. and {Sanders}, N. and {Satyavolu}, S. and {Schlegel}, D. and {Schubnell}, M. and {Seo}, H. and {Shafieloo}, A. and {Sharples}, R. and {Silber}, J. and {Sinigaglia}, F. and {Sprayberry}, D. and {Tan}, T. and {Tarl{\'e}}, G. and {Taylor}, P. and {Turner}, W. and {Valdes}, F. and {Vargas-Maga{\~n}a}, M. and {Walther}, M. and {Weaver}, B.~A. and {Wolfson}, M. and {Y{\`e}che}, C. and {Zarrouk}, P. and {Zhou}, R. and {Zou}, H.},
        title = "{DESI DR2 results. I. Baryon acoustic oscillations from the Lyman alpha forest}",
      journal = {\prd},
         year = 2025,
        month = oct,
       volume = {112},
       number = {8},
          eid = {083514},
        pages = {083514},
          doi = {10.1103/2wwn-xjm5},
archivePrefix = {arXiv},
       eprint = {2503.14739},
 primaryClass = {astro-ph.CO},
       adsurl = {https://ui.adsabs.harvard.edu/abs/2025PhRvD.112h3514A}
}

@ARTICLE{Casas2025,
       author = {{Casas}, L. and {Herrera-Alcantar}, H.~K. and {Chaves-Montero}, J. and {Cuceu}, A. and {Font-Ribera}, A. and {Lokken}, M. and {Abdul-Karim}, M. and {Ram{\'\i}rez-P{\'e}rez}, C. and {Alonso}, D. and {Aguilar}, J. and {Ahlen}, S. and {Andrade}, U. and {Armengaud}, E. and {Aviles}, A. and {Bailey}, S. and {BenZvi}, S. and {Bianchi}, D. and {Brodzeller}, A. and {Brooks}, D. and {Canning}, R. and {Rosell}, A. Carnero and {Charles}, M. and {Chaussidon}, E. and {Claybaugh}, T. and {Dawson}, K.~S. and {de la Macorra}, A. and {de Mattia}, A. and {Dey}, Arjun and {Dey}, Biprateep and {Ding}, Z. and {Doel}, P. and {Eisenstein}, D.~J. and {Elbers}, W. and {Ferraro}, S. and {Forero-Romero}, J.~E. and {Garcia-Quintero}, C. and {Garrison}, Lehman H. and {Gazta{\~n}aga}, E. and {Gil-Mar{\'\i}n}, H. and {Gontcho}, S. Gontcho A. and {Gonzalez-Morales}, A.~X. and {Gordon}, C. and {Gutierrez}, G. and {Guy}, J. and {Herbold}, M. and {Honscheid}, K. and {Howlett}, C. and {Huterer}, D. and {Ishak}, M. and {Juneau}, S. and {Kehoe}, R. and {Kirkby}, D. and {Kisner}, T. and {Kremin}, A. and {Lahav}, O. and {Landriau}, M. and {Le Goff}, J.~M. and {Le Guillou}, L. and {Leauthaud}, A. and {Levi}, M.~E. and {Li}, Q. and {Manera}, M. and {Martini}, P. and {Meisner}, A. and {Mena-Fern{\'a}ndez}, J. and {Miquel}, R. and {Moustakas}, J. and {Santos}, D. Mu{\~n}oz and {Myers}, A.~D. and {Nadathur}, S. and {Napolitano}, L. and {Niz}, G. and {Noriega}, H.~E. and {Paillas}, E. and {Palanque-Delabrouille}, N. and {Percival}, W.~J. and {Pieri}, Matthew M. and {Poppett}, C. and {Prada}, F. and {P{\'e}rez-R{\`a}fols}, I. and {Ravoux}, C. and {Rossi}, G. and {Sanchez}, E. and {Schlegel}, D. and {Schubnell}, M. and {Seo}, H. and {Sinigaglia}, F. and {Sprayberry}, D. and {Tan}, T. and {Tarl{\'e}}, G. and {Taylor}, P. and {Turner}, W. and {Vargas-Maga{\~n}a}, M. and {Walther}, M. and {Weaver}, B.~A. and {Wolfson}, M. and {Y{\`e}che}, C. and {Zarrouk}, P. and {Zhou}, R. and {DESI Collaboration}},
        title = "{Validation of the DESI DR2 Ly{\ensuremath{\alpha}} BAO analysis using synthetic datasets}",
      collaboration = {DESI Collaboration},
      journal = {\prd},
      volume = {113},
      issue = {2},
      pages = {023520},
      numpages = {24},
      year = {2026},
      month = {Jan},
      publisher = {American Physical Society},
      doi = {10.1103/fvgh-kswf},
      url = {https://link.aps.org/doi/10.1103/fvgh-kswf},
      eid = {023520},
      archivePrefix = {arXiv},
      eprint = {2503.14741},
      primaryClass = {astro-ph.IM}
}

@ARTICLE{brodzeller2025,
       author = {{Brodzeller}, A. and {Wolfson}, M. and {Santos}, D.~M. and {Ho}, M. and {Tan}, T. and {Pieri}, M.~M. and {Cuceu}, A. and {Abdul-Karim}, M. and {Aguilar}, J. and {Ahlen}, S. and {Anand}, A. and {Andrade}, U. and {Armengaud}, E. and {Aviles}, A. and {Bailey}, S. and {Bault}, A. and {Bianchi}, D. and {Brooks}, D. and {Canning}, R. and {Casas}, L. and {Charles}, M. and {Chaussidon}, E. and {Chaves-Montero}, J. and {Chebat}, D. and {Claybaugh}, T. and {Dawson}, K.~S. and {de Belsunce}, R. and {de la Macorra}, A. and {de Mattia}, A. and {Dey}, Arjun and {Dey}, Biprateep and {Doel}, P. and {Doshi}, M. and {Elbers}, W. and {Ferraro}, S. and {Font-Ribera}, A. and {Forero-Romero}, J.~E. and {Garcia-Quintero}, C. and {Garrison}, L.~H. and {Gazta{\~n}aga}, E. and {Gontcho A Gontcho}, S. and {Gonzalez-Morales}, A.~X. and {Green}, D. and {Gutierrez}, G. and {Guy}, J. and {Hahn}, C. and {Herbold}, M. and {Herrera-Alcantar}, H.~K. and {Honscheid}, K. and {Howlett}, C. and {Huterer}, D. and {Ishak}, M. and {Juneau}, S. and {Kehoe}, R. and {Kisner}, T. and {Kremin}, A. and {Lahav}, O. and {Lamman}, C. and {Landriau}, M. and {Le Goff}, J.~M. and {Le Guillou}, L. and {Leauthaud}, A. and {Levi}, M.~E. and {Li}, Q. and {Manera}, M. and {Martini}, P. and {Meisner}, A. and {Mena-Fern{\'a}ndez}, J. and {Miquel}, R. and {Moustakas}, J. and {Mu{\~n}oz-Guti{\'e}rrez}, A. and {Myers}, A.~D. and {Nadathur}, S. and {Napolitano}, L. and {Noriega}, H.~E. and {Paillas}, E. and {Palanque-Delabrouille}, N. and {Percival}, W.~J. and {Poppett}, C. and {Prada}, F. and {P{\'e}rez-R{\`a}fols}, I. and {Ram{\'\i}rez-P{\'e}rez}, C. and {Ravoux}, C. and {Rohlf}, J. and {Rossi}, G. and {Sanchez}, E. and {Schlegel}, D. and {Schubnell}, M. and {Sinigaglia}, F. and {Sprayberry}, D. and {Tarl{\'e}}, G. and {Taylor}, P. and {Turner}, W. and {Walther}, M. and {Weaver}, B.~A. and {Y{\`e}che}, C. and {Zhou}, R. and {Zou}, H. and {Zou}, S. and {DESI Collaboration}},
        title = "{Construction of the damped Ly{\ensuremath{\alpha}} absorber catalog for DESI DR2 Ly{\ensuremath{\alpha}} BAO}",
      journal = {\prd},
         year = 2025,
        month = oct,
       volume = {112},
       number = {8},
          eid = {083510},
        pages = {083510},
          doi = {10.1103/wxyv-46kb},
archivePrefix = {arXiv},
       eprint = {2503.14740},
 primaryClass = {astro-ph.CO},
       adsurl = {https://ui.adsabs.harvard.edu/abs/2025PhRvD.112h3510B}
}

@ARTICLE{DESI2022.KP1.Instr,
       author = {{DESI Collaboration} and {Abareshi}, B. and {Aguilar}, J. and {Ahlen}, S. and {Alam}, Shadab and {Alexander}, David M. and {Alfarsy}, R. and {Allen}, L. and {Allende Prieto}, C. and {Alves}, O. and {Ameel}, J. and {Armengaud}, E. and {Asorey}, J. and {Aviles}, Alejandro and {Bailey}, S. and {Balaguera-Antol{\'\i}nez}, A. and {Ballester}, O. and {Baltay}, C. and {Bault}, A. and {Beltran}, S.~F. and {Benavides}, B. and {BenZvi}, S. and {Berti}, A. and {Besuner}, R. and {Beutler}, Florian and {Bianchi}, D. and {Blake}, C. and {Blanc}, P. and {Blum}, R. and {Bolton}, A. and {Bose}, S. and {Bramall}, D. and {Brieden}, S. and {Brodzeller}, A. and {Brooks}, D. and {Brownewell}, C. and {Buckley-Geer}, E. and {Cahn}, R.~N. and {Cai}, Z. and {Canning}, R. and {Capasso}, R. and {Carnero Rosell}, A. and {Carton}, P. and {Casas}, R. and {Castander}, F.~J. and {Cervantes-Cota}, J.~L. and {Chabanier}, S. and {Chaussidon}, E. and {Chuang}, C. and {Circosta}, C. and {Cole}, S. and {Cooper}, A.~P. and {da Costa}, L. and {Cousinou}, M. -C. and {Cuceu}, A. and {Davis}, T.~M. and {Dawson}, K. and {de la Cruz-Noriega}, R. and {de la Macorra}, A. and {de Mattia}, A. and {Della Costa}, J. and {Demmer}, P. and {Derwent}, M. and {Dey}, A. and {Dey}, B. and {Dhungana}, G. and {Ding}, Z. and {Dobson}, C. and {Doel}, P. and {Donald-McCann}, J. and {Donaldson}, J. and {Douglass}, K. and {Duan}, Y. and {Dunlop}, P. and {Edelstein}, J. and {Eftekharzadeh}, S. and {Eisenstein}, D.~J. and {Enriquez-Vargas}, M. and {Escoffier}, S. and {Evatt}, M. and {Fagrelius}, P. and {Fan}, X. and {Fanning}, K. and {Fawcett}, V.~A. and {Ferraro}, S. and {Ereza}, J. and {Flaugher}, B. and {Font-Ribera}, A. and {Forero-Romero}, J.~E. and {Frenk}, C.~S. and {Fromenteau}, S. and {G{\"a}nsicke}, B.~T. and {Garcia-Quintero}, C. and {Garrison}, L. and {Gazta{\~n}aga}, E. and {Gerardi}, F. and {Gil-Mar{\'\i}n}, H. and {Gontcho a Gontcho}, S. and {Gonzalez-Morales}, Alma X. and {Gonzalez-de-Rivera}, G. and {Gonzalez-Perez}, V. and {Gordon}, C. and {Graur}, O. and {Green}, D. and {Grove}, C. and {Gruen}, D. and {Gutierrez}, G. and {Guy}, J. and {Hahn}, C. and {Harris}, S. and {Herrera}, D. and {Herrera-Alcantar}, Hiram K. and {Honscheid}, K. and {Howlett}, C. and {Huterer}, D. and {Ir{\v{s}}i{\v{c}}}, V. and {Ishak}, M. and {Jelinsky}, P. and {Jiang}, L. and {Jimenez}, J. and {Jing}, Y.~P. and {Joyce}, R. and {Jullo}, E. and {Juneau}, S. and {Kara{\c{c}}ayl{\i}}, N.~G. and {Karamanis}, M. and {Karcher}, A. and {Karim}, T. and {Kehoe}, R. and {Kent}, S. and {Kirkby}, D. and {Kisner}, T. and {Kitaura}, F. and {Koposov}, S.~E. and {Kov{\'a}cs}, A. and {Kremin}, A. and {Krolewski}, Alex and {L'Huillier}, B. and {Lahav}, O. and {Lambert}, A. and {Lamman}, C. and {Lan}, Ting-Wen and {Landriau}, M. and {Lane}, S. and {Lang}, D. and {Lange}, J.~U. and {Lasker}, J. and {Le Guillou}, L. and {Leauthaud}, A. and {Le Van Suu}, A. and {Levi}, Michael E. and {Li}, T.~S. and {Magneville}, C. and {Manera}, M. and {Manser}, Christopher J. and {Marshall}, B. and {Martini}, Paul and {McCollam}, W. and {McDonald}, P. and {Meisner}, Aaron M. and {Mena-Fern{\'a}ndez}, J. and {Meneses-Rizo}, J. and {Mezcua}, M. and {Miller}, T. and {Miquel}, R. and {Montero-Camacho}, P. and {Moon}, J. and {Moustakas}, J. and {Mueller}, E. and {Mu{\~n}oz-Guti{\'e}rrez}, Andrea and {Myers}, Adam D. and {Nadathur}, S. and {Najita}, J. and {Napolitano}, L. and {Neilsen}, E. and {Newman}, Jeffrey A. and {Nie}, J.~D. and {Ning}, Y. and {Niz}, G. and {Norberg}, P. and {Noriega}, Hern{\'a}n E. and {O'Brien}, T. and {Obuljen}, A. and {Palanque-Delabrouille}, N. and {Palmese}, A. and {Zhiwei}, P. and {Pappalardo}, D. and {PENG}, X. and {Percival}, W.~J. and {Perruchot}, S. and {Pogge}, R. and {Poppett}, C. and {Porredon}, A. and {Prada}, F. and {Prochaska}, J. and {Pucha}, R. and {P{\'e}rez-Fern{\'a}ndez}, A. and {P{\'e}rez-R{\`a}fols}, I. and {Rabinowitz}, D. and {Raichoor}, A. and {Ramirez-Solano}, S. and {Ram{\'\i}rez-P{\'e}rez}, C{\'e}sar and {Ravoux}, C. and {Reil}, K. and {Rezaie}, M. and {Rocher}, A. and {Rockosi}, C. and {Roe}, N.~A. and {Roodman}, A. and {Ross}, A.~J. and {Rossi}, G. and {Ruggeri}, R. and {Ruhlmann-Kleider}, V. and {Sabiu}, C.~G. and {Safonova}, S. and {Said}, K. and {Saintonge}, A. and {Salas Catonga}, Javier and {Samushia}, L. and {Sanchez}, E. and {Saulder}, C. and {Schaan}, E. and {Schlafly}, E. and {Schlegel}, D. and {Schmoll}, J. and {Scholte}, D. and {Schubnell}, M. and {Secroun}, A. and {Seo}, H. and {Serrano}, S. and {Sharples}, Ray M. and {Sholl}, Michael J. and {Silber}, Joseph Harry and {Silva}, D.~R. and {Sirk}, M. and {Siudek}, M. and {Smith}, A. and {Sprayberry}, D. and {Staten}, R. and {Stupak}, B. and {Tan}, T. and {Tarl{\'e}}, Gregory and {Tie}, Suk Sien and {Tojeiro}, R. and {Ure{\~n}a-L{\'o}pez}, L.~A. and {Valdes}, F. and {Valenzuela}, O. and {Valluri}, M. and {Vargas-Maga{\~n}a}, M. and {Verde}, L. and {Walther}, M. and {Wang}, B. and {Wang}, M.~S. and {Weaver}, B.~A. and {Weaverdyck}, C. and {Wechsler}, R. and {Wilson}, Michael J. and {Yang}, J. and {Yu}, Y. and {Yuan}, S. and {Y{\`e}che}, Christophe and {Zhang}, H. and {Zhang}, K. and {Zhao}, Cheng and {Zhou}, Rongpu and {Zhou}, Zhimin and {Zou}, H. and {Zou}, J. and {Zou}, S. and {Zu}, Y. and {DESI Collaboration}},
        title = "{Overview of the Instrumentation for the Dark Energy Spectroscopic Instrument}",
      journal = {\aj},
         year = 2022,
        month = nov,
       volume = {164},
       number = {5},
          eid = {207},
        pages = {207},
          doi = {10.3847/1538-3881/ac882b},
archivePrefix = {arXiv},
       eprint = {2205.10939},
 primaryClass = {astro-ph.IM},
       adsurl = {https://ui.adsabs.harvard.edu/abs/2022AJ....164..207A}
}

@ARTICLE{DESI2023a.KP1.SV,
       author = {{DESI Collaboration} and {Adame}, A.~G. and {Aguilar}, J. and {Ahlen}, S. and {Alam}, S. and {Aldering}, G. and {Alexander}, D.~M. and {Alfarsy}, R. and {Allende Prieto}, C. and {Alvarez}, M. and {Alves}, O. and {Anand}, A. and {Andrade-Oliveira}, F. and {Armengaud}, E. and {Asorey}, J. and {Avila}, S. and {Aviles}, A. and {Bailey}, S. and {Balaguera-Antol{\'\i}nez}, A. and {Ballester}, O. and {Baltay}, C. and {Bault}, A. and {Bautista}, J. and {Behera}, J. and {Beltran}, S.~F. and {BenZvi}, S. and {Beraldo e Silva}, L. and {Bermejo-Climent}, J.~R. and {Berti}, A. and {Besuner}, R. and {Beutler}, F. and {Bianchi}, D. and {Blake}, C. and {Blum}, R. and {Bolton}, A.~S. and {Brieden}, S. and {Brodzeller}, A. and {Brooks}, D. and {Brown}, Z. and {Buckley-Geer}, E. and {Burtin}, E. and {Cabayol-Garcia}, L. and {Cai}, Z. and {Canning}, R. and {Cardiel-Sas}, L. and {Carnero Rosell}, A. and {Castander}, F.~J. and {Cervantes-Cota}, J.~L. and {Chabanier}, S. and {Chaussidon}, E. and {Chaves-Montero}, J. and {Chen}, S. and {Chen}, X. and {Chuang}, C. and {Claybaugh}, T. and {Cole}, S. and {Cooper}, A.~P. and {Cuceu}, A. and {Davis}, T.~M. and {Dawson}, K. and {de Belsunce}, R. and {de la Cruz}, R. and {de la Macorra}, A. and {de Mattia}, A. and {Demina}, R. and {Demirbozan}, U. and {DeRose}, J. and {Dey}, A. and {Dey}, B. and {Dhungana}, G. and {Ding}, J. and {Ding}, Z. and {Doel}, P. and {Doshi}, R. and {Douglass}, K. and {Edge}, A. and {Eftekharzadeh}, S. and {Eisenstein}, D.~J. and {Elliott}, A. and {Escoffier}, S. and {Fagrelius}, P. and {Fan}, X. and {Fanning}, K. and {Fawcett}, V.~A. and {Ferraro}, S. and {Ereza}, J. and {Flaugher}, B. and {Font-Ribera}, A. and {Forero-S{\'a}nchez}, D. and {Forero-Romero}, J.~E. and {Frenk}, C.~S. and {G{\"a}nsicke}, B.~T. and {Garc{\'\i}a}, L. {\'A}. and {Garc{\'\i}a-Bellido}, J. and {Garcia-Quintero}, C. and {Garrison}, L.~H. and {Gil-Mar{\'\i}n}, H. and {Golden-Marx}, J. and {Gontcho A Gontcho}, S. and {Gonzalez-Morales}, A.~X. and {Gonzalez-Perez}, V. and {Gordon}, C. and {Graur}, O. and {Green}, D. and {Gruen}, D. and {Guy}, J. and {Hadzhiyska}, B. and {Hahn}, C. and {Han}, J.~J. and {Hanif}, M.~M.~S. and {Herrera-Alcantar}, H.~K. and {Honscheid}, K. and {Hou}, J. and {Howlett}, C. and {Huterer}, D. and {Ir{\v{s}}i{\v{c}}}, V. and {Ishak}, M. and {Jana}, A. and {Jiang}, L. and {Jimenez}, J. and {Jing}, Y.~P. and {Joudaki}, S. and {Jullo}, E. and {Joyce}, R. and {Juneau}, S. and {Kizhuprakkat}, N. and {Kara{\c{c}}ayl{\i}}, N.~G. and {Karim}, T. and {Kehoe}, R. and {Kent}, S. and {Khederlarian}, A. and {Kim}, S. and {Kirkby}, D. and {Kisner}, T. and {Kitaura}, F. and {Kneib}, J. and {Koposov}, S.~E. and {Kov{\'a}cs}, A. and {Kremin}, A. and {Krolewski}, A. and {L'Huillier}, B. and {Lahav}, O. and {Lambert}, A. and {Lamman}, C. and {Lan}, T. -W. and {Landriau}, M. and {Lang}, D. and {Lange}, J.~U. and {Lasker}, J. and {Le Guillou}, L. and {Leauthaud}, A. and {Levi}, M.~E. and {Li}, T.~S. and {Linder}, E. and {Lyons}, A. and {Magneville}, C. and {Manera}, M. and {Manser}, C.~J. and {Margala}, D. and {Martini}, P. and {McDonald}, P. and {Medina}, G.~E. and {Medina-Varela}, L. and {Meisner}, A. and {Mena-Fern{\'a}ndez}, J. and {Meneses-Rizo}, J. and {Mezcua}, M. and {Miquel}, R. and {Montero-Camacho}, P. and {Moon}, J. and {Moore}, S. and {Moustakas}, J. and {Mueller}, E. and {Mundet}, J. and {Mu{\~n}oz-Guti{\'e}rrez}, A. and {Myers}, A.~D. and {Nadathur}, S. and {Napolitano}, L. and {Neveux}, R. and {Newman}, J.~A. and {Nie}, J. and {Niz}, G. and {Norberg}, P. and {Noriega}, H.~E. and {Paillas}, E. and {Palanque-Delabrouille}, N. and {Palmese}, A. and {Zhiwei}, P. and {Parkinson}, D. and {Penmetsa}, S. and {Percival}, W.~J. and {P{\'e}rez-Fern{\'a}ndez}, A. and {P{\'e}rez-R{\`a}fols}, I. and {Pieri}, M. and {Poppett}, C. and {Porredon}, A. and {Prada}, F. and {Pucha}, R. and {Raichoor}, A. and {Ram{\'\i}rez-P{\'e}rez}, C. and {Ramirez-Solano}, S. and {Rashkovetskyi}, M. and {Ravoux}, C. and {Rocher}, A. and {Rockosi}, C. and {Ross}, A.~J. and {Rossi}, G. and {Ruggeri}, R. and {Ruhlmann-Kleider}, V. and {Sabiu}, C.~G. and {Said}, K. and {Saintonge}, A. and {Samushia}, L. and {Sanchez}, E. and {Saulder}, C. and {Schaan}, E. and {Schlafly}, E.~F. and {Schlegel}, D. and {Scholte}, D. and {Schubnell}, M. and {Seo}, H. and {Shafieloo}, A. and {Sharples}, R. and {Sheu}, W. and {Silber}, J. and {Sinigaglia}, F. and {Siudek}, M. and {Slepian}, Z. and {Smith}, A. and {Sprayberry}, D. and {Stephey}, L. and {Su{\'a}rez-P{\'e}rez}, J. and {Sun}, Z. and {Tan}, T. and {Tarl{\'e}}, G. and {Tojeiro}, R. and {Ure{\~n}a-L{\'o}pez}, L.~A. and {Vaisakh}, R. and {Valcin}, D. and {Valdes}, F. and {Valluri}, M. and {Vargas-Maga{\~n}a}, M. and {Variu}, A. and {Verde}, L. and {Walther}, M. and {Wang}, B. and {Wang}, M.~S. and {Weaver}, B.~A. and {Weaverdyck}, N. and {Wechsler}, R.~H. and {White}, M. and {Xie}, Y. and {Yang}, J. and {Y{\`e}che}, C. and {Yu}, J. and {Yuan}, S. and {Zhang}, H. and {Zhang}, Z. and {Zhao}, C. and {Zheng}, Z. and {Zhou}, R. and {Zhou}, Z. and {Zou}, H. and {Zou}, S. and {Zu}, Y. and {DESI Collaboration}},
        title = "{Validation of the Scientific Program for the Dark Energy Spectroscopic Instrument}",
      journal = {\aj},
         year = 2024,
        month = feb,
       volume = {167},
       number = {2},
          eid = {62},
        pages = {62},
          doi = {10.3847/1538-3881/ad0b08},
archivePrefix = {arXiv},
       eprint = {2306.06307},
 primaryClass = {astro-ph.CO},
       adsurl = {https://ui.adsabs.harvard.edu/abs/2024AJ....167...62A}
}

@ARTICLE{DESI2024.II.KP3,
       author = {{Adame}, A.~G. and {Aguilar}, J. and {Ahlen}, S. and {Alam}, S. and {Alexander}, D.~M. and {Alvarez}, M. and {Alves}, O. and {Anand}, A. and {Andrade}, U. and {Armengaud}, E. and {Avila}, S. and {Aviles}, A. and {Awan}, H. and {Bailey}, S. and {Baltay}, C. and {Bault}, A. and {Behera}, J. and {BenZvi}, S. and {Beutler}, F. and {Bianchi}, D. and {Blake}, C. and {Blum}, R. and {Brieden}, S. and {Brodzeller}, A. and {Brooks}, D. and {Brown}, Z. and {Buckley-Geer}, E. and {Burtin}, E. and {Calderon}, R. and {Canning}, R. and {Carnero Rosell}, A. and {Cereskaite}, R. and {Cervantes-Cota}, J.~L. and {Chabanier}, S. and {Chaussidon}, E. and {Chaves-Montero}, J. and {Chen}, S. and {Chen}, X. and {Claybaugh}, T. and {Cole}, S. and {Cuceu}, A. and {Davis}, T.~M. and {Dawson}, K. and {de la Macorra}, A. and {de Mattia}, A. and {Deiosso}, N. and {Demina}, R. and {Dey}, A. and {Dey}, B. and {Ding}, Z. and {Doel}, P. and {Edelstein}, J. and {Eftekharzadeh}, S. and {Eisenstein}, D.~J. and {Elliott}, A. and {Fagrelius}, P. and {Fanning}, K. and {Ferraro}, S. and {Ereza}, J. and {Findlay}, N. and {Flaugher}, B. and {Font-Ribera}, A. and {Forero-S{\'a}nchez}, D. and {Forero-Romero}, J.~E. and {Frenk}, C.~S. and {Garcia-Quintero}, C. and {Gazta{\~n}aga}, E. and {Gil-Mar{\'\i}n}, H. and {Gontcho}, S. Gontcho A. and {Gonzalez-Morales}, A.~X. and {Gonzalez-Perez}, V. and {Gordon}, C. and {Green}, D. and {Gruen}, D. and {Gsponer}, R. and {Gutierrez}, G. and {Guy}, J. and {Hadzhiyska}, B. and {Hahn}, C. and {Hanif}, M.~M.~S. and {Herrera-Alcantar}, H.~K. and {Honscheid}, K. and {Hou}, J. and {Howlett}, C. and {Huterer}, D. and {Ir{\v{s}}i{\v{c}}}, V. and {Ishak}, M. and {Juneau}, S. and {Kara{\c{c}}ayl{\i}}, N.~G. and {Kehoe}, R. and {Kent}, S. and {Kirkby}, D. and {Kitaura}, F.-S. and {Kong}, H. and {Kremin}, A. and {Krolewski}, A. and {Lai}, Y. and {Lan}, T.-W. and {Landriau}, M. and {Lang}, D. and {Lasker}, J. and {Le Goff}, J.~M. and {Le Guillou}, L. and {Leauthaud}, A. and {Levi}, M.~E. and {Li}, T.~S. and {Lodha}, K. and {Magneville}, C. and {Manera}, M. and {Margala}, D. and {Martini}, P. and {Maus}, M. and {McDonald}, P. and {Medina-Varela}, L. and {Meisner}, A. and {Mena-Fern{\'a}ndez}, J. and {Miquel}, R. and {Moon}, J. and {Moore}, S. and {Moustakas}, J. and {Mudur}, N. and {Mueller}, E. and {Mu{\~n}oz-Guti{\'e}rrez}, A. and {Myers}, A.~D. and {Nadathur}, S. and {Napolitano}, L. and {Neveux}, R. and {Newman}, J.~A. and {Nguyen}, N.~M. and {Nie}, J. and {Niz}, G. and {Noriega}, H.~E. and {Padmanabhan}, N. and {Paillas}, E. and {Palanque-Delabrouille}, N. and {Pan}, J. and {Penmetsa}, S. and {Percival}, W.~J. and {Pieri}, M.~M. and {Pinon}, M. and {Poppett}, C. and {Porredon}, A. and {Prada}, F. and {P{\'e}rez-Fern{\'a}ndez}, A. and {P{\'e}rez-R{\`a}fols}, I. and {Rabinowitz}, D. and {Raichoor}, A. and {Ram{\'\i}rez-P{\'e}rez}, C. and {Ramirez-Solano}, S. and {Rashkovetskyi}, M. and {Ravoux}, C. and {Rezaie}, M. and {Rich}, J. and {Rocher}, A. and {Rockosi}, C. and {Roe}, N.~A. and {Rosado-Marin}, A. and {Ross}, A.~J. and {Rossi}, G. and {Ruggeri}, R. and {Ruhlmann-Kleider}, V. and {Samushia}, L. and {Sanchez}, E. and {Saulder}, C. and {Schlafly}, E.~F. and {Schlegel}, D. and {Scholte}, D. and {Schubnell}, M. and {Seo}, H. and {Sharples}, R. and {Silber}, J. and {Slosar}, A. and {Smith}, A. and {Sprayberry}, D. and {Tan}, T. and {Tarl{\'e}}, G. and {Trusov}, S. and {Vaisakh}, R. and {Valcin}, D. and {Valdes}, F. and {Vargas-Maga{\~n}a}, M. and {Verde}, L. and {Walther}, M. and {Wang}, B. and {Wang}, M.~S. and {Weaver}, B.~A. and {Weaverdyck}, N. and {Wechsler}, R.~H. and {Weinberg}, D.~H. and {White}, M. and {Wilson}, M.~J. and {Yu}, J. and {Yu}, Y. and {Yuan}, S. and {Y{\`e}che}, C. and {Zaborowski}, E.~A. and {Zarrouk}, P. and {Zhang}, H. and {Zhao}, C. and {Zhao}, R.},
        title = "{DESI 2024 II: sample definitions, characteristics, and two-point clustering statistics}",
      journal = {\jcap},
         year = 2025,
        month = jul,
       volume = {2025},
       number = {7},
          eid = {017},
        pages = {017},
          doi = {10.1088/1475-7516/2025/07/017},
archivePrefix = {arXiv},
       eprint = {2411.12020},
 primaryClass = {astro-ph.CO},
       adsurl = {https://ui.adsabs.harvard.edu/abs/2025JCAP...07..017A}
}

@ARTICLE{DESI2024.III.KP4,
      author={{DESI Collaboration} and A. G. Adame and J. Aguilar and S. Ahlen and S. Alam and D. M. Alexander and M. Alvarez and O. Alves and A. Anand and U. Andrade and E. Armengaud and S. Avila and A. Aviles and H. Awan and S. Bailey and C. Baltay and A. Bault and J. Behera and S. BenZvi and F. Beutler and D. Bianchi and C. Blake and R. Blum and S. Brieden and A. Brodzeller and D. Brooks and E. Buckley-Geer and E. Burtin and R. Calderon and R. Canning and A. Carnero Rosell and R. Cereskaite and J. L. Cervantes-Cota and S. Chabanier and E. Chaussidon and J. Chaves-Montero and S. Chen and X. Chen and T. Claybaugh and S. Cole and A. Cuceu and T. M. Davis and K. Dawson and A. de la Macorra and A. de Mattia and N. Deiosso and A. Dey and B. Dey and Z. Ding and P. Doel and J. Edelstein and S. Eftekharzadeh and D. J. Eisenstein and A. Elliott and P. Fagrelius and K. Fanning and S. Ferraro and J. Ereza and N. Findlay and B. Flaugher and A. Font-Ribera and D. Forero-Sánchez and J. E. Forero-Romero and C. Garcia-Quintero and E. Gaztañaga and H. Gil-Marín and S. Gontcho A Gontcho and A. X. Gonzalez-Morales and V. Gonzalez-Perez and C. Gordon and D. Green and D. Gruen and R. Gsponer and G. Gutierrez and J. Guy and B. Hadzhiyska and C. Hahn and M. M. S Hanif and H. K. Herrera-Alcantar and K. Honscheid and C. Howlett and D. Huterer and V. Iršič and M. Ishak and S. Juneau and N. G. Karaçaylı and R. Kehoe and S. Kent and D. Kirkby and A. Kremin and A. Krolewski and Y. Lai and T. -W. Lan and M. Landriau and D. Lang and J. Lasker and J. M. Le Goff and L. Le Guillou and A. Leauthaud and M. E. Levi and T. S. Li and E. Linder and K. Lodha and C. Magneville and M. Manera and D. Margala and P. Martini and M. Maus and P. McDonald and L. Medina-Varela and A. Meisner and J. Mena-Fernández and R. Miquel and J. Moon and S. Moore and J. Moustakas and N. Mudur and E. Mueller and A. Muñoz-Gutiérrez and A. D. Myers and S. Nadathur and L. Napolitano and R. Neveux and J. A. Newman and N. M. Nguyen and J. Nie and G. Niz and H. E. Noriega and N. Padmanabhan and E. Paillas and N. Palanque-Delabrouille and J. Pan and S. Penmetsa and W. J. Percival and M. Pieri and M. Pinon and C. Poppett and A. Porredon and F. Prada and A. Pérez-Fernández and I. Pérez-Ràfols and D. Rabinowitz and A. Raichoor and C. Ramírez-Pérez and S. Ramirez-Solano and M. Rashkovetskyi and M. Rezaie and J. Rich and A. Rocher and C. Rockosi and N. A. Roe and A. Rosado-Marin and A. J. Ross and G. Rossi and R. Ruggeri and V. Ruhlmann-Kleider and L. Samushia and E. Sanchez and C. Saulder and E. F. Schlafly and D. Schlegel and M. Schubnell and H. Seo and R. Sharples and J. Silber and A. Slosar and A. Smith and D. Sprayberry and J. Swanson and T. Tan and G. Tarlé and S. Trusov and R. Vaisakh and D. Valcin and F. Valdes and M. Vargas-Magaña and L. Verde and M. Walther and B. Wang and M. S. Wang and B. A. Weaver and N. Weaverdyck and R. H. Wechsler and D. H. Weinberg and M. White and J. Yu and Y. Yu and S. Yuan and C. Yèche and E. A. Zaborowski and P. Zarrouk and H. Zhang and C. Zhao and R. Zhao and R. Zhou and H. Zou},
        title = "{DESI 2024 III: baryon acoustic oscillations from galaxies and quasars}",
      journal = {\jcap},
         year = 2025,
        month = apr,
       volume = {2025},
       number = {4},
          eid = {012},
        pages = {012},
          doi = {10.1088/1475-7516/2025/04/012},
archivePrefix = {arXiv},
       eprint = {2404.03000},
 primaryClass = {astro-ph.CO},
       adsurl = {https://ui.adsabs.harvard.edu/abs/2025JCAP...04..012A}
}

@ARTICLE{DESI2024.V.KP5,
       author = {{Adame}, A.~G. and {Aguilar}, J. and {Ahlen}, S. and {Alam}, S. and {Alexander}, D.~M. and {Alvarez}, M. and {Alves}, O. and {Anand}, A. and {Andrade}, U. and {Armengaud}, E. and {Avila}, S. and {Aviles}, A. and {Awan}, H. and {Bailey}, S. and {Baltay}, C. and {Bault}, A. and {Behera}, J. and {BenZvi}, S. and {Beutler}, F. and {Bianchi}, D. and {Blake}, C. and {Blum}, R. and {Brieden}, S. and {Brodzeller}, A. and {Brooks}, D. and {Buckley-Geer}, E. and {Burtin}, E. and {Calderon}, R. and {Canning}, R. and {Carnero Rosell}, A. and {Cereskaite}, R. and {Cervantes-Cota}, J.~L. and {Chabanier}, S. and {Chaussidon}, E. and {Chaves-Montero}, J. and {Chen}, S. and {Chen}, X. and {Claybaugh}, T. and {Cole}, S. and {Cuceu}, A. and {Davis}, T.~M. and {Dawson}, K. and {de la Macorra}, A. and {de Mattia}, A. and {Deiosso}, N. and {Dey}, A. and {Dey}, B. and {Ding}, Z. and {Doel}, P. and {Edelstein}, J. and {Eftekharzadeh}, S. and {Eisenstein}, D.~J. and {Elliott}, A. and {Fagrelius}, P. and {Fanning}, K. and {Ferraro}, S. and {Ereza}, J. and {Findlay}, N. and {Flaugher}, B. and {Font-Ribera}, A. and {Forero-S{\'a}nchez}, D. and {Forero-Romero}, J.~E. and {Garcia-Quintero}, C. and {Garrison}, L.~H. and {Gazta{\~n}aga}, E. and {Gil-Mar{\'\i}n}, H. and {Gontcho}, S. Gontcho A. and {Gonzalez-Morales}, A.~X. and {Gonzalez-Perez}, V. and {Gordon}, C. and {Green}, D. and {Gruen}, D. and {Gsponer}, R. and {Gutierrez}, G. and {Guy}, J. and {Hadzhiyska}, B. and {Hahn}, C. and {Hanif}, M.~M.~S. and {Herrera-Alcantar}, H.~K. and {Honscheid}, K. and {Howlett}, C. and {Huterer}, D. and {Ir{\v{s}}i{\v{c}}}, V. and {Ishak}, M. and {Juneau}, S. and {Kara{\c{c}}ayl{\i}}, N.~G. and {Kehoe}, R. and {Kent}, S. and {Kirkby}, D. and {Kong}, H. and {Koposov}, S.~E. and {Kremin}, A. and {Krolewski}, A. and {Lai}, Y. and {Lan}, T.-W. and {Landriau}, M. and {Lang}, D. and {Lasker}, J. and {Le Goff}, J.~M. and {Le Guillou}, L. and {Leauthaud}, A. and {Levi}, M.~E. and {Li}, T.~S. and {Lodha}, K. and {Magneville}, C. and {Manera}, M. and {Margala}, D. and {Martini}, P. and {Maus}, M. and {McDonald}, P. and {Medina-Varela}, L. and {Meisner}, A. and {Mena-Fern{\'a}ndez}, J. and {Miquel}, R. and {Moon}, J. and {Moore}, S. and {Moustakas}, J. and {Mueller}, E. and {Mu{\~n}oz-Guti{\'e}rrez}, A. and {Myers}, A.~D. and {Nadathur}, S. and {Napolitano}, L. and {Neveux}, R. and {Newman}, J.~A. and {Nguyen}, N.~M. and {Nie}, J. and {Niz}, G. and {Noriega}, H.~E. and {Padmanabhan}, N. and {Paillas}, E. and {Palanque-Delabrouille}, N. and {Pan}, J. and {Penmetsa}, S. and {Percival}, W.~J. and {Pieri}, M.~M. and {Pinon}, M. and {Poppett}, C. and {Porredon}, A. and {Prada}, F. and {P{\'e}rez-Fern{\'a}ndez}, A. and {P{\'e}rez-R{\`a}fols}, I. and {Rabinowitz}, D. and {Raichoor}, A. and {Ram{\'\i}rez-P{\'e}rez}, C. and {Ramirez-Solano}, S. and {Rashkovetskyi}, M. and {Ravoux}, C. and {Rezaie}, M. and {Rich}, J. and {Rocher}, A. and {Rockosi}, C. and {Rodr{\'\i}guez-Mart{\'\i}nez}, F. and {Roe}, N.~A. and {Rosado-Marin}, A. and {Ross}, A.~J. and {Rossi}, G. and {Ruggeri}, R. and {Ruhlmann-Kleider}, V. and {Samushia}, L. and {Sanchez}, E. and {Saulder}, C. and {Schlafly}, E.~F. and {Schlegel}, D. and {Schubnell}, M. and {Seo}, H. and {Sharples}, R. and {Silber}, J. and {Slosar}, A. and {Smith}, A. and {Sprayberry}, D. and {Tan}, T. and {Tarl{\'e}}, G. and {Trusov}, S. and {Vaisakh}, R. and {Valcin}, D. and {Valdes}, F. and {Vargas-Maga{\~n}a}, M. and {Verde}, L. and {Walther}, M. and {Wang}, B. and {Wang}, M.~S. and {Weaver}, B.~A. and {Weaverdyck}, N. and {Wechsler}, R.~H. and {Weinberg}, D.~H. and {White}, M. and {Wilson}, M.~J. and {Yu}, J. and {Yu}, Y. and {Yuan}, S. and {Y{\`e}che}, C. and {Zaborowski}, E.~A. and {Zarrouk}, P. and {Zhang}, H. and {Zhao}, C. and {Zhao}, R. and {Zhou}, R. and {Zou}, H. and {The DESI collaboration}},
        title = "{DESI 2024 V: Full-Shape galaxy clustering from galaxies and quasars}",
      journal = {\jcap},
         year = 2025,
        month = sep,
       volume = {2025},
       number = {9},
          eid = {008},
        pages = {008},
          doi = {10.1088/1475-7516/2025/09/008},
archivePrefix = {arXiv},
       eprint = {2411.12021},
 primaryClass = {astro-ph.CO},
       adsurl = {https://ui.adsabs.harvard.edu/abs/2025JCAP...09..008A}
}

@ARTICLE{DESI2024.VI.KP7A,
      author={{DESI Collaboration} and A. G. Adame and J. Aguilar and S. Ahlen and S. Alam and D. M. Alexander and M. Alvarez and O. Alves and A. Anand and U. Andrade and E. Armengaud and S. Avila and A. Aviles and H. Awan and B. Bahr-Kalus and S. Bailey and C. Baltay and A. Bault and J. Behera and S. BenZvi and A. Bera and F. Beutler and D. Bianchi and C. Blake and R. Blum and S. Brieden and A. Brodzeller and D. Brooks and E. Buckley-Geer and E. Burtin and R. Calderon and R. Canning and A. Carnero Rosell and R. Cereskaite and J. L. Cervantes-Cota and S. Chabanier and E. Chaussidon and J. Chaves-Montero and S. Chen and X. Chen and T. Claybaugh and S. Cole and A. Cuceu and T. M. Davis and K. Dawson and A. de la Macorra and A. de Mattia and N. Deiosso and A. Dey and B. Dey and Z. Ding and P. Doel and J. Edelstein and S. Eftekharzadeh and D. J. Eisenstein and A. Elliott and P. Fagrelius and K. Fanning and S. Ferraro and J. Ereza and N. Findlay and B. Flaugher and A. Font-Ribera and D. Forero-Sánchez and J. E. Forero-Romero and C. S. Frenk and C. Garcia-Quintero and E. Gaztañaga and H. Gil-Marín and S. Gontcho A Gontcho and A. X. Gonzalez-Morales and V. Gonzalez-Perez and C. Gordon and D. Green and D. Gruen and R. Gsponer and G. Gutierrez and J. Guy and B. Hadzhiyska and C. Hahn and M. M. S Hanif and H. K. Herrera-Alcantar and K. Honscheid and C. Howlett and D. Huterer and V. Iršič and M. Ishak and S. Juneau and N. G. Karaçaylı and R. Kehoe and S. Kent and D. Kirkby and A. Kremin and A. Krolewski and Y. Lai and T. -W. Lan and M. Landriau and D. Lang and J. Lasker and J. M. Le Goff and L. Le Guillou and A. Leauthaud and M. E. Levi and T. S. Li and E. Linder and K. Lodha and C. Magneville and M. Manera and D. Margala and P. Martini and M. Maus and P. McDonald and L. Medina-Varela and A. Meisner and J. Mena-Fernández and R. Miquel and J. Moon and S. Moore and J. Moustakas and N. Mudur and E. Mueller and A. Muñoz-Gutiérrez and A. D. Myers and S. Nadathur and L. Napolitano and R. Neveux and J. A. Newman and N. M. Nguyen and J. Nie and G. Niz and H. E. Noriega and N. Padmanabhan and E. Paillas and N. Palanque-Delabrouille and J. Pan and S. Penmetsa and W. J. Percival and M. Pieri and M. Pinon and C. Poppett and A. Porredon and F. Prada and A. Pérez-Fernández and I. Pérez-Ràfols and D. Rabinowitz and A. Raichoor and C. Ramírez-Pérez and S. Ramirez-Solano and M. Rashkovetskyi and M. Rezaie and J. Rich and A. Rocher and C. Rockosi and N. A. Roe and A. Rosado-Marin and A. J. Ross and G. Rossi and R. Ruggeri and V. Ruhlmann-Kleider and L. Samushia and E. Sanchez and C. Saulder and E. F. Schlafly and D. Schlegel and M. Schubnell and H. Seo and A. Shafieloo and R. Sharples and J. Silber and A. Slosar and A. Smith and D. Sprayberry and T. Tan and G. Tarlé and P. Taylor and S. Trusov and L. A. Ureña-López and R. Vaisakh and D. Valcin and F. Valdes and M. Vargas-Magaña and L. Verde and M. Walther and B. Wang and M. S. Wang and B. A. Weaver and N. Weaverdyck and R. H. Wechsler and D. H. Weinberg and M. White and J. Yu and Y. Yu and S. Yuan and C. Yèche and E. A. Zaborowski and P. Zarrouk and H. Zhang and C. Zhao and R. Zhao and R. Zhou and T. Zhuang and H. Zou},
        title = "{DESI 2024 VI: cosmological constraints from the measurements of baryon acoustic oscillations}",
      journal = {\jcap},
         year = 2025,
        month = feb,
       volume = {2025},
       number = {2},
          eid = {021},
        pages = {021},
          doi = {10.1088/1475-7516/2025/02/021},
archivePrefix = {arXiv},
       eprint = {2404.03002},
 primaryClass = {astro-ph.CO},
       adsurl = {https://ui.adsabs.harvard.edu/abs/2025JCAP...02..021A}
}

@ARTICLE{KP6s5-Guy,
       author = {{Guy}, J. and {Gontcho}, S. Gontcho A. and {Armengaud}, E. and {Brodzeller}, A. and {Cuceu}, A. and {Font-Ribera}, A. and {Herrera-Alcantar}, H.~K. and {Kara{\c{c}}ayl{\i}}, N.~G. and {Mu{\~n}oz-Guti{\'e}rrez}, A. and {Pieri}, M.~M. and {P{\'e}rez-R{\`a}fols}, I. and {Ram{\'\i}rez-P{\'e}rez}, C. and {Ravoux}, C. and {Rich}, J. and {Walther}, M. and {Abdul Karim}, M. and {Aguilar}, J. and {Ahlen}, S. and {Bault}, A. and {Brooks}, D. and {Claybaugh}, T. and {de la Cruz}, R. and {de la Macorra}, A. and {Doel}, P. and {Fanning}, K. and {Forero-Romero}, J.~E. and {Gazta{\~n}aga}, E. and {Gonzalez-Morales}, A.~X. and {Gutierrez}, G. and {Hahn}, C. and {Honscheid}, K. and {Juneau}, S. and {Kehoe}, R. and {Kirkby}, D. and {Kisner}, T. and {Kremin}, A. and {Lambert}, A. and {Landriau}, M. and {Le Guillou}, L. and {Manera}, M. and {Martini}, P. and {Meisner}, A. and {Miquel}, R. and {Montero-Camacho}, P. and {Moustakas}, J. and {Mueller}, E. and {Myers}, A.~D. and {Nie}, J. and {Niz}, G. and {Palanque-Delabrouille}, N. and {Percival}, W.~J. and {Poppett}, C. and {Rezaie}, M. and {Rossi}, G. and {Sanchez}, E. and {Schlegel}, D. and {Schubnell}, M. and {Seo}, H. and {Silber}, J. and {Sprayberry}, D. and {Tan}, T. and {Tarl{\'e}}, G. and {Vargas-Maga{\~n}a}, M. and {Zou}, H.},
        title = "{Characterization of contaminants in the Lyman-alpha forest auto-correlation with DESI}",
      journal = {\jcap},
         year = 2025,
        month = jan,
       volume = {2025},
       number = {1},
          eid = {140},
        pages = {140},
          doi = {10.1088/1475-7516/2025/01/140},
archivePrefix = {arXiv},
       eprint = {2404.03003},
 primaryClass = {astro-ph.CO},
       adsurl = {https://ui.adsabs.harvard.edu/abs/2025JCAP...01..140G}
}

@ARTICLE{KP6s9-Martini,
       author = {{Martini}, P. and {Cuceu}, A. and {Ennesser}, L. and {Brodzeller}, A. and {Aguilar}, J. and {Ahlen}, S. and {Brooks}, D. and {Claybaugh}, T. and {de Belsunce}, R. and {de la Macorra}, A. and {Dey}, Arjun and {Doel}, P. and {Forero-Romero}, J.~E. and {Gazta{\~n}aga}, E. and {Gontcho}, S. Gontcho A. and {Guy}, J. and {Herrera-Alcantar}, H.~K. and {Honscheid}, K. and {Kara{\c{c}}ayl{\i}}, N.~G. and {Kisner}, T. and {Kremin}, A. and {Lambert}, A. and {Le Guillou}, L. and {Manera}, M. and {Meisner}, A. and {Miquel}, R. and {Montero-Camacho}, P. and {Moustakas}, J. and {Niz}, G. and {Palanque-Delabrouille}, N. and {Percival}, W.~J. and {P{\'e}rez-R{\`a}fols}, I. and {Poppett}, C. and {Prada}, F. and {Ravoux}, C. and {Rezaie}, M. and {Rossi}, G. and {Sanchez}, E. and {Schlegel}, D. and {Schubnell}, M. and {Seo}, H. and {Sprayberry}, D. and {Tan}, T. and {Tarl{\'e}}, G. and {Walther}, M. and {Weaver}, B.~A. and {Zou}, H.},
        title = "{Validation of the DESI 2024 Lyman alpha forest BAL masking strategy}",
      journal = {\jcap},
         year = 2025,
        month = jan,
       volume = {2025},
       number = {1},
          eid = {137},
        pages = {137},
          doi = {10.1088/1475-7516/2025/01/137},
archivePrefix = {arXiv},
       eprint = {2405.09737},
 primaryClass = {astro-ph.CO},
       adsurl = {https://ui.adsabs.harvard.edu/abs/2025JCAP...01..137M}
}

@ARTICLE{Snowmass2013.Levi,
       author = {{Levi}, Michael and {Bebek}, Chris and {Beers}, Timothy and {Blum}, Robert and {Cahn}, Robert and {Eisenstein}, Daniel and {Flaugher}, Brenna and {Honscheid}, Klaus and {Kron}, Richard and {Lahav}, Ofer and {McDonald}, Patrick and {Roe}, Natalie and {Schlegel}, David and {representing the DESI collaboration}},
        title = "{The DESI Experiment, a whitepaper for Snowmass 2013}",
      journal = {arXiv e-prints},
         year = 2013,
        month = aug,
          eid = {arXiv:1308.0847},
        pages = {arXiv:1308.0847},
          doi = {10.48550/arXiv.1308.0847},
archivePrefix = {arXiv},
       eprint = {1308.0847},
 primaryClass = {astro-ph.CO},
       adsurl = {https://ui.adsabs.harvard.edu/abs/2013arXiv1308.0847L}
}

@ARTICLE{Corrector.Miller.2023,
       author = {{Miller}, Timothy N. and {Doel}, Peter and {Gutierrez}, Gaston and {Besuner}, Robert and {Brooks}, David and {Gallo}, Giuseppe and {Heetderks}, Henry and {Jelinsky}, Patrick and {Kent}, Stephen M. and {Lampton}, Michael and {Levi}, Michael E. and {Liang}, Ming and {Meisner}, Aaron and {Sholl}, Michael J. and {Silber}, Joseph Harry and {Sprayberry}, David and {Aguilar}, Jessica Nicole and {de la Macorra}, Axel and {Eisenstein}, Daniel and {Fanning}, Kevin and {Font-Ribera}, Andreu and {Gazta{\~n}aga}, Enrique and {Gontcho A Gontcho}, Satya and {Honscheid}, Klaus and {Jimenez}, Jorge and {Joyce}, Dick and {Kehoe}, Robert and {Kisner}, Theodore and {Kremin}, Anthony and {Landriau}, Martin and {Le Guillou}, Laurent and {Magneville}, Christophe and {Martini}, Paul and {Miquel}, Ramon and {Moustakas}, John and {Nie}, Jundan and {Percival}, Will and {Poppett}, Claire and {Prada}, Francisco and {Rossi}, Graziano and {Schlegel}, David and {Schubnell}, Michael and {Seo}, Hee-Jong and {Sharples}, Ray and {Tarl{\'e}}, Gregory and {Vargas-Maga{\~n}a}, Mariana and {Zhou}, Zhimin and {the DESI Collaboration}},
        title = "{The Optical Corrector for the Dark Energy Spectroscopic Instrument}",
      journal = {\aj},
         year = 2024,
        month = aug,
       volume = {168},
       number = {2},
          eid = {95},
        pages = {95},
          doi = {10.3847/1538-3881/ad45fe},
archivePrefix = {arXiv},
       eprint = {2306.06310},
 primaryClass = {astro-ph.IM},
       adsurl = {https://ui.adsabs.harvard.edu/abs/2024AJ....168...95M}
}

@ARTICLE{TS.Pipeline.Myers.2023,
       author = {{Myers}, Adam D. and {Moustakas}, John and {Bailey}, Stephen and {Weaver}, Benjamin A. and {Cooper}, Andrew P. and {Forero-Romero}, Jaime E. and {Abolfathi}, Bela and {Alexander}, David M. and {Brooks}, David and {Chaussidon}, Edmond and {Chuang}, Chia-Hsun and {Dawson}, Kyle and {Dey}, Arjun and {Dey}, Biprateep and {Dhungana}, Govinda and {Doel}, Peter and {Fanning}, Kevin and {Gazta{\~n}aga}, Enrique and {A Gontcho}, Satya Gontcho and {Gonzalez-Morales}, Alma X. and {Hahn}, ChangHoon and {Herrera-Alcantar}, Hiram K. and {Honscheid}, Klaus and {Ishak}, Mustapha and {Karim}, Tanveer and {Kirkby}, David and {Kisner}, Theodore and {Koposov}, Sergey E. and {Kremin}, Anthony and {Lan}, Ting-Wen and {Landriau}, Martin and {Lang}, Dustin and {Levi}, Michael E. and {Magneville}, Christophe and {Napolitano}, Lucas and {Martini}, Paul and {Meisner}, Aaron and {Newman}, Jeffrey A. and {Palanque-Delabrouille}, Nathalie and {Percival}, Will and {Poppett}, Claire and {Prada}, Francisco and {Raichoor}, Anand and {Ross}, Ashley J. and {Schlafly}, Edward F. and {Schlegel}, David and {Schubnell}, Michael and {Tan}, Ting and {Tarle}, Gregory and {Wilson}, Michael J. and {Y{\`e}che}, Christophe and {Zhou}, Rongpu and {Zhou}, Zhimin and {Zou}, Hu},
        title = "{The Target-selection Pipeline for the Dark Energy Spectroscopic Instrument}",
      journal = {\aj},
         year = 2023,
        month = feb,
       volume = {165},
       number = {2},
          eid = {50},
        pages = {50},
          doi = {10.3847/1538-3881/aca5f9},
archivePrefix = {arXiv},
       eprint = {2208.08518},
 primaryClass = {astro-ph.IM},
       adsurl = {https://ui.adsabs.harvard.edu/abs/2023AJ....165...50M}
}

@ARTICLE{Spectro.Pipeline.Guy.2023,
       author = {{Guy}, J. and {Bailey}, S. and {Kremin}, A. and {Alam}, Shadab and {Alexander}, D.~M. and {Allende Prieto}, C. and {BenZvi}, S. and {Bolton}, A.~S. and {Brooks}, D. and {Chaussidon}, E. and {Cooper}, A.~P. and {Dawson}, K. and {de la Macorra}, A. and {Dey}, A. and {Dey}, Biprateep and {Dhungana}, G. and {Eisenstein}, D.~J. and {Font-Ribera}, A. and {Forero-Romero}, J.~E. and {Gazta{\~n}aga}, E. and {Gontcho A Gontcho}, S. and {Green}, D. and {Honscheid}, K. and {Ishak}, M. and {Kehoe}, R. and {Kirkby}, D. and {Kisner}, T. and {Koposov}, Sergey E. and {Lan}, Ting-Wen and {Landriau}, M. and {Le Guillou}, L. and {Levi}, Michael E. and {Magneville}, C. and {Manser}, Christopher J. and {Martini}, P. and {Meisner}, Aaron M. and {Miquel}, R. and {Moustakas}, J. and {Myers}, Adam D. and {Newman}, Jeffrey A. and {Nie}, Jundan and {Palanque-Delabrouille}, N. and {Percival}, W.~J. and {Poppett}, C. and {Prada}, F. and {Raichoor}, A. and {Ravoux}, C. and {Ross}, A.~J. and {Schlafly}, E.~F. and {Schlegel}, D. and {Schubnell}, M. and {Sharples}, Ray M. and {Tarl{\'e}}, Gregory and {Weaver}, B.~A. and {Y{\'e}che}, Christophe and {Zhou}, Rongpu and {Zhou}, Zhimin and {Zou}, H.},
        title = "{The Spectroscopic Data Processing Pipeline for the Dark Energy Spectroscopic Instrument}",
      journal = {\aj},
         year = 2023,
        month = apr,
       volume = {165},
       number = {4},
          eid = {144},
        pages = {144},
          doi = {10.3847/1538-3881/acb212},
archivePrefix = {arXiv},
       eprint = {2209.14482},
 primaryClass = {astro-ph.IM},
       adsurl = {https://ui.adsabs.harvard.edu/abs/2023AJ....165..144G}
}

@ARTICLE{Wang2022,
       author = {{Wang}, Ben and {Zou}, Jiaqi and {Cai}, Zheng and {Prochaska}, J. Xavier and {Sun}, Zechang and {Ding}, Jiani and {Font-Ribera}, Andreu and {Gonzalez}, Alma and {Herrera-Alcantar}, Hiram K. and {Irsic}, Vid and {Lin}, Xiaojing and {Brooks}, David and {Chabanier}, Sol{\'e}ne and {de Belsunce}, Roger and {Palanque-Delabrouille}, Nathalie and {Tarle}, Gregory and {Zhou}, Zhimin},
        title = "{Deep Learning of Dark Energy Spectroscopic Instrument Mock Spectra to Find Damped Ly{\ensuremath{\alpha}} Systems}",
      journal = {\apjs},
         year = 2022,
        month = mar,
       volume = {259},
       number = {1},
          eid = {28},
        pages = {28},
          doi = {10.3847/1538-4365/ac4504},
       adsurl = {https://ui.adsabs.harvard.edu/abs/2022ApJS..259...28W}
      }

@misc{2006_dark_energy_report,
      title={Report of the Dark Energy Task Force}, 
      author={Andreas Albrecht and Gary Bernstein and Robert Cahn and Wendy L. Freedman and Jacqueline Hewitt and Wayne Hu and John Huth and Marc Kamionkowski and Edward W. Kolb and Lloyd Knox and John C. Mather and Suzanne Staggs and Nicholas B. Suntzeff},
      year={2006},
      eprint={astro-ph/0609591},
      archivePrefix={arXiv},
      primaryClass={astro-ph},
      url={https://arxiv.org/abs/astro-ph/0609591}, 
}

\end{document}